\documentclass[preprint,showpacs,preprintnumbers]{revtex4}
\usepackage{amssymb}
\usepackage{amsmath}
\usepackage{graphicx}
\usepackage{dcolumn}
\usepackage{bm}
\usepackage{epsfig}
\usepackage[T1]{fontenc}
\usepackage{ae,aecompl}
\usepackage{epstopdf, epsfig}
\usepackage{color}
\usepackage{appendix}
\usepackage[export]{adjustbox}
\usepackage{graphicx}
\usepackage{epstopdf, epsfig}
\usepackage[section]{placeins}
\usepackage[utf8]{inputenc}

\begin{document}

\title{Kinetics of Rayleigh-Taylor instability in van der Waals fluid: the influence of compressibility}
\author{Jie Chen$^{1,2}$, Aiguo Xu$^{2,3,4,5}$\footnote{
Corresponding author. E-mail: Xu\_Aiguo@iapcm.ac.cn},Yudong Zhang$^6$,  Dawei Chen$^2$, Zhihua Chen$^1$}
\affiliation{1, National Key Laboratory of Transient Physics, Nanjing University of Science and Technology, Nanjing 210094, China \\
2, National Key Laboratory of Computational Physics, Institute of Applied Physics and Computational Mathematics, P. O. Box 8009-26, Beijing 100088, China \\
3, State Key Laboratory of Explosion Science and Safety Protection, Beijing Institute of Technology, Beijing 100081, China \\
4, HEDPS, Center for Applied Physics and Technology, and College of Engineering, Peking University, Beijing 100871, China \\
5, National Key Laboratory of Shock Wave and Detonation Physics, Mianyang 621999, China \\
6, School of Mechanics and Safety Engineering, Zhengzhou University, Zhengzhou 450001, China }
\date{\today }

\begin{abstract}
Early studies on Rayleigh-Taylor instability (RTI) primarily relied on the Navier-Stokes (NS) model. As research progresses, it becomes increasingly evident that the kinetic information that the NS model failed to capture is of great value for identifying and even controlling the RTI process; simultaneously, the lack of analysis techniques for complex physical fields results in a significant waste of data information. In addition, early RTI studies mainly focused on the incompressible case and the weakly compressible case.
In the case of strong compressibility, the density of the fluid from the upper layer (originally heavy fluid) may become smaller than that of the surrounding (originally light) fluid, thus invalidating the early method of distinguishing light and heavy fluids based on density.  
In this paper, tracer particles are incorporated into a single-fluid discrete Boltzmann method (DBM) model that considers the van der Waals potential. By using tracer particles to label the matter-particle sources, a careful study of the matter-mixing and energy-mixing processes of the RTI evolution is realized in the single-fluid framework.
The effects of compressibility on the evolution of RTI are examined mainly through the analysis of bubble and spike velocities, the ratio of area occupied by heavy fluid, and various entropy generation rates of the system. 
It is demonstrated that: (\romannumeral 1) compressibility has a suppressive effect on the spike velocity, and this suppressive impact diminishes as the Atwood number ($At$) increases. The influence of compressibility on bubble velocity shows a staged behavior with increasing $At$. (\romannumeral 2) The impact of compressibility on the entropy production rate associated with the heat flow (${{\dot{S}}_{NOEF}}$) is related to the stages of RTI evolution. 
Moreover, this staged impact of compressibility on ${{\dot{S}}_{NOEF}}$ varies with $At$. Compressibility exhibits an inhibitory effect on the entropy production rate associated with viscous stresses (${{\dot{S}}_{NOMF}}$). (\romannumeral 3) By incorporating the morphological parameter of the proportion of area occupied by heavy fluid (${{A}_{h}}$), it is observed that the first minimum point of ${d{{A}_{h}}}/{dt}$ can serve as a criterion for identifying the point at which bubble velocity reaches its first maximum value. The series of physical cognition provides a more accurate understanding of the RTI kinetics and a helpful reference for the development of corresponding regulation techniques.

\end{abstract}

\maketitle

\preprint{APS/123-QED}

\section{Introduction}

The interface instability that arises when a heavy medium is supported or accelerated by a light-medium is known as Rayleigh-Taylor instability (RTI) ~\cite{rayleigh1882investigation,taylor1950instability}. RTI is found in a wide range of natural sciences and engineering applications, such as supernova explosions, inertial confinement fusion (ICF), and super combustion ram engines~\cite{zhou2024hydrodynamic,ZHOU20171,zhou2017rayleigh,zhou2021rayleigh}. In these processes, the flow is strongly compressible. However, most of the current research on RTI is based on the incompressible or weakly compressible case. At present, research on RTI in the case of strong compressibility remains limited, especially in the later stages of material and energy mixing. It is of great significance to study the influence of compressibility on RTI to control the evolution of interface instability in these processes.

Due to the importance of compressible RTI, this problem has attracted considerable attention from scholars, prompting extensive research in this field. However, the complexity of the issue, coupled with the evolving comprehension of the phenomenon, has resulted in varying or even opposite conclusions about the influence of compressibility on RTI. For example, in the early days, some scholars proposed that compressibility enhances RTI~\cite{Bernstein1983Effect,yang1993general}, while others argued that compressibility inhibits it~\cite{blake1972fluid}. Moreover, there are findings suggesting that both effects coexist~\cite{baker1983compressible}.

As research advances, it has become evident that a single parameter cannot determine the influence of compressibility on RTI. For instance, Livescu proposed that the impact of compressibility on RTI depends on the specific heat ratio and the equilibrium pressure at the interface, which have opposite effects on the development of RTI~\cite{livescu2004compressibility}. Xue and Ye proposed that the influence of compressibility on RTI depends on three factors: density profile, pressure at the interface, and the specific heat ratio~\cite{xue2010destabilizing}. In fact, compressibility in RTI research can be separated into two categories: fluid compressibility and flow compressibility~\cite{livescu2004compressibility, lafay2007compressibility, reckinger2016comprehensive, gauthier2017compressible}. The former is related to the inherent properties of fluids (associated with the equation of state and the specific heat ratio difference between the two fluids). The latter, or flow compressibility, is connected to the system's thermodynamic state. Its static effect leads to background stratification, while its dynamic effect results in an expansion-compression impact on the flow process~\cite{luo2020effects}.

Existing research indicates that flow compressibility shows different effects on bubbles depending on the Atwood numbers ($At$). There is a critical value of $At$, below which the development of RTI is inhibited by flow compressibility. Conversely, when $At$ exceeds this critical value, flow compressibility promotes the growth of RTI~\cite{reckinger2016comprehensive, luo2020effects, fu2022nonlinear}. Luo et al. and Fu et al. suggested that this is due to the competition between density stratification and expansion-compression effects~\cite{QinChengsen2001, QinChengsen2004, luo2020effects, fu2022nonlinear}. Fu et al. proposed a modified buoyancy-drag model for stratified compressible RTI. They predicted the critical $At$ using this model and obtained good agreement with results from direct numerical simulation (DNS). Simultaneously, they discovered that a similar nonlinear saturation is shown for the bubble evolution at $At=0.9$ when the pistonlike effect~\cite{Olson2007, reckinger2012simulations}, which characterizes the compressive forces exerted by the rising bubble of light fluid on the heavy fluid in front of it, is completely extracted. The piston effect of the rising bubble causes the acceleration of the heavy fluid compressed in front of the bubble at later stages to show an inverse power law of $A{{t}^{2.5}}M{{a}^{2}}$. For the RTI of compressible fluid, bubble reacceleration is caused not only by vorticity deposition in bubbles but also by compressibility. Under strong stratification and high $At$, flow compressibility dominates bubble reacceleration~\cite{fu2022nonlinear}. Using the vortex transport equation, Wieland et al. explained the asymmetric effect of weak stratification on bubble and spike growth at small $At$ and the inhibitory effect of strong stratification on RTI growth at small $At$~\cite{wieland2019effects}. By introducing dilatation into the classical model, which only considers vortex deposition, Fu et al. proposed a new model that can reliably describe the reacceleration of stratified compressible RTI~\cite{fu2023bubble}.

The evolution of hydrodynamic instabilities is a complex process, especially for those exhibiting strong compressibility. In previous studies, the evolution of the RTI system was often described by the velocity and amplitude of bubbles and spikes, as well as the mixing width. These physical quantities are very helpful and intuitive. However, it is far from enough to describe the evolution of the RTI system only based on these physical quantities. They lose a lot of information; that is to say, even if we know these physical quantities, there is still considerable uncertainty in the actual process of material and energy mixing. The incorporation of complex physical field analysis techniques, such as morphological quantities, can be a good complement to the description of RTI evolution. For example, Chen et al. found that the interface length of light and heavy fluid ($L$) can be used to measure the ratio of buoyancy to tangential force intensity and to quantitatively judge the primary mechanism in the early development of the coupled Rayleigh–Taylor–Kelvin–Helmholtz instability (RTKHI) system~\cite{chen2020morphological}. Actually, the RTI system is a complex, non-equilibrium system, and the physical quantities, such as the velocities and amplitudes of bubbles and spikes, as well as the length of the interface between light and heavy fluids, are only manifestations of the system's non-equilibrium from a specific perspective. Various perspectives on non-equilibrium evolution are interrelated and complementary, and they can form a more complete image of the system by combining each other.

In recent years, based on statistical physics, Xu et al. have established and developed the discrete Boltzmann method (DBM)~\cite{Xu2022-Complex, xu2024advances}. From a historical perspective, DBM is developed from the physical modeling branch of Lattice Boltzmann Method (LBM)~\cite{succi2001lattice, osborn1995lattice,swift1995lattice,liang2014phase,liang2016lattice,liang2021late}.
As a method of physical modeling and complex physical field analysis, DBM mainly focuses on constructing physical models and extracting valuable information from extensive datasets. In addition to the three conservation moments of mass, momentum, and energy, which were concerned in the Navier-Stokes (NS) model, DBM describes the evolution of complex systems from multiple perspectives with the help of complex physical field analysis methods, such as the evolution of some non-conservation moments closely related to system evolution and the morphological analysis. This enables DBM to provide some insight that NS methods cannot or are not convenient to obtain. In the study of hydrodynamic instabilities systems, a series of investigations based on DBM have contributed to our understanding of the physical processes and mechanisms involved in the development of hydrodynamic instabilities~\cite{chen2018collaboration, chen2020morphological, gan2019nonequilibrium,lai2016nonequilibrium,lin2017discrete}. In the study of compressible RTI, with the help of DBM, Lai et al. discovered that the global thermodynamic non-equilibrium (TNE) intensity shows opposite trends in the initial and later stages of the RTI~\cite{lai2016nonequilibrium}. The local TNE provided a useful physical observable value for tracking the interface of light and heavy fluids. Lin et al. proposed a DBM model for two-component compressible flows with an adjustable specific heat ratio and independent discrete velocity models for each component~\cite{lin2017discrete}. Using this model, they studied the invariants of tensors for nonequilibrium effects and the mixed entropy of compressible RTI systems under various Reynolds conditions. These studies describe the evolution of compressible RTI systems from different non-equilibrium perspectives, which are good supplements to the previous research results and help us to understand the physical process in the development of compressible RTI more deeply, but they are all weakly compressible.

In cases of strong compressibility, there may be some novel behaviors and mechanisms in the evolution of the RTI system. For example, \textit{the density of the fluid from the upper layer (which was originally heavy fluid) may become lower than that of the fluid around it (which was originally light fluid), thus causing the identity exchange of light and heavy fluids and triggering anti-RTI behavior. The appearance of these novel behaviors and mechanisms makes the early method of judging the source of material particles based on density invalid.} It is well known that particle imaging velocimeter (PIV) and planar laser-induced fluorescence (PLIF) technologies have been widely used in experimental fluid mechanics. \textit{Inspired by this, under the framework of a single fluid, we can incorporate tracer particles into the numerical simulation to identify and trace the source of material particles in the mixed region.} Zhang et al. introduced tracer particles into DBM for the first time. They studied the flow and mixing of compressible RTI, and the description method based on the velocity phase space of tracer particles opened up a new perspective for analyzing and profoundly understanding the flow system~\cite{zhang2021delineation}. Later, Li et al. extended the phase space description of tracer particles from simple velocity phase space to position-velocity phase space and studied the evolution of multimode RTI~\cite{li2022rayleigh}. The distribution of tracer particles in position space provides a clear interface for the evolution of the light-heavy fluid interface. The distribution of tracer particles in position-velocity phase space further opens a new perspective for studying complex flow systems. 

In previous studies, hydrodynamic instability problems have often been investigated by using ideal gas models to simplify the problems. However, in a real fluid, intermolecular interaction potentials are present. When the intermolecular interaction force is negligible compared to the external force acting on the medium during the physical processes, ideal gas models can help us capture the problem's primary characteristics and provide a relatively satisfactory answer. Unfortunately, in many cases, intermolecular interactions cannot be negligible compared to external forces. For example, many physical phenomena observed in processes like ICF, such as microjets, tensile failure, melting phase transitions, and perturbation stabilization, are all related to the presence of interaction forces between medium molecules or atoms~\cite{miles1966taylor, piriz2009linear, bi2020experimental}. In such instances, the effects of intermolecular interaction forces must be considered.
For compressible RTI systems, the presence of intermolecular interaction potential induces a modification in the fluid media state equation, which directly affects the effect of compressibility on RTI's evolution. Currently, research on this aspect is relatively limited due to its complexity. The van der Waals model represents the simplest form of intermolecular interaction potential.
As a preliminary study, this paper considers the van der Waals interaction potential and investigates the effect of compressibility on the evolution of the van der Waals fluid single-mode RTI system from a statistical physics perspective with the help of tracer particles and morphological analysis. This paper is structured as follows: Section II introduces the numerical method, Section III covers numerical simulation and result analysis, and Section IV presents the conclusion and discussion.

\section{NUMERCAL METHODS}

The DBM is a mesoscale physical model-building method and a complex physical field analysis method that was established and developed in recent years\cite{Xu2022-Complex, xu2024advances, 2015Progess,Xu2018-Chap2,xu2021progress,xu2021Progressofmesoscale,xu2021modeling,GanXuLai2022,Zhang2023SBI}. 
Unlike the NS method, DBM is based on the Boltzmann equation, making it convenient for DBM to describe the non-equilibrium behaviors of the system with more non-conservative moments and statistical mechanical quantities. At the same time, since it is not based on the assumption of a continuous medium and the higher-order non-equilibrium moment of the distribution function can be preserved as needed, DBM can study the near-continuous or discontinuous problems or higher-order non-equilibrium problems that the NS methods cannot or are not convenient to study.

It is worth pointing out that as the degree of discretization and non-equilibrium increases, the evolution of fluid instability systems is often accompanied by the emergence of small-scale structures and fast modes. The small-scale structure may lead to the failure of the continuous medium assumption. At this time, the validity of macroscopic hydrodynamic models such as NS is challenged.
The flow of small-scale structures often exhibits behavior that appears to be “anomalous” (as opposed to the macroscopic continuum), such as anomalies in heat conduction\cite{LiuZongh2011}. Although a wealth of progress has been made on heat and mass transfer in small systems\cite{Wang2010PRL}, the small systems extensively studied in the literature of statistical physics and nonlinear science are often much smaller than the whole hydrodynamic instability system. In the face of this dilemma, DBM can better demonstrate its advantages.

\begin{figure}[tbp]
\includegraphics[width=0.90\textwidth,trim=0.1 0.1 0.1 0.1,clip]{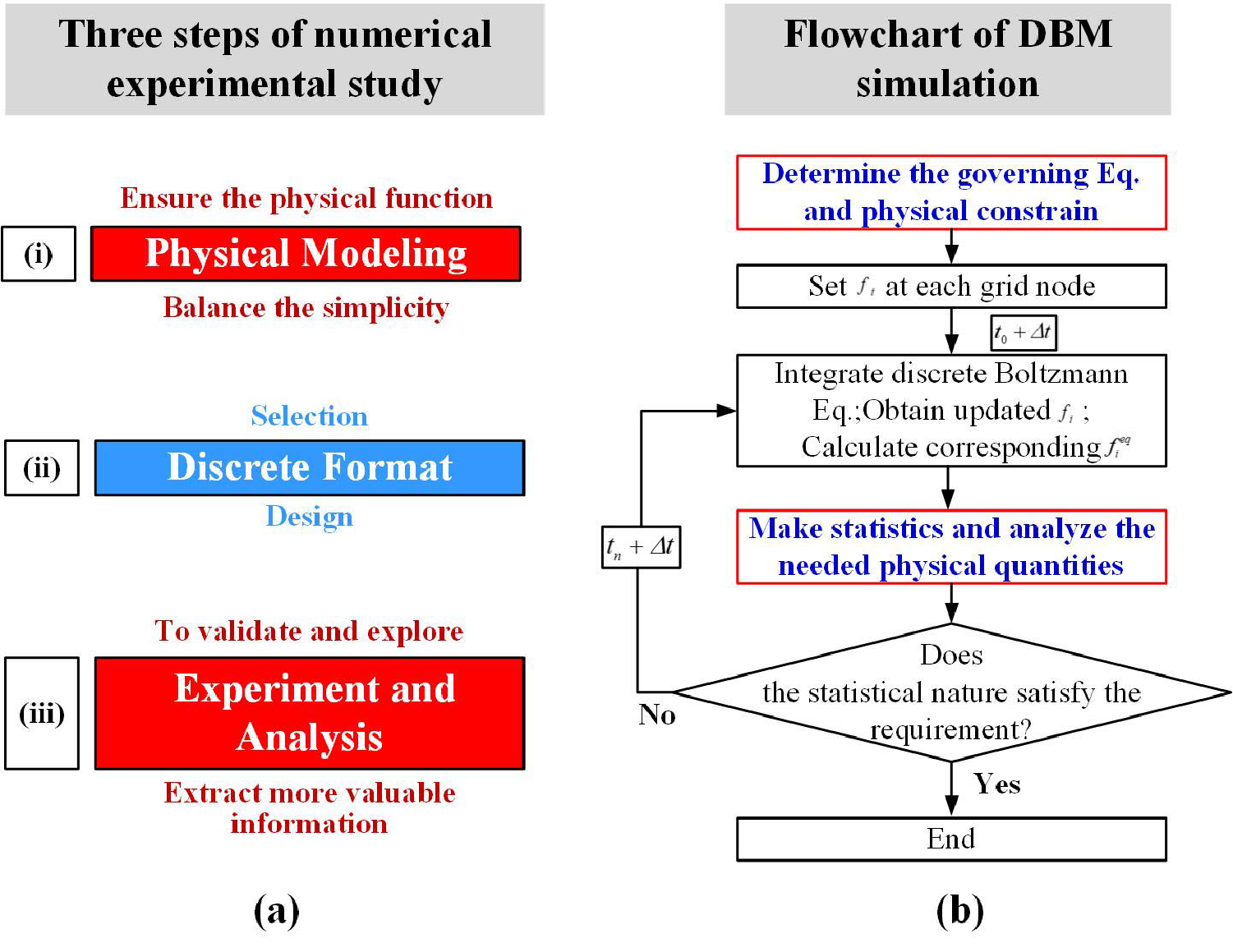}
\caption{(a) DBM simulation study and (b) Flowchart of DBM simulation.} \label{Fig1}
\end{figure}

It is known that, as illustrated in Fig. \ref{Fig1}(a), numerical simulation research can be divided into three parts: (\romannumeral 1) physical modeling; (\romannumeral 2) algorithm design; and (\romannumeral 3) numerical experimentation and complex physical field analysis. DBM mainly focuses on (\romannumeral 1) and (\romannumeral 3). For algorithm design, the focus of DBM is not to build and design new numerical formats and discrete methods for velocity but to select existing numerical formats and methods for discrete velocity as a user. The construction of a physical model is the basis of DBM. The purpose and core of DBM are to effectively extract and analyze massive amounts of information in complex physical fields. The DBM numerical simulation flow chart is shown in Fig. \ref{Fig1}(b).

\subsection{Physical modeling}

Unlike the NS method, the model established by DBM is a mesoscale model based on the Boltzmann equation. The original Boltzmann equation is a high-dimensional differential integral equation, as shown in Eq.(\ref{Eq.1})\cite{Xu2022-Complex}:
\begin{equation}
\frac{\partial f}{\partial t}+\mathbf{v}\cdot \frac{\partial f}{\partial \mathbf{r}}+\mathbf{a}\cdot \frac{\partial f}{\partial \mathbf{v}}=\int_{-\infty }^{+\infty }{\int_{0}^{4\pi }{\left( {{f}^{*}}f_{1}^{*}-f{{f}_{1}} \right){{v}_{r}}\sigma d\Omega d{{\mathbf{v}}_{1}}}}\text{.}  \label{Eq.1}
\end{equation}%

Evidently, the collision term in the original Boltzmann equation is quite complex, and the direct solution presents significant challenges. Furthermore, the original Boltzmann equation only considers the hardball collision between molecules, a premise suited only for ideal gases. Thus, there arises a necessity to modify and simplify the original Boltzmann equation for a convenient and efficient description of fluid systems. For the van der Waals fluid, by introducing the intermolecular interaction potential and simplifying the collision term with the BGK model, the Boltzmann equation of the van der Waals fluid system in BGK form can be obtained:
\begin{equation}
\frac{\partial f}{\partial t}+\mathbf{v}\cdot \frac{\partial f}{\partial \mathbf{r}}+\mathbf{a}\cdot \frac{\partial f}{\partial \mathbf{v}}=-\frac{1}{\tau }(f-{{f}^{eq}})+I\text{,}  \label{Eq.2}
\end{equation}%
where$f,t,\mathbf{v},\mathbf{r},\mathbf{a}$ and $\tau $ are molecular distribution function, time, velocity, spatial position, acceleration, and relaxation time, respectively. ${{f}^{eq}}=\rho {{\left( \frac{1}{2\pi RT} \right)}^{{D}/{2}\;}}\exp \left[ -\frac{{{\left( \mathbf{v}-\mathbf{u} \right)}^{2}}}{2RT} \right]$ is the Maxwell distribution function. $I=-\left[ A+\mathbf{B}\cdot \left( \mathbf{v}-\mathbf{u} \right)+\left( C+{{C}_{q}} \right){{\left| \mathbf{v}-\mathbf{u} \right|}^{2}} \right]{{f}^{eq}}$ represents the intermolecular interaction\cite{gonnella2007lattice, Xu2022-Complex}, where
\begin{equation}
A=-2\left( C\text{+}{{C}_{q}} \right)T\text{,}  \label{Eq.3}
\end{equation}%
\begin{equation}
\mathbf{B}=\frac{1}{\rho T}\nabla \cdot \left[ \left( P-\rho T \right)\mathbf{I}+{\bm{\Lambda}} \right]\text{,}  \label{Eq.4}
\end{equation}%
\begin{equation}
\begin{aligned}
 C=&\frac{1}{2\rho {{T}^{2}}}\left[ \left( P-\rho T \right)\nabla \cdot \mathbf{u} \right.+\bm{\Lambda }:\nabla \mathbf{u}+a{{\rho }^{2}}\nabla \cdot \mathbf{u} \\
 -  &\left. M(\frac{1}{2}\nabla \rho \cdot \nabla \rho \nabla \cdot \mathbf{u}+\nabla \rho \cdot \nabla \left( \nabla \cdot \mathbf{u} \right)+\nabla \rho \cdot \nabla \mathbf{u}\cdot \nabla \rho ) \right]\text{,}  
\end{aligned} \label{Eq.5}
\end{equation}%
\begin{equation}
{{C}_{q}}=\frac{1}{\rho {{T}^{2}}}\nabla \cdot \left( q\rho T\nabla T \right)\text{.}  \label{Eq.6}
\end{equation}%
Where
\begin{equation}
{\bm{\Lambda}}={{p}_{1}}\mathbf{I}+M\nabla \rho\nabla \rho\text{,}  \label{Eq.7}
\end{equation}%
\begin{equation}
{{p}_{1}}=-\rho T\nabla \frac{M}{T}\cdot \nabla \rho -\rho M{{\nabla }^{2}}\rho +(\rho {M}'-M)\frac{1}{2}{{\left| \nabla \rho  \right|}^{2}}\text{,}  \label{Eq.8}
\end{equation}%
\begin{equation}
P=\frac{\rho T}{1-b\rho }-a{{\rho }^{2}}\text{.}  \label{Eq.9}
\end{equation}%
Where
$M=K+HT$, ${M}'={{\left( \frac{\partial M}{\partial \rho} \right)}_{T}}$, $H$ and $K$ are functions of $\rho$, $a$ and $b$ are parameters related to the intermolecular interaction potential. 
It should be pointed out that the BGK-like linearization models in a series of current kinetic methods are not the models with the same name that were directly simplified by adding constraints to Boltzmann equation, they can only be the products of combining the kinetic theory and the mean field theory according to specific problems\cite{Xu2022-Complex, xu2024advances, Gan2022JFM}.
After discretizing the velocity space of Eq.(\ref{Eq.2}), the discrete model can be obtained as\cite{chen2022discrete}:
\begin{equation}
{{\partial }_{t}}{{f}_{ji}}+{{\mathbf{v}}_{ji}}\cdot \nabla {{f}_{ji}}-\frac{\mathbf{a}\cdot \left( {{\mathbf{v}}_{ji}}-\mathbf{u} \right)}{RT}f_{ji}^{eq}=-\frac{1}{\tau }\left[ {{f}_{ji}}-f_{ji}^{eq} \right]+{{I}_{ji}}\text{.}  \label{Eq.10}
\end{equation}%
Where ${{f}_{ji}}$ and $f_{ji}^{eq}$ are discrete distribution functions and discrete Maxwell distribution functions, respectively. Here, the distribution function $f$ in the third term at the left end of Eq.(\ref{Eq.2}) is approximately considered as ${{f}^{eq}}$ during the discretization process\cite{Song2024}. The schematic of discrete velocities used in this paper is shown in Fig. \ref{FigA1}.

\subsection{Complex physical field analysis}

Establishing an accurate and satisfactory physical model is just the first step in numerical simulation. We can obtain a wealth of information and outcomes by accurately calculating the specific problems using the established model. How to extract useful information from massive results and conduct in-depth analyses of them is critical for us to know and understand physical problems. Unfortunately, due to the non-equilibrium system's complexity, very little information has been effectively used, and it can be said that most of the information is dormant.

For the analysis of complex physical fields, the task of DBM is to extract more and more effective information from massive amounts of information. In addition to the well-known macroscopic quantities such as the pressure gradient $\nabla p$, temperature gradient $\nabla T$, density gradient $\nabla \rho $, and Knudsen number $Kn$ in macroscopic fluid mechanics, DBM describes how and to what extent the system deviates from equilibrium by using more non-conservative kinetic moments of distribution functions:
\begin{equation}
{{\bm{\Delta }}_{m,n}}={{\mathbf{M}}_{m,n}}\left( f \right)-{{\mathbf{M}}_{m,n}}\left( {{f}^{EQ}} \right)\text{,}  \label{Eq.11}
\end{equation}%
\begin{equation}
{{\bm{{\Delta }'}}_{m,n}}={{\mathbf{{M}'}}_{m,n}}\left( f \right)-{{\mathbf{{M}'}}_{m,n}}\left( {{f}^{EQ}} \right)\text{.}  \label{Eq.12}
\end{equation}%
In Eq.(\ref{Eq.11}), ${{\mathbf{M}}_{m,n}}\left( f \right)$ and ${{\mathbf{M}}_{m,n}}\left( {{f}^{EQ}} \right)$ are the n-order tensors contracted from the m-order kinetic moments of the distribution function $f$ and the equilibrium distribution function ${{f}^{EQ}}$ (is equal to ${{f}^{eq}}$ for ideal gas) for the molecular velocity $\mathbf{v}$, respectively. ${{\bm{\Delta }}_{m,n}}$ is called the thermo-hydrodynamic non-equilibrium (THNE) characteristic quantity, which describes the THNE characteristics of the system. 
${{\mathbf{{M}'}}_{m,n}}\left( f \right)$ and ${{\mathbf{{M}'}}_{m,n}}\left( {{f}^{EQ}} \right)$ in Eq.(\ref{Eq.12}) are the n-order tensors contracted from the m-order kinetic central moments of $f$ and ${{f}^{EQ}}$ for the thermal fluctuation velocity $\mathbf{v}-\mathbf{u}$, respectively. ${{\bm{{\Delta }'}}_{m,n}}$ is called the thermodynamic non-equilibrium(TNE) characteristic quantity, which describes the TNE characteristics of the system. The TNE behavior neglected by NS is attracting more attention with time and is becoming one of the current hot research topics\cite{Xu2022-Complex, xu2024advances, 2021Hybrid, 2018Experimental, e26030200}.

The non-conservative moments mentioned above describe the non-equilibrium behaviors and characteristics of the system from their individual perspectives. These non-conservative moments are related and complementary. Together, they form a relatively more complete picture of the system. However, because of the complexity of non-equilibrium systems, DBM incorporates morphological and thermodynamic characteristics, in addition to non-conservative moments, to study and describe complex physical fields.

Entropy is a crucial physical quantity in compression science, ICF, aerospace, and other fields and has an essential influence on the system's evolution. In ICF, for instance, an increase in entropy will, on the one hand, make it harder to compact the target pill; however, an increase in entropy can prevent instability from developing on the other hand. As a result, it is critical to control the entropy increase reasonably for ignition to succeed. With DBM, we can conveniently analyze the system's entropy. Considering the contribution of the density gradient to entropy and internal energy, we can get the generalized form of the total entropy of the system\cite{gonnella2007lattice, onuki2005dynamic}:
\begin{equation}
{{S}_{b}}=\int{\left[ ns\left( n,e \right)-\frac{1}{2}H{{\left| \nabla n \right|}^{2}} \right]}d\mathbf{r}\text{,}  \label{Eq.13}
\end{equation}%
and the total internal energy of the system:
\begin{equation}
{{E}_{b}}=\int{(e+\frac{1}{2}K|\nabla n{{|}^{2}})}d\mathbf{r}\text{,}  \label{Eq.114}
\end{equation}%
where $n$ is the particle-number density, $s$ is the entropy per particle, and $e$ is the internal energy per particle. The gradient terms in Eqs. (\ref{Eq.13}) and (\ref{Eq.114}) represent a decrease in entropy and an increase in internal energy because of the inhomogeneity of $n$, respectively; the space integrals extend to the entire computational domain. For van der Waals fluids, we can get\cite{Xu2022-Complex}:
\begin{equation}
\frac{d{{S}_{b}}}{dt}=\int{\left( \frac{1}{T}\nabla \cdot \mathbf{j}+\frac{1}{T}\bm{\Pi }:\nabla \mathbf{u} \right)}d\mathbf{r}\text{,}  \label{Eq.14}
\end{equation}%
where $\mathbf{j}$ and $\bm{\Pi }$ are heat flux and viscous stress, respectively. Through the relationship between the material derivative and the temporal partial derivative, we can get the temporal partial derivative of the entropy density:
\begin{equation}
\frac{\partial {{s}_{b}}}{\partial t}=-\nabla \cdot \left( \mathbf{u}{{s}_{b}}-\frac{1}{T}\mathbf{j} \right)-\mathbf{j}\cdot \nabla \left( \frac{1}{T} \right)+\frac{1}{T}\bm{\Pi }:\nabla \mathbf{u}\text{.}  \label{Eq.15}
\end{equation}%
It can be seen that the entropy production rate can be divided into two parts. One part is contributed by heat flux:
\begin{equation}
{{\dot{S}}_{NOEF}}=\int{\mathbf{j}\cdot \nabla \frac{1}{T}}d\mathbf{r}\text{;}  \label{Eq.16}
\end{equation}%
the other part is contributed by viscous stress:
\begin{equation}
{{\dot{S}}_{NOMF}}=\int{-\frac{1}{T}\bm{\Pi }:\nabla \mathbf{u}}d\mathbf{r}\text{.}  \label{Eq.17}
\end{equation}%
\noindent
Under the continuous limit, $\mathbf{j}\text{=}-\kappa\nabla T$ and $\bm{\Pi }=-\mu \left[ \nabla \mathbf{u}+{{\left( \nabla \mathbf{u} \right)}^{T}}-\left( \nabla \cdot \mathbf{u} \right)\mathbf{I} \right]$, where $\kappa$ is the heat conductivity and $\mu $ is the viscosity coefficient. Substituting them into Eq.(\ref{Eq.16}) and Eq.(\ref{Eq.17}), respectively, we can get
\begin{equation}
{{\dot{S}}_{NOEF}}=\int{\frac{\kappa{{\left| \nabla T \right|}^{2}}}{{{T}^{2}}}}d\mathbf{r}\text{,}  \label{Eq.18}
\end{equation}%

\begin{equation}
{{\dot{S}}_{NOMF}}=\int{\frac{\mu \left[ \nabla \mathbf{u}:\nabla \mathbf{u}+{{\left( \nabla \mathbf{u} \right)}^{T}}:\nabla \mathbf{u}-{{\left| \nabla \cdot \mathbf{u} \right|}^{2}} \right]}{T}}d\mathbf{r}\text{.}  \label{Eq.19}
\end{equation}%

The morphological analysis method has gradually become an effective data analysis and information extraction technology for complex physical fields. By incorporating different morphological quantities into DBM, a series of new insights into the study of hydrodynamic instability systems have been obtained\cite{gan2019nonequilibrium,chen2020morphological}. In this paper, we introduce the ratio of the area occupied by heavy fluids (${{A}_{h}}$) and the length of the interface between light and heavy fluids ($L$) to provide a supplementary description of the evolution of the RTI.

In actuality, rather than relying only on variables with the same unit or dimension to describe the non-equilibrium state and behavior of the system together, DBM defines a more generalized non-equilibrium intensity, $\mathbf{D}$, to describe the system's non-equilibrium for the extraction and analysis of complex physical field information. This generalized non-equilibrium intensity can include various non-equilibrium parameters of the system, such as:
\begin{equation}
\begin{aligned}
\mathbf{D}=&\{ Kn, \lvert\nabla \rho\rvert, \lvert\nabla T\rvert, \lvert\nabla p\rvert, L, {{A}_{h}}, \lvert{{\bm{\Delta }}_{m,n}}\rvert, \\
 &\lvert{{{\bm{{\Delta }'}}}_{m,n}}\rvert, \lvert{{{\dot{S}}}_{NOEF}}\rvert, \lvert{{{\dot{S}}}_{NOMF}}\rvert,\cdots  \}\text{.}   \\ 
\end{aligned}\label{Eq.20}
\end{equation}%
The specific parameters and the number of parameters contained in $\mathbf{D}$ can be selected according to the needs of specific problems.

\subsection{Introduction of tracer particles}
In cases of strong compressibility, the early approach of determining the fluid source based on density is invalid because the density of fluid from the upper layer (original weight) may be smaller than that of the surrounding (original light) fluid. Taking inspiration from PIV and PILF technologies in experimental fluid mechanics, introducing tracer particles into numerical simulation allows for the tracking and identification of fluid sources in mixed regions within the framework of a single fluid.

As is well known, the motion state of moving particles in fluid media can be described by the Stokes number ($Stk$):
\begin{equation}
Stk=\frac{{{u}_{0}}\cdot {{\tau }_{P}}}{{{l}_{0}}}\text{.}  \label{Eq.21}
\end{equation}%
Where ${{u}_{0}}$ is the local flow velocity, ${{\tau }_{P}}$ is the relaxation time of particles, and ${{l}_{0}}$ is the characteristic size of particles. The Stokes number represents the inertia of particles moving in a fluid medium. The smaller the ${{\tau }_{P}}$ is, the easier it is for particles to follow the surrounding fluid. When ${{\tau }_{P}}$ is close to 0, $Stk<<1$, the particles completely follow the fluid's movement. Therefore, if we set the volume and mass of moving particles very small in numerical simulation, the momentum exchange between particles and fluid medium will be completed instantly. The particle velocity will be completely determined by the local flow velocity\cite{zhang2021delineation}, as shown in Eq.(\ref{Eq.22}). Then, the function of tracing particles can be realized numerically by marking the moving particles to distinguish the source fluid where the particles come from.
\begin{equation}
{{\mathbf{u}}_{P}}\left( {{\mathbf{r}}_{k}} \right)=\int_{Z}{\mathbf{u}\left( \mathbf{r} \right)}\cdot \delta \left( \left| \mathbf{r}-{{\mathbf{r}}_{k}} \right| \right)d\mathbf{r}\text{,}  \label{Eq.22}
\end{equation}%
where ${{\mathbf{u}}_{p}}$ is the particle velocity, ${{\mathbf{r}}_{k}}$ is the position of the particle, $\mathbf{u}\left( \mathbf{r} \right)$ is the local velocity at the spatial position $\mathbf{r}$, $Z$ is the region where the particle is affected by the local velocity of the fluid medium, $\delta$ is the Dirac function, and its discrete form $\psi$ is usually used in numerical simulation. In the two-dimensional case, $\psi$ can be expressed as: 
\begin{equation}
\psi \left( {{\mathbf{r}}_{i,j}},{{\mathbf{r}}_{k}} \right)=\psi \left( \left| {{\mathbf{r}}_{i,j}}-{{\mathbf{r}}_{k}} \right| \right)=\varphi (\Delta {{r}_{x}})\cdot \varphi (\Delta {{r}_{y}})\text{,}  \label{Eq.23}
\end{equation}%
where $\varphi$ is the weight function. In this paper, $\varphi$ is selected as\cite{zhang2021delineation}:
\begin{equation}
\varphi (\Delta {{r}_{x}})=\left\{ \begin{matrix}\left\{ 1+\cos \left[ (\Delta {{r}_{x}}/\Delta x)\cdot \pi /2 \right] \right\}/4, & \Delta {{r}_{x}}\le 2\Delta x \\
0, & \Delta {{r}_{x}}>2\Delta x \\ \end{matrix} \right.\text{.}  \label{Eq.24}
\end{equation}%

\section{\protect\bigskip SIMULATIONS AND RESULTS}
By introducing tracer particles into the DBM model with van der Waals potential, we study the evolution of a two-dimensional single-mode compressible RTI system under different compressibility conditions in a single fluid framework. The research details and analysis of the results are presented in this section.

\subsection{Initialization and numerical specifications}
\begin{figure}[tbp]
\includegraphics[width=0.90\textwidth,trim=0.1 0.1 0.1 0.1,clip]{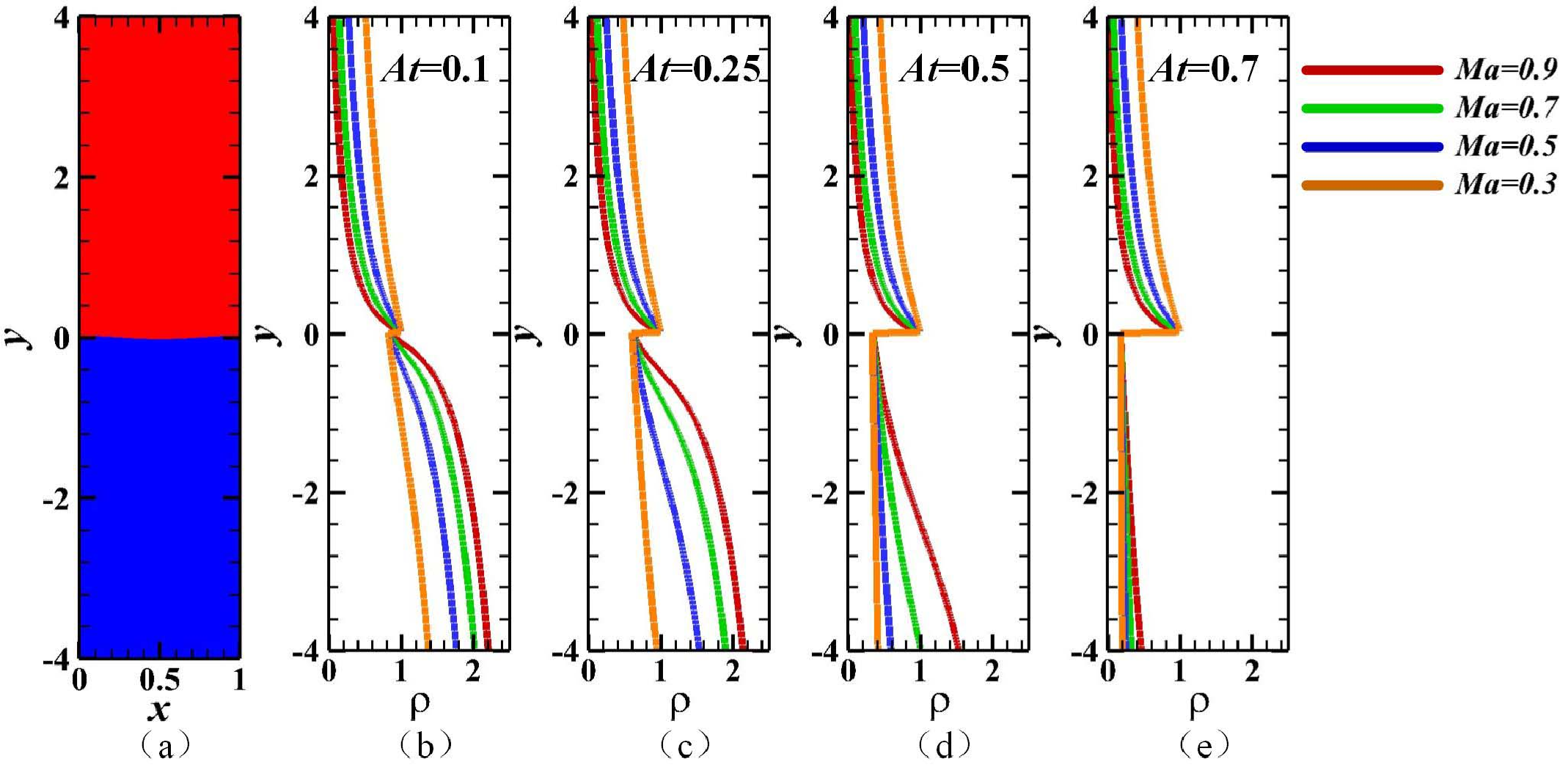}
\caption {(a) The computational domain with $\left[ 0,1 \right]\times \left[ -4,4 \right]$. Red represents heavy fluid, and blue represents light fluid. (b)$-$(e) The density profile at $x=0$ across various $Ma$ conditions for $At=0.1, 0.25, 0.5 $, and $0.7$, respectively.} \label{Fig2}
\end{figure}

Figure \ref{Fig2}(a) shows the two-dimensional single-mode RTI system studied in this paper. In a computational domain with the width of $\left[ 0,1 \right]$ and the height of $\left[ -4,4 \right]$, the upper part is heavy fluid, and the lower part is light fluid. At the initial time, an initial disturbance ${{y}_{c}}\left( x \right)={{y}_{0}}\cos \left( kx \right)$ of amplitude ${{y}_{0}}=0.02$ exists at the light-heavy fluid interface that positioned at $y = 0$. $k=2\pi /\lambda $ is the wave number, and $\lambda =1$ is the wavelength of the disturbance. The density of heavy fluid at the interface is
\begin{equation}
{{\rho }_{0}}\left( 0,y_{\text{c}}^{+}\left( 0 \right) \right)=1\text{.}  \label{Eq.25}
\end{equation}%
The density of light fluid at the interface is
\begin{equation}
{{\rho }_{0}}\left( 0,y_{c}^{-}\left( 0 \right) \right)=1\times \left[ {\left(1-At\right)}/{\left( 1+At \right)}\; \right]\text{.}  \label{Eq.26}
\end{equation}%
$y_{\text{c}}^{+}$ represents the upper side of the disturbance interface and $y_{\text{c}}^{-}$ represents the lower side of the disturbance interface. The temperatures of the light and heavy fluids are set to be:
\begin{equation}
\begin{aligned}
 \text{ }{{T}_{0}}\left( y \right)=&1\text{,}y\ge {{y}_{c}}\left( x \right)\text{,}  \\
 \text{ }{{T}_{0}}\left( y \right)=&{\left[ 1+a\cdot {{\rho }_{0}}\left( 0,y_{c}^{-}\left( 0 \right) \right) \right]\cdot}{\left[ 1-b\cdot {{\rho }_{0}}\left( 0,y_{c}^{-}\left( 0 \right) \right) \right]} \\
 &/{{{\rho }_{0}}\left( 0,y_{c}^{-}\left( 0 \right) \right)}\text{,}y<{{y}_{c}}\left( x \right)\text{.} \\ 
\end{aligned} \label{Eq.27}
\end{equation}%
Fluid in the flow field satisfies hydrostatic equilibrium, i.e.,
\begin{equation}
{{\partial }_{y}}{{P}_{0}}\left( y \right)=-g{{\rho }_{0}}\left( y \right)\text{.}  \label{Eq.28}
\end{equation}%
Substituting the van der Waals state equation, Eq.(\ref{Eq.9}), into Eq.(\ref{Eq.28}) gives:
\begin{equation}
{{\partial }_{y}}{{\rho }_{0}}\left( y \right)={g}/{2a}\;-\frac{1}{2a}\cdot \frac{Tg}{T-2a{{\rho }_{0}}\left( y \right){{\left[ 1-b{{\rho }_{0}}\left( y \right) \right]}^{2}}}\text{,}  \label{Eq.29}
\end{equation}%
where $a=\frac{9}{8}$ and $b=\frac{1}{3}$ in this paper.

Following the previous studies\cite{fu2023bubble, fu2022nonlinear, wieland2019effects}, the parameter settings and research results presented in this paper are dimensionless. The characteristic scales selected are the perturbation wave length ${{\lambda }^{*}}$, initial temperature ${{T}^{I*}}$, density ${{\rho }^{I*}}$, and the isothermal speed of sound ${{U}^{*}}\equiv \sqrt{{{{P}^{I*}}}/{{{\rho }^{I*}}}\;}$ at the interface. The superscript ‘$*$’ means the dimensional physical quantities and the superscript ‘$I$’ means the quantities at the initial interface. Several dimensionless numbers used in this paper are defined as follows\cite{fu2023bubble, fu2022nonlinear}:
\begin{equation}
\begin{aligned}
  & At=\frac{\rho {{_{h}^{I}}^{*}}-\rho {{_{l}^{I}}^{*}}}{\rho {{_{h}^{I}}^{*}}+\rho {{_{l}^{I}}^{*}}},Ma=\sqrt{\frac{{{g}^{*}}{{\lambda }^{*}}}{{{{P}^{I}}^{*}}/{{{\rho }^{*}}}\;}},Re=\frac{{{\rho }^{*}}{{U}^{*}}{{\lambda }^{*}}}{{{\mu }^{*}}}, \\ 
 & {{Re}_{P}}={{{\rho }^{*}}{{\lambda }^{*}}\sqrt{{At}/{\left( 1+At \right){{g}^{*}}{{\lambda }^{*}}}\;}}/{{{\mu }^{*}}}\;,Pr ={{C}_{P}}\frac{{{\mu }^{*}}}{{{k}^{*}}}. \\ 
\end{aligned}
  \label{Eq.30}
\end{equation}%
Here, $At$ represents the light and heavy fluid density ratio at the interface, and $Ma$ characterizes the strength of the isothermal background stratification\cite{reckinger2016comprehensive, livescu2004compressibility}.

Our current study examined the evolution of RTI across a range of $At$ from 0.1 to 0.7 and $Ma$ from 0.3 to 0.9. Figs. \ref{Fig2}(b)$-$\ref{Fig2}(e) represent the density profile for different $At$ and $Ma$ at $x=0$. It can be seen that with the increase in $At$, the density difference between light and heavy fluids at the interface increases, and the density stratification becomes less pronounced. Other parameters are set as follows: Prandtl number $Pr =0.72$ and disturbed Reynolds number ${{Re}_{P}}=1500$. The heights of the bubble and spike are defined as the distance from the tips of the bubble and spike to $y=0$, respectively. The bubble and spike velocities are defined as the derivatives of the bubble height and spike height over time, respectively. As done by Fu et al. and Bian et al., the time scale $\sqrt{{\lambda }/{\left( Atg \right)}\;}$ and the saturated velocity $\sqrt{{Atg\lambda }/{\left( 1+At \right) }\;}$ are used to scale time and the bubble and spike velocities, respectively\cite{fu2023bubble, fu2022nonlinear, bian2020revisiting}.

In this paper, the first-order forward Euler finite difference scheme is used for the discretization of the temporal derivative. The second-order non-oscillatory non-free dissipative (NND) scheme and the nine-point stencil scheme\cite{zhang2019entropy, tiribocchi2009hybrid} are used to discretize spatial derivative and the intermolecular interaction item $I$, respectively. The variables $\rho $, $\mathbf{u}$, and $e$ are fixed at their initial values at the borders of the y-direction, and periodic boundary conditions are applied along the x-direction\cite{fu2023bubble, luo2020effects, reckinger2016comprehensive}. The grid length is set to be $\Delta x=\Delta y=0.0039$.

\subsection{RESULTS AND DISCUSSION}
\subsubsection{Bubble and spike velocity}

\begin{figure}[tbp]
\center{\epsfig{file=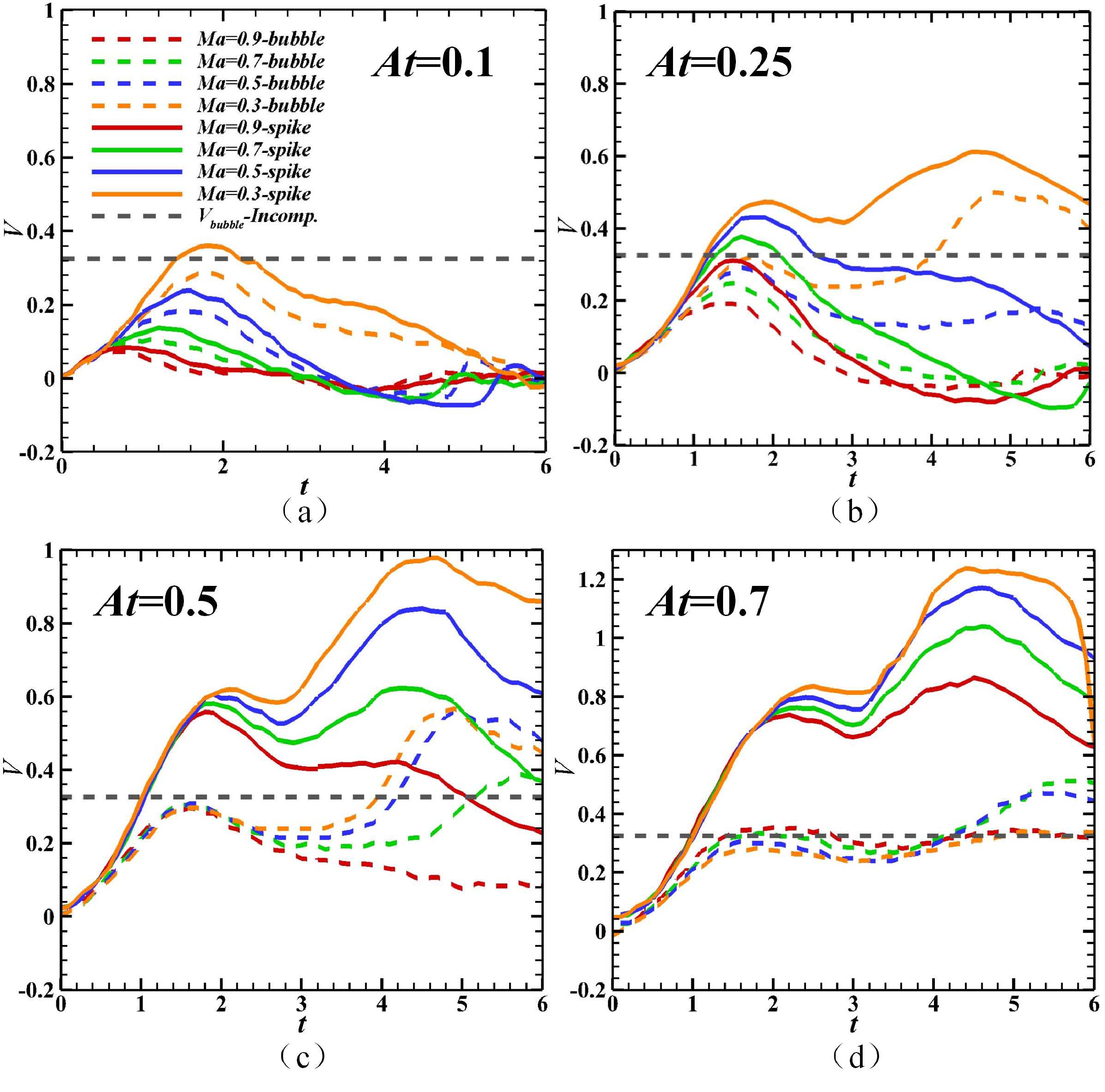,
width=0.8\textwidth,trim=0 0 0 0, clip }}
\caption{The spike and bubble velocity under various $Ma$ conditions for (a) $At=0.1$, (b) $At=0.25$, (c) $At=0.5$, and (d) $At=0.7$. The horizontal gray dashed line represents the bubble-saturated velocity predicted by potential flow models.} \label{Fig3}
\end{figure}

Figures \ref{Fig3}(a)$-$\ref{Fig3}(d) show the spike and bubble velocity under different $Ma$ conditions for $At=0.1, 0.25, 0.5,$ and 0.7, respectively. The horizontal gray dashed line in each panel represents the bubble-saturated velocity predicted by the potential flow model. 

As indicated in Fig. \ref{Fig3}(a), when $At=0.1$, all bubble velocities are lower than the value predicted by the potential flow model. Following the initial acceleration, the velocities of the bubble and spike did not reach and maintain a saturated velocity but instead began to decrease continuously. As $Ma$ increases, the velocities of the bubble and spike decrease noticeably. Compressibility shows a significant inhibitory effect on bubble and spike velocity.
When $At=0.25$, the bubble and spike velocities are greater than those under the same Ma condition and the condition of $At=0.1$, as depicted in Fig. \ref{Fig3}(b). When $Ma=0.3$, both bubbles and spikes are observed to reaccelerate. The inhibitory effect of compressibility on bubble and spike velocity diminished with the increase in $At$.
With the further increase of $At$ to 0.5, except for the case of $Ma=0.9$, both bubble and spike exhibit reacceleration behavior, as shown in Fig. \ref{Fig3}(c). 
When $At=0.7$, the spike reaccelerates under all the considered $Ma$ values. It is noteworthy that the bubble only exhibits reacceleration at $Ma=0.5$ and $Ma=0.7$. The bubble has no reacceleration at a greater $Ma$ value of 0.9 and a lower $Ma$ value of 0.3. Before the RTI enters the bubble reacceleration stage, compressibility exhibits an inhibitory effect on spike velocity but a facilitating effect on bubble velocity.

In a compressible system, velocity can be divided into two parts: the compressible (irrotational) component caused by the fluid's expansion and compression, and the solenoidal component, independent of the fluid's density change. To determine the reason for the variation in the effect of compressibility on bubble and spike velocity, we employ the Helmholtz decomposition of the velocity field\cite{wang2013cascade}:
\begin{equation}
\mathbf{u}={{\mathbf{u}}_{C}}+{{\mathbf{u}}_{S}}\text{.}  \label{Eq.31}
\end{equation}%
Where $\mathbf{u}=\left( U, V \right)$, $U$ represents the velocity component in the $x$ direction, and $V$ represents the velocity component in the $y$ direction. ${{\mathbf{u}}_{C}}=\left( {{U}_{C}},{{V}_{C}} \right)$ is the compressible (irrotational) component, and ${{\mathbf{u}}_{S}}=\left( {{U}_{S}},{{V}_{S}} \right)$ is the solenoidal component.

\begin{figure}[tbp]
\center{\epsfig{file=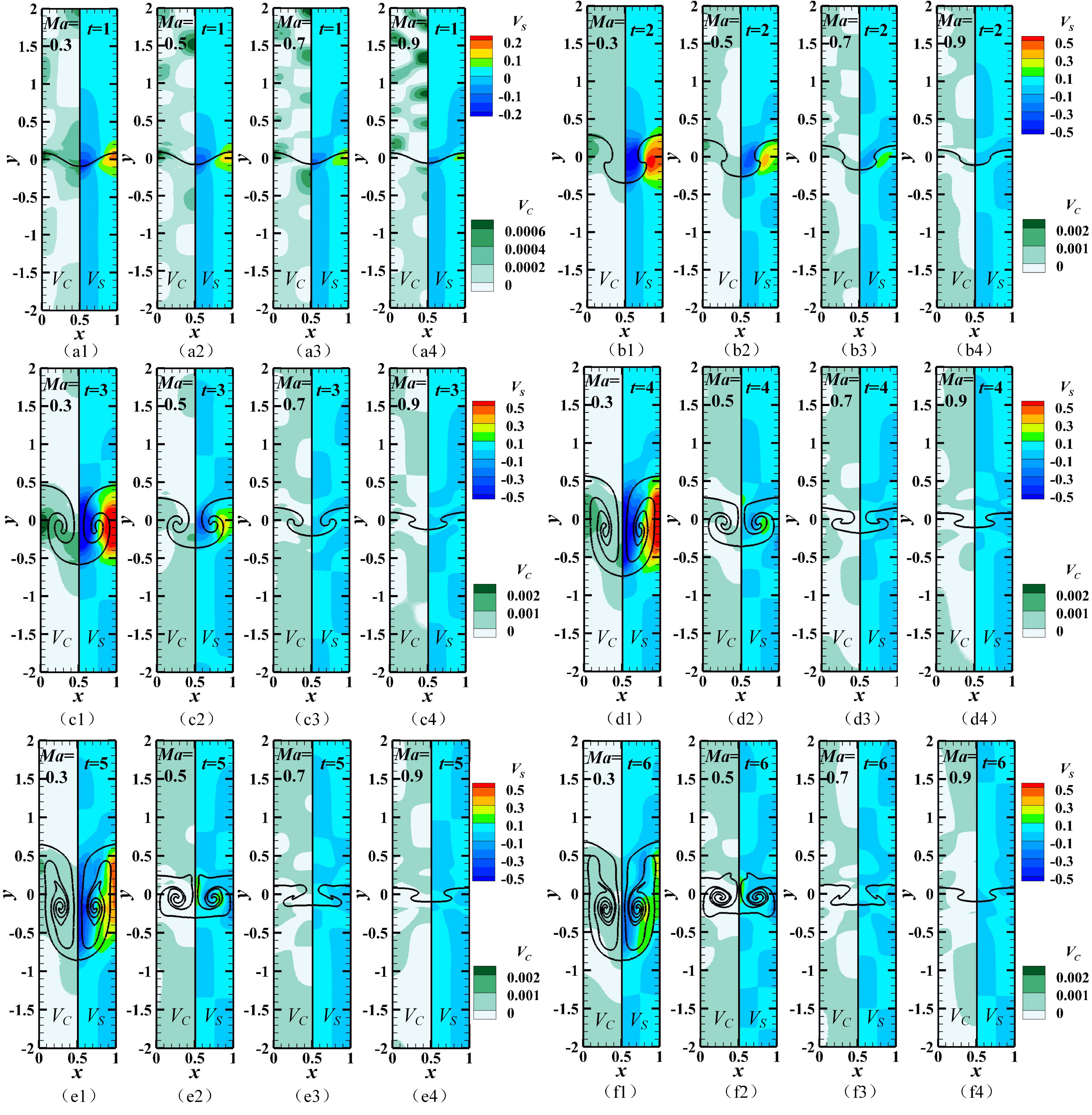,
width=0.95\textwidth,trim=0 0 0 0, clip }}
\caption{The contour plots of the compressive (${{V}_{C}}$, left half) and solenoidal (${{V}_{S}}$, right half) components of vertical velocity for $At=0.1$. From (a) to (f): $t=1, 2, 3, 4, 5$, and $6$, respectively. From (1) to (4): $Ma=0.3, 0.5, 0.7$, and $0.9$, respectively. The dark lines represent the interface of the heavy and light fluid.} \label{Fig4}
\end{figure}

Figure \ref{Fig4} shows the contour plots of the compressive (${{V}_{C}}$, left half) and solenoidal (${{V}_{S}}$, right half) components of vertical velocity under the condition of $At=0.1$. It can be seen that ${{V}_{C}}$ is two orders of magnitude smaller than ${{V}_{S}}$. This indicates that in the current movement, divergence-free movement is dominant. The velocity change caused by the fluid medium's expansion and compression is minimal. 
It can be seen that with the increase in $Ma$, ${{V}_{S}}$ decreases significantly. Compressibility exerts a significant inhibitory effect on ${{V}_{S}}$. From a stress perspective, the primary reason for this phenomenon is that under the condition of $At=0.1$, the density difference between the light and heavy fluids at the interface is slight. As the heavy fluid moves into the light fluid, the density of the light fluid around the spike quickly exceeds that of the heavy fluid inside the spike. The buoyancy of the heavy fluid increases rapidly. 
With the increase in $Ma$, the density stratification intensifies, causing the spike to experience a stronger upward force while descending an equivalent distance. This resulted in a lower ${{V}_{S}}$ at a higher $Ma$. It can be observed that at higher $Ma$, the pressure severely distorts the bubble and the spike, limiting their movement to a narrow region near the original interface.

Interestingly, although ${{V}_{C}}$ is caused by the fluid's expansion and compression, it decreases with an increase in $Ma$ at $At=0.1$. This is primarily due to the fact that the range of motion for both light and heavy fluids is significantly limited. Because of this limitation, the pressure changes of light and heavy fluids that occur during movement are also restricted, resulting in the restricted expansion and compression of both fluids.
\begin{figure}[tbp]
\center{\epsfig{file=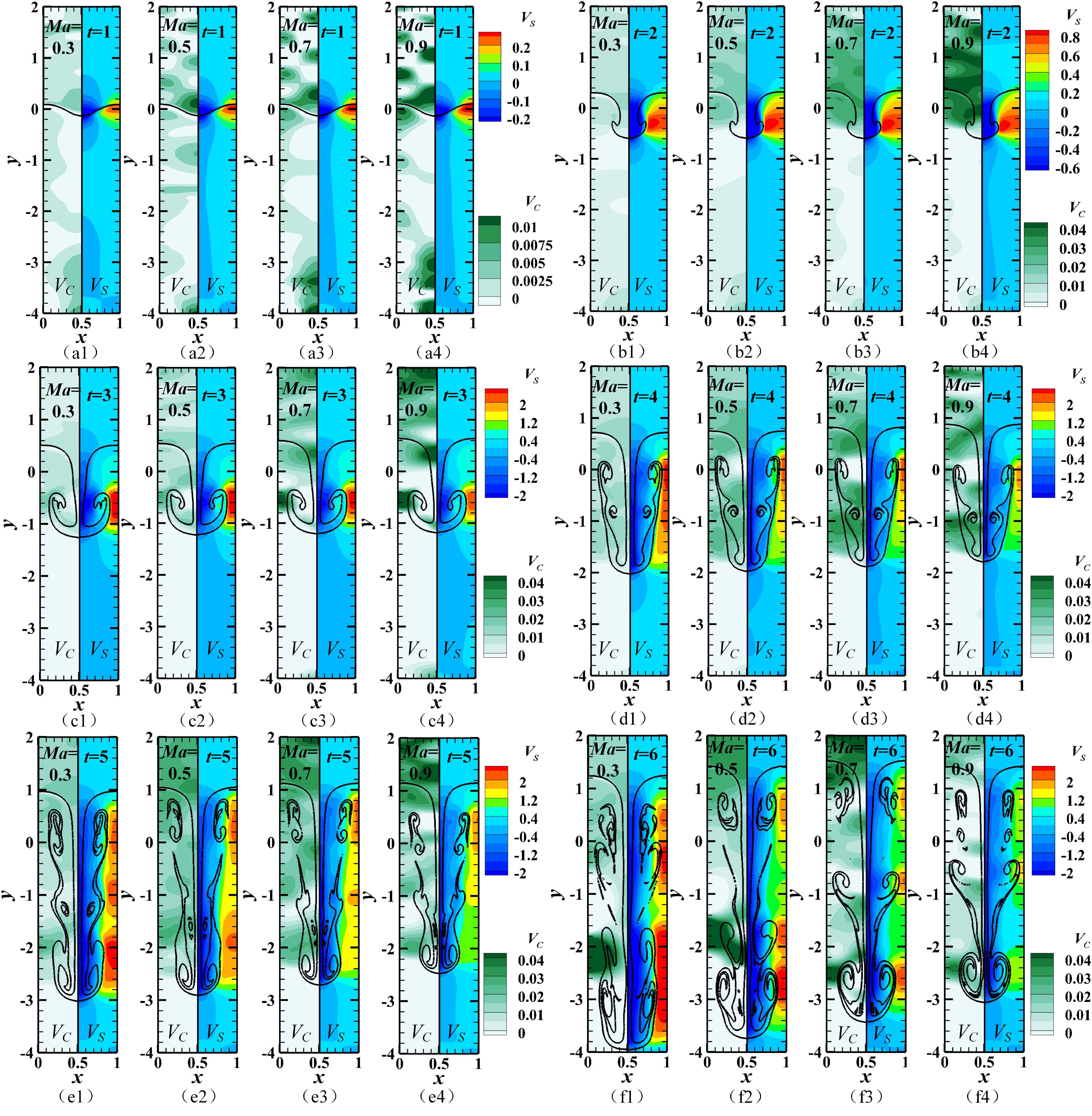,
width=0.95\textwidth,trim=0 0 0 0, clip }}
\caption{The contour plots of the compressive (${{V}_{C}}$, left half) and solenoidal (${{V}_{S}}$, right half) components of vertical velocity for $At=0.7$. From (a) to (f): $t=1, 2, 3, 4, 5$, and $6$, respectively. From (1) to (4): $Ma=0.3, 0.5, 0.7$, and $0.9$, respectively. The dark lines represent the interface of the heavy and light fluid.} \label{Fig5}
\end{figure}

With the increase in $At$, the inhibitory effect of compressibility on the bubble and spike velocity is weakened. When $At=0.7$, compressibility even shows a promoting effect on bubble velocity. Fig. \ref{Fig5} shows the ${{V}_{C}}$ and ${{V}_{S}}$ contour plots for $At=0.7$. It is demonstrated that ${{V}_{C}}$ is one order of magnitude smaller than ${{V}_{S}}$. This means that the movement of the bubble and spike is still dominated by ${{V}_{S}}$. However, with the increase in $Ma$, the spike and bubble are not restricted in a narrow range near the initial interface under the condition of $At=0.7$. And there was no obvious difference in the spike amplitude across various $Ma$ conditions before $t=3$. 

We can try to understand why the inhibitory effect of compressibility was weakened with the help of Fig. \ref{Fig2}(e). As shown in Fig. \ref{Fig2}(e),  there is a significant disparity in density between the light and heavy fluids at the interface at the initial time. The density of the light fluid at the interface is significantly lower than that of the heavy fluid at the interface. 
Thus, compared with the density of the heavy fluid at the initial interface, the density increase of light fluid with the decrease in height is slight, and the difference in density stratification of light fluid across various $Ma$ conditions is not apparent. Therefore, as the heavy fluid moves downward, the pressure increase is slight compared with the gravity of the spike. The difference in the blocking effect of light fluid on spikes under different $Ma$ conditions is not apparent, especially in the early stages. This is why the early movements of the spikes under different conditions of $Ma$ are almost the same, and the movements of the bubble and spike are not limited to a narrow range near the initial interface with the increase in $Ma$. 

With the further evolution of RTI, the heavy fluid keeps converging to the spike. As shown in Fig. \ref{Fig2}(e), it is evident that with the increase in $Ma$, there is a noticeable decrease in the density of the heavy fluid as the height increases. This causes the total mass of the heavy fluid accumulated within the spike to decrease with the increase in $Ma$. Thus, during the downward motion of the spike, while the buoyancy of the spike remains relatively constant as $Ma$ increases, the spike's gravitational force noticeably reduces as $Ma$ increases. This is the main reason there is an obvious decrease in spike velocity in the later period as $Ma$ increases.

\begin{figure}[tbp]
\center{\epsfig{file=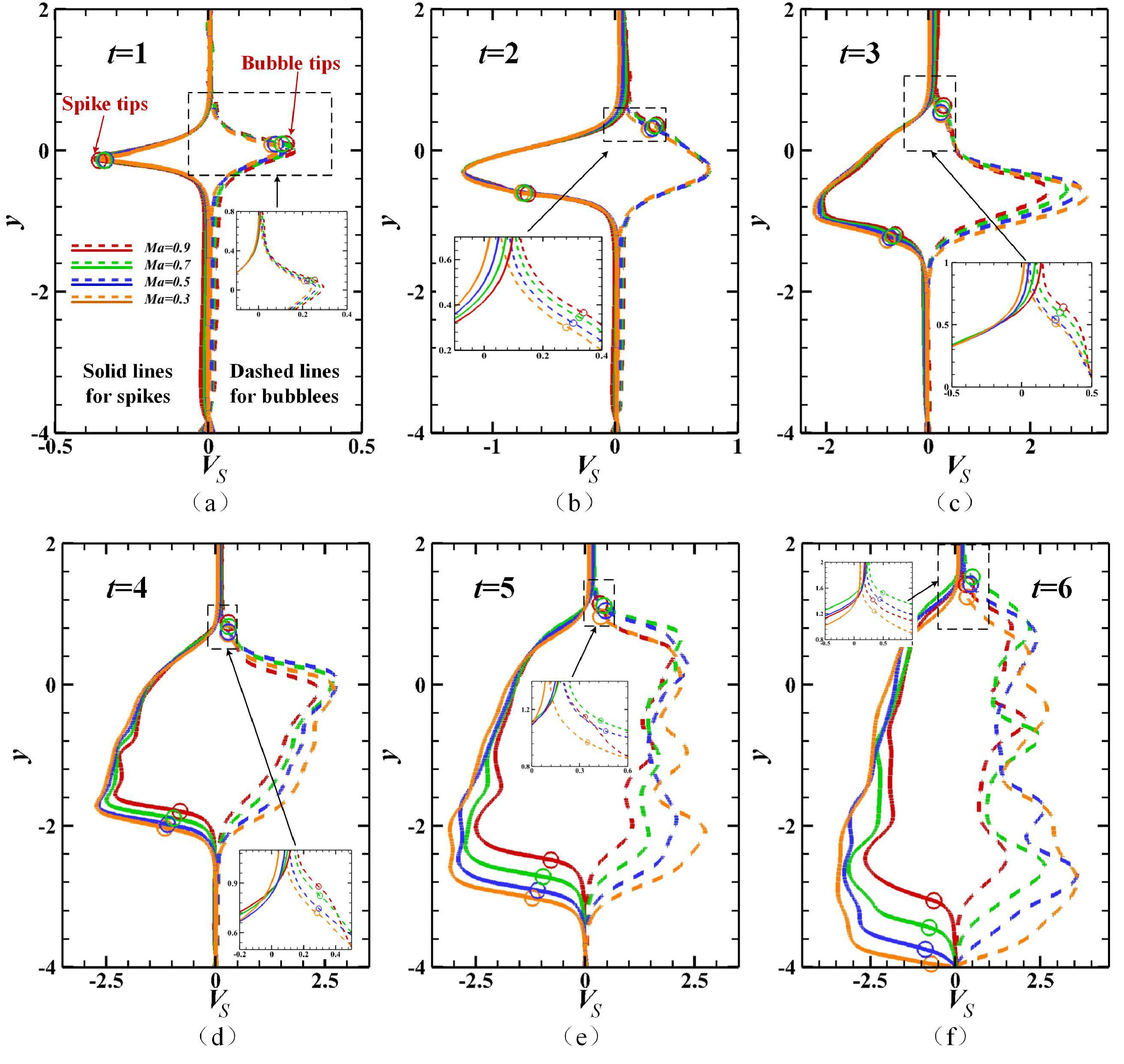,
width=1.0\textwidth,trim=0 0 0 0, clip }}
\caption{The ${{V}_{S}}$ profiles along the vertical lines across the bubble ($x=1$ , denoted by dash lines) and spike ($x=0.5$ , denoted by solid lines) tips for $At=0.7$. From (a) to (f): $t=1, 2, 3, 4, 5$, and $6$, respectively. The circles marked on the ${{V}_{S}}$ profile denote the vertical position of the bubble and spike tips at each moment.} \label{Fig6}
\end{figure}

Interestingly, in the case of $At=0.7$, the compressibility shows a promotion effect on the bubble velocity before the bubble enters the reacceleration stage. The bubble's reacceleration phenomenon is only observed under the middle $Ma$ conditions of $Ma=0.5$ and $Ma=0.7$, while it is not observed when $Ma$ decreases to 0.3 and increases to 0.9. Fig. \ref{Fig5} illustrates that when $At=0.7$, the motion of the bubble and spike is still dominated by ${{V}_{S}}$. To conduct a more comprehensive examination of the reasons contributing to this intriguing phenomenon, we graph the ${{V}_{S}}$ curves along the vertical lines across the tips of the bubble ($x=1$, denoted by the dashed line) and the spike ($x=0.5$, denoted by the solid line) at $t=1, 2, 3, 4, 5$, and 6, as shown in Fig. \ref{Fig6}. The circles on the ${{V}_{S}}$ profile indicate the vertical position of the bubble and spike tips at each moment.

\begin{figure}[tbp]
\center{\epsfig{file=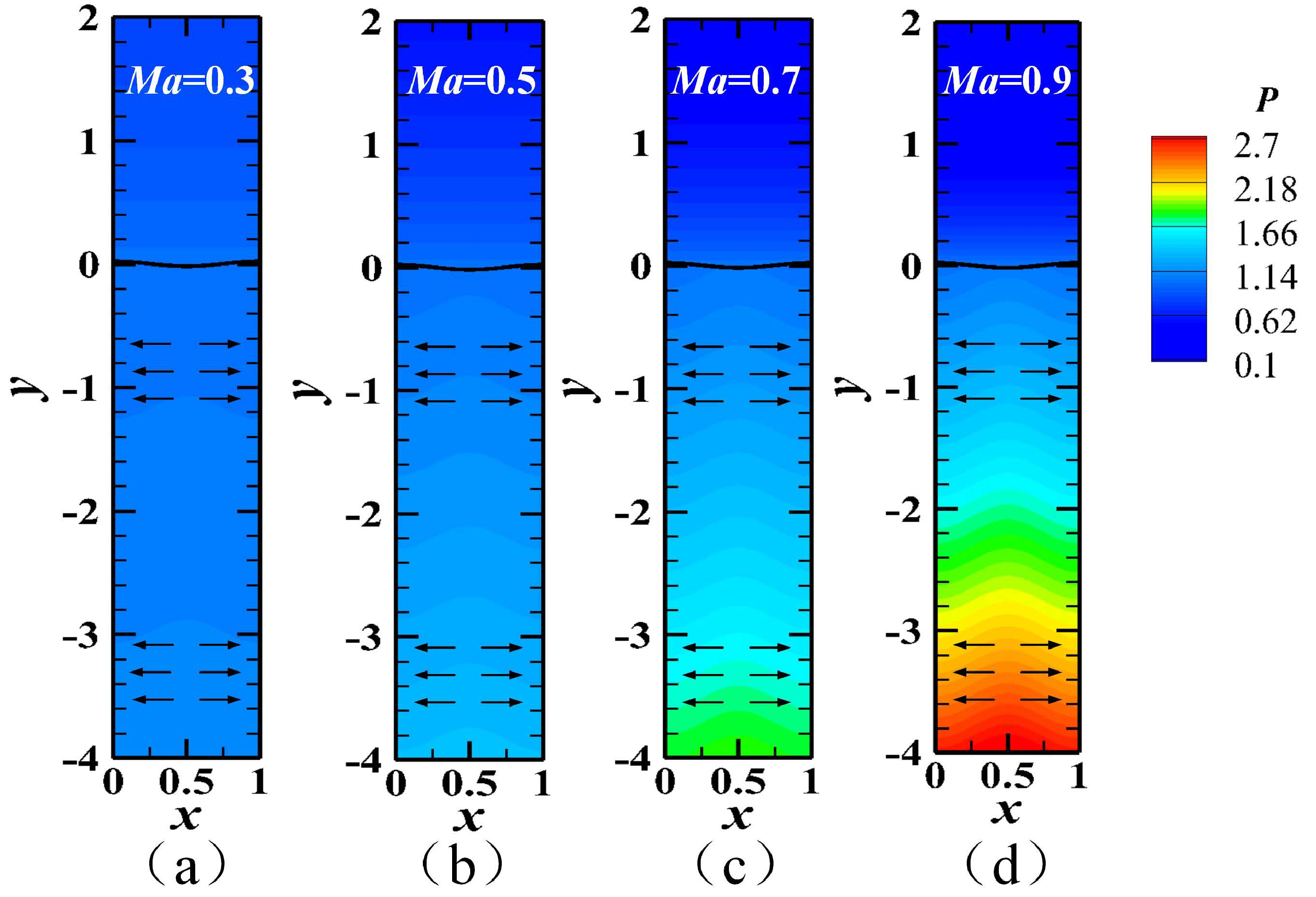,
width=0.8\textwidth,trim=0 0 0 0.5, clip }}
\caption{The pressure contour at the initial time for $At=0.7$. From (a) to (d): $Ma=0.3, 0.5, 0.7$, and 0.9, respectively. The dark lines represent the interface of the heavy and light fluids} \label{Fig7}
\end{figure}

As illustrated in Fig. 6(a), at $t=1$, the bubble and spike ${{V}_{S}}$ increase as $Ma$ increases. This can be understood with the help of Fig. \ref{Fig7}. Fig. \ref{Fig7} shows the pressure contour at the initial time for $At=0.7$ under various $Ma$ conditions. The dark line represents the interface of the heavy and light fluids. Initially, the pressure of the heavy fluid above the interface is uniform in the x direction. However, below the interface, the pressure in the middle of the light fluid is higher than that on both sides. This causes the heavy fluid in the spike and the light fluid below the spike to press and move toward the light fluid on both sides. With the increase in $Ma$, this pressure difference increases, and the tendency of the middle zone fluid to press and move to both sides of the fluid increases. This causes the spike's ${{V}_{S}}$ to rise as $Ma$ increases. Simultaneously, the fluids on both sides move upward due to compression from the middle zone fluid. So, as $Ma$ increases, the ${{V}_{S}}$ below the bubble also increases.

At $t=2$ and $t=3$, the bubble ${{V}_{S}}$ still increases as $Ma$ increases, as shown in Figs. \ref{Fig6}(b) and \ref{Fig6}(c). However, unlike at time $t=1$, the ${{V}_{S}}$ below the tip of the bubble under different $Ma$ conditions is nearly the same at $t=2$ and $t=3$. Therefore, the increase in ${{V}_{S}}$ at the bubble's tip with an increase in $Ma$ at this time is not due to a difference in the compression of the bubble by the fluid below. In Fig. \ref{Fig2}(e), we can see that as $Ma$ increases, the density of heavy fluid decreases rapidly with height. So, the pressure on the bubble tip decreases rapidly during the bubble-rising process. This is the primary cause of the growth in bubble ${{V}_{S}}$ as $Ma$ increases.

After $t=4$, the bubble ${{V}_{S}}$ increases first, then decreases as $Ma$ increases. At this period, the bubble ${{V}_{S}}$ is gradually influenced by the vorticity deposition at the tail of the Kelvin-Helmholtz (KH) vortex created by the upward curl of the spike. Compressibility has two effects on the transport of vorticity deposited at the tail of the KH vortex to the tip of the bubble. On the one hand, with the increase in $Ma$, the length of the KH vortex (the distance from the tip of the spike to the tail of the KH vortex) decreases, which causes the tail of the KH vortex to be farther away from the tip of the bubble, as shown in Figs. \ref{Fig5}(d)$-$\ref{Fig5}(f). On the other hand, with the increase in $Ma$, the downward motion distance of the spike decreases, which causes the tail of the KH vortex to be closer to the tip of the bubble. Due to the influence of these two factors, when $Ma$ is high or low, the vorticity deposited at the tail of the KH vortex cannot be effectively transported to the tip of the bubble and thus fails to cause the reacceleration of the bubble. As a result, the bubble has no reacceleration in cases of high compressibility ($Ma = 0.9$) and low compressibility ($Ma = 0.3$). This is the reason why the bubble ${{V}_{S}}$ initially rises and subsequently falls as $Ma$ increases.

\subsubsection{Proportion of area occupied by the heavy fluid}

The amplitude and velocity of the bubble and spike are a convenient and intuitive way to describe the evolution of the RTI. However, they represent the evolution of RTI in a single direction and dimension; their depiction of the system is constrained. The non-equilibrium evolution of RTI systems, especially compressible RTI systems, is complex and multidimensional. There will be complex physical processes in the evolution of the compressible RTI system, such as the transformation among the expansion and compression work, the internal energy, the kinetic energy, and the potential energy. As a result, the fluid's density, temperature, and velocity changed. By paying attention to the evolution of fluid density, we can better understand the physical mechanics of these complex processes. In the two-dimensional case, we can use the proportion of area occupied by the heavy fluid to describe the density variation of the heavy and light fluids.

\begin{figure}[tbp]
\center{\epsfig{file=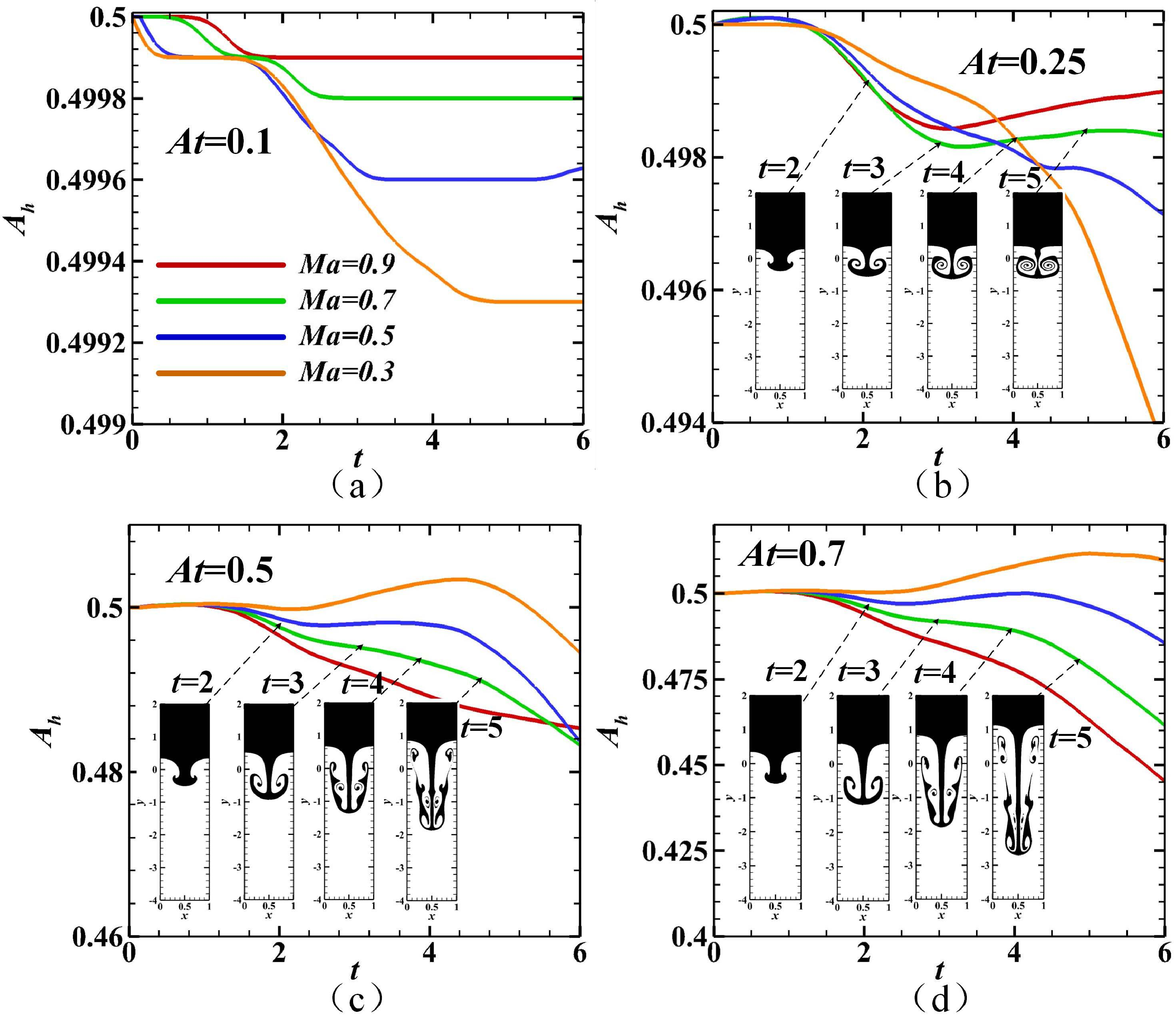,
width=0.85\textwidth,trim=0 0 0 0, clip }}
\caption{The evolution of the proportion of area occupied by the heavy fluid ${{A}_{h}}$ for (a) $At=0.1$, (b) $At=0.25$, (c) $At=0.5$, and (d) $At=0.7$.} \label{Fig8}
\end{figure}

Figures \ref{Fig8}(a)$-$\ref{Fig8}(d) illustrate the evolution of the proportion of area occupied by the heavy fluid ${{A}_{h}}$ under varying $Ma$ for $At = 0.1, 0.25, 0.5$, and 0.7, respectively. To intuitively correlate the evolution of ${{A}_{h}}$ with the evolution stages of RTI, we take the case of $Ma=0.7$ as an example and insert the flow field contour at the corresponding time point in Figs. \ref{Fig8}(a)$-$\ref{Fig8}(d).

\begin{figure}[tbp]
\center{\epsfig{file=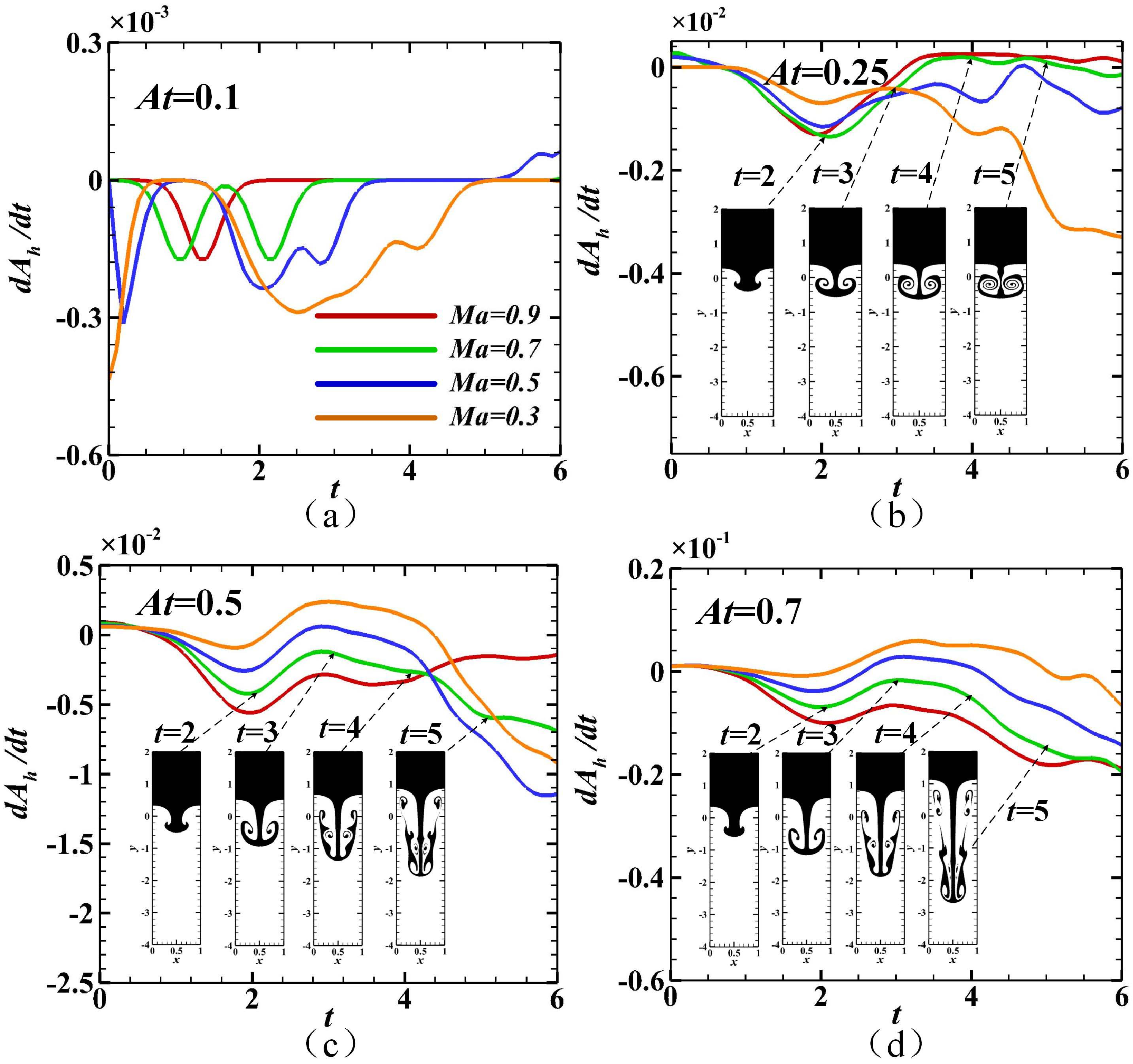,
width=0.85\textwidth,trim=0 0 0 0, clip }}
\caption{The evolution of ${d{{A}_{h}}}/{dt}$ for (a) $At=0.1$, (b) $At=0.25$, (c) $At=0.5$, and (d) $At=0.7$.} \label{Fig9}
\end{figure}

As shown in Fig. \ref{Fig8}, the evolution of ${{A}_{h}}$ over time presents specific stages across all $At$ conditions. The effect of compressibility on ${{A}_{h}}$ shows clear differences across various $At$ conditions. To provide a clearer depiction of the temporal evolution stages of ${{A}_{h}}$, the evolution of ${d{{A}_{h}}}/{dt}$ is derived by taking the derivative of ${{A}_{h}}$ to time, as illustrated in Fig. \ref{Fig9}. Similarly, we inserted the flow field contour under $ Ma=0.7$ at the corresponding time point in the corresponding positions in Figs. \ref{Fig9}(b)$-$\ref{Fig9}(d).

In the following, with the help of Figs. \ref{Fig8} and \ref{Fig9}, let’s try to understand the physical process of the compressible RTI's evolution under various $At$ and $Ma$ conditions from the ${{A}_{h}}$ perspective. Fig. \ref{Fig8}(a) shows the evolution of ${{A}_{h}}$ under different $Ma$ conditions when $At=0.1$. It can be seen that when $At=0.1$, the variation of ${{A}_{h}}$ over time is minimal. This is because the fluid's expansion and compression are closely related to the mutual flow between light and heavy fluids. During RTI development, the heavy fluid moves downward into the light fluid, causing an increase in pressure that compresses the heavy fluid. Conversely, the light fluid flows upward into the heavy fluid, causing a gradual decrease in pressure and an outward expansion of the light fluid.
In Fig. \ref{Fig2}(b), we can see that when $At=0.1$, the density difference between the light and heavy fluids at the initial interface is slight. Thus, during the downward movement of the heavy fluid, the density of the light fluid around the spike quickly exceeds that of the heavy fluid inside the spike. The spike's buoyant force will soon exceed the gravitational force, forming an upward resultant force that inhibits the spike's downward movement. As $Ma$ increases, the density and pressure gradients intensify. This led to a stronger inhibition effect on the spike's downward movement.
As shown in Fig. \ref{Fig4}, under the conditions of $Ma=0.5, 0.7$, and 0.9, the bubble's motion and the spike's motion are confined to a very narrow range near the initial interface. Thus, the pressure change of the spike and bubble during the movement is minimal. Therefore, the expansion and compression of light and heavy fluids are limited. This explains why the ${{A}_{h}}$ variation is so small. In the case of $Ma=0.3$, it is evident that the spike moves downward a certain distance. However, due to the light fluid's weak density stratification, the change in pressure within the range of the spike's movement is still relatively small. So, under these circumstances, the expansion and compression of light and heavy fluids are also small.

When $At=0.25$, the density difference between the light and heavy fluids at the initial interface increases, causing the bubble and spike's motion range to no longer be as severely suppressed as when $At=0.1$. As a result, the ${{A}_{h}}$ variation range increases.

As shown in Fig. \ref{Fig8}(b), the effect of compressibility on ${{A}_{h}}$ varies over time. This can be understood with the help of Fig. \ref{Fig9}(b). According to Fig. \ref{Fig9}(b), ${d{{A}_{h}}}/{dt}$ is positive and increases with increasing $Ma$ before $t=1$. This is mainly because, in the early stage, the heavy fluid within the spike expands to the surrounding light fluid due to the higher pressure, as shown in Fig. \ref{Fig7}. This led to a positive value of ${d{{A}_{h}}}/{dt}$ during this process. Furthermore, the higher the $Ma$, the greater the pressure difference and expansion. So, ${d{{A}_{h}}}/{dt}$ increases with increasing $Ma$.
With the development of RTI, between $t=1$ and 2, the spike accelerates downward into the light fluid. During this process, the pressure surrounding the spike accelerates to become larger, which makes the heavy fluid compressed and the ${{A}_{h}}$ smaller. Moreover, as $Ma$ increases, so does the downward pressure gradient. As a result, the ${d{{A}_{h}}}/{dt}$ is negative during this period and decreases as $Ma$ increases.
After $t=2$, an upward-curling KH vortex starts forming at the spike's head, as shown in the flow field contour in Fig. \ref{Fig9}(b). A portion of the heavy fluid expands through the upward-curling KH vortex, and its density decreases. At this stage, the evolution of ${{A}_{h}}$ is impacted by two factors. On the one hand, the spike's downward movement causes the heavy fluid to be constantly compressed, reducing ${{A}_{h}}$. On the other hand, the formation of the KH vortex causes heavy fluids to curl and expand upward, which increases ${{A}_{h}}$. There is a competition between the two influences.
During $t=2$ to 3, the spike velocity gradually decreases, as shown in Fig. \ref{Fig3}(b). This leads to a decreased influence of compression on the heavy fluid. So, ${d{{A}_{h}}}/{dt}$ gradually increases at this stage, mainly due to the upward expansion of the heavy fluid through the KH vortex.
After $t=3$, the evolutionary trend of ${d{{A}_{h}}}/{dt}$ under different $Ma$ conditions exhibits differences. For $Ma=0.3$, ${d{{A}_{h}}}/{dt}$ begins to decrease. However, in the cases of $Ma=0.5, 0.7$, and 0.9, ${d{{A}_{h}}}/{dt}$ continues to increase. This is because, after $t=3$, the evolution of the spike velocity under different $Ma$ conditions appears different. In the case of $Ma=0.3$, the spike velocity starts to increase again, and the effect of the compression on the spike due to its accelerating downward motion gradually increases so that ${d{{A}_{h}}}/{dt}$ decreases again. For cases of $Ma=0.5, 0.7$, and 0.9, the spike velocity keeps decelerating. This caused ${d{{A}_{h}}}/{dt}$ to be primarily affected by the expansion of the heavy fluid through the KH vortex, resulting in a gradual increase.
After $t=4$, for the cases of $Ma=0.7$ and 0.9, the spike and bubble velocity becomes negative. This is because the motion of the bubble and spike is limited to a specific range in these circumstances, as depicted in the flow field diagrams for $t=4$ and 5 in Fig. 9(b). Therefore, the pressure variation on the spike and bubble over time is minimal, and ${d{{A}_{h}}}/{dt}$ is nearly zero. 

\begin{figure}[tbp]
\center{\epsfig{file=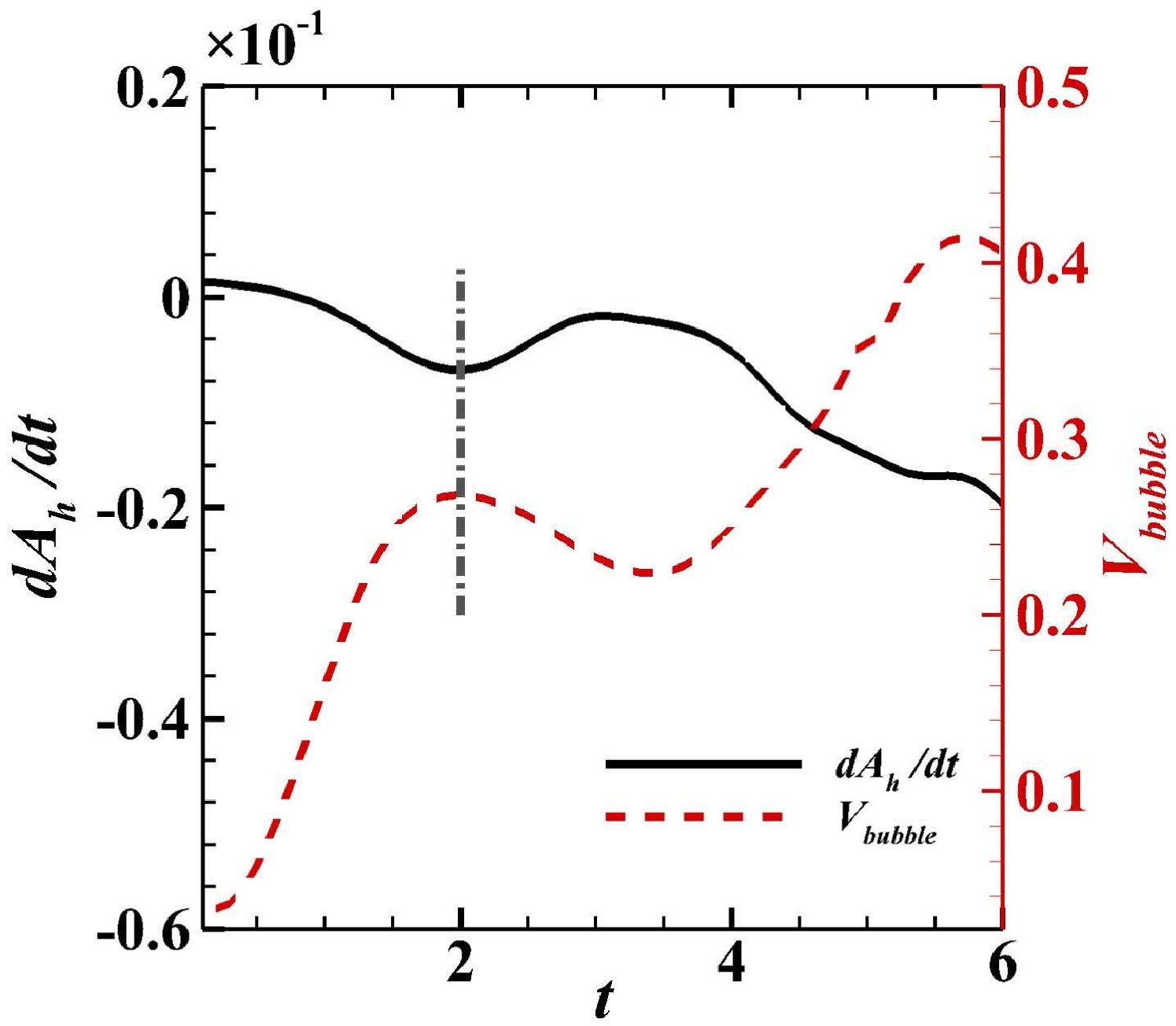,
width=0.65\textwidth,trim=2.5 2.5 2.5 2.5, clip }}
\caption{Evolution of ${d{{A}_{h}}}/{dt}$ and ${{V}_{bubble}}$ for $At=0.7$ and $Ma=0.7$. The vertical grey dash-dot line represents the moment $t=2.0$.} \label{Fig10}
\end{figure}

As $At$ increases further, the density difference between the light and heavy fluids at the initial interface becomes more pronounced. As a result, the restricted impact of compressibility on bubble and spike movement is further reduced.
It can be seen that when $At=0.5$, only in the case of $Ma=0.9$ does ${d{{A}_{h}}}/{dt}$ gradually increase after $t=3$. This is because the spike is inhibited from reaccelerating after $t=3$, only in the case of $Ma=0.9$, as shown in Fig. \ref{Fig3}.
When $At=0.7$, under all $Ma$ values considered here, ${d{{A}_{h}}}/{dt}$ gradually decreases at the later stages due to the reacceleration of the spike.
Through the investigation of ${{A}_{h}}$ and ${d{{A}_{h}}}/{dt}$, it was found that the first minimum value of ${d{{A}_{h}}}/{dt}$ can serve as a criterion for the bubble velocity to reach the first maximal value.
Here, we take the case under the conditions of $At=0.7$ and $Ma=0.7$ as an example to show the evolution of ${d{{A}_{h}}}/{dt}$ and bubble velocity, as depicted in Fig. \ref{Fig10}.

The evolution of non-equilibrium systems is multifaceted; as the saying goes, "It's a range viewed from the face and peaks viewed from the side, assuming different shapes viewed from far and wide."
The descriptions of the evolution of non-equilibrium systems from different perspectives are often complementary and related, but not interchangeable. The evolution of ${{A}_{h}}$ and ${d{{A}_{h}}}/{dt}$ is closely linked to the developmental stage of RTI. Although the evolution of the bubble and spike velocity and the evolution of ${{A}_{h}}$ and ${d{{A}_{h}}}/{dt}$ are interrelated and consistent, it can be seen that the evolution of ${{A}_{h}}$ and ${d{{A}_{h}}}/{dt}$ enhanced our understanding of the physical process of the system's evolution.

\subsubsection{Entropy production of the system}
An increase in entropy is an important product of the evolution of a system from a non-equilibrium state to an equilibrium state. The entropy production rate is closely related to the non-equilibrium state of the system and its evolution. As shown in Eq.(\ref{Eq.15}), the entropy production rate of the system can be divided into two parts: the part related to heat flow and the part related to viscous stress, denoted by ${{\dot{S}}_{NOEF}}$ and ${{\dot{S}}_{NOMF}}$, respectively.

\begin{figure}[tbp]
\center{\epsfig{file=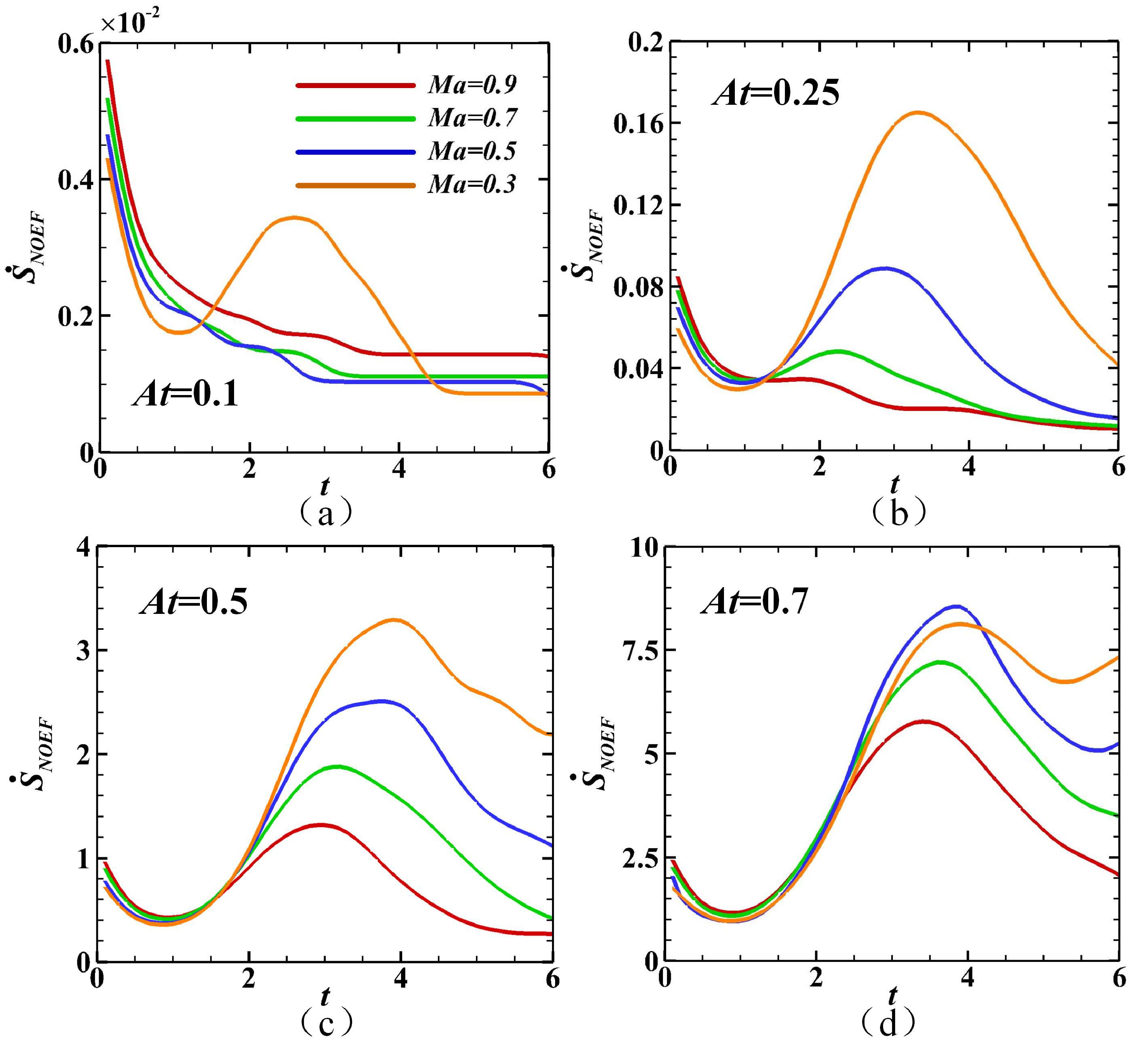,
width=0.8\textwidth,trim=0 0 0 0, clip }}
\caption{The evolution of ${{\dot{S}}_{NOEF}}$ for (a) $At=0.1$, (b) $At=0.25$, (c) $At=0.5$, and (d) $At=0.7$.} \label{Fig11}
\end{figure}

\begin{figure}[tbp]
\center{\epsfig{file=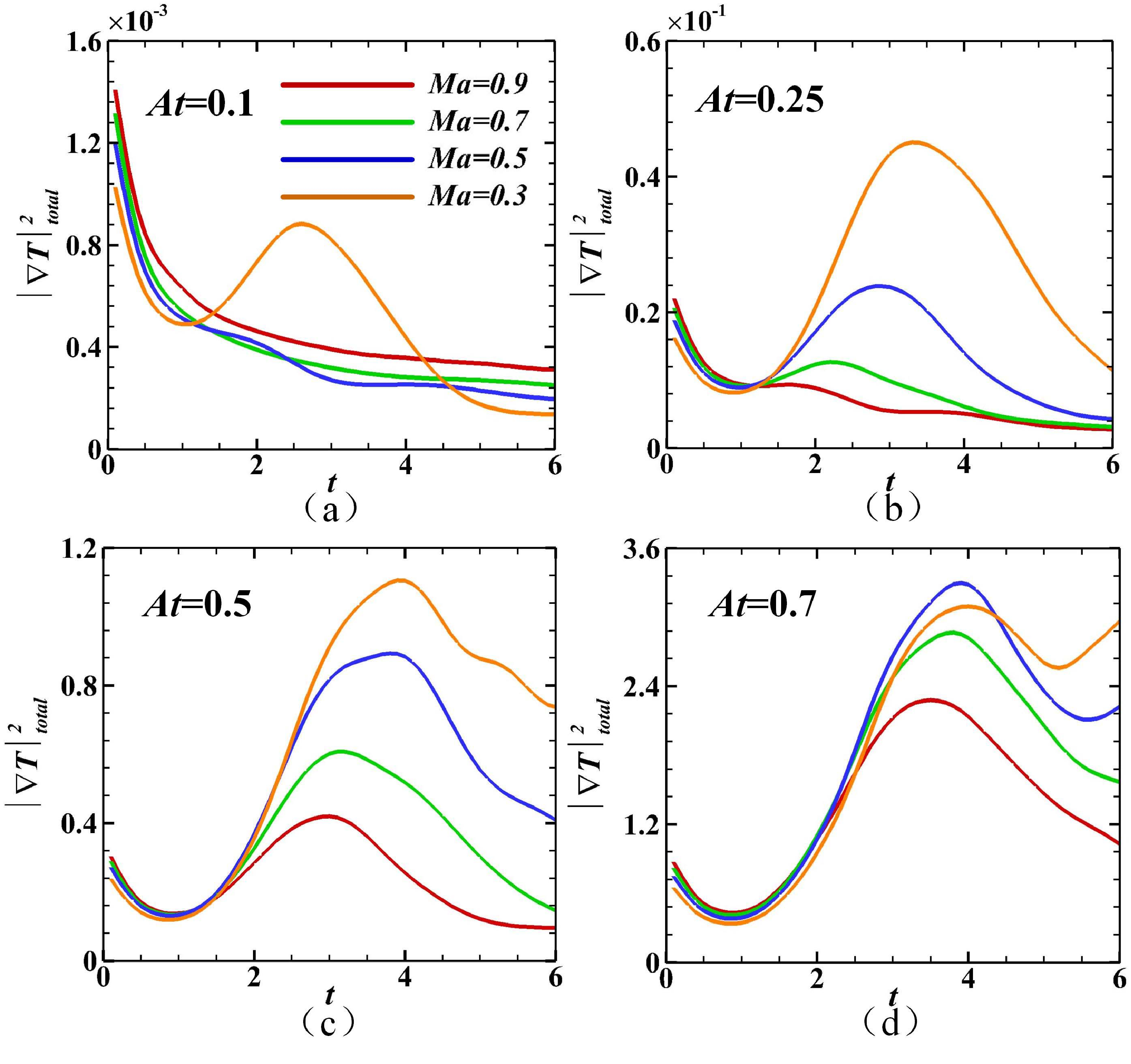,
width=0.8\textwidth,trim=0 0 0 0, clip }}
\caption{The evolution of $\left| \nabla T \right|_{total}^{2}$ for (a) $At=0.1$, (b) $At=0.25$, (c) $At=0.5$, and (d) $At=0.7$.} \label{Fig12}
\end{figure}

Figures \ref{Fig11}(a)$-$\ref{Fig11}(d) show the evolution of ${{\dot{S}}_{NOEF}}$ under different $Ma$ for $At=0.1, 0.25, 0.5$, and 0.7, respectively. It can be seen that the effect of $At$ and $Ma$ on the evolution of ${{\dot{S}}_{NOEF}}$ is complicated. As $At$ increases, the evolutionary trend of ${{\dot{S}}_{NOEF}}$ over time changes, and so does the effect of $Ma$ on ${{\dot{S}}_{NOEF}}$. As shown in Eq.(\ref{Eq.18}), ${{\dot{S}}_{NOEF}}$ can be written as a function of the square of the temperature gradient mode ${{\left| \nabla T \right|}^{2}}$. So the evolution of the total flow field ${{\left| \nabla T \right|}^{2}}$ can, to a certain extent, represent the evolution of ${{\dot{S}}_{NOEF}}$. Figs. \ref{Fig12}(a)$-$\ref{Fig12}(d) show the evolution of $\left| \nabla T \right|_{total}^{2}$ under different $Ma$ for $At=0.1, 0.25, 0.5$, and 0.7, respectively. It is evident that the evolution of ${{\dot{S}}_{NOEF}}$ is very consistent with the evolution of $\left| \nabla T \right|_{total}^{2}$. Therefore, we can try to figure out what causes the evolutionary pattern of ${{\dot{S}}_{NOEF}}$ by analyzing the evolution of $\left| \nabla T \right|_{total}^{2}$.

\begin{figure}[tbp]
\center{\epsfig{file=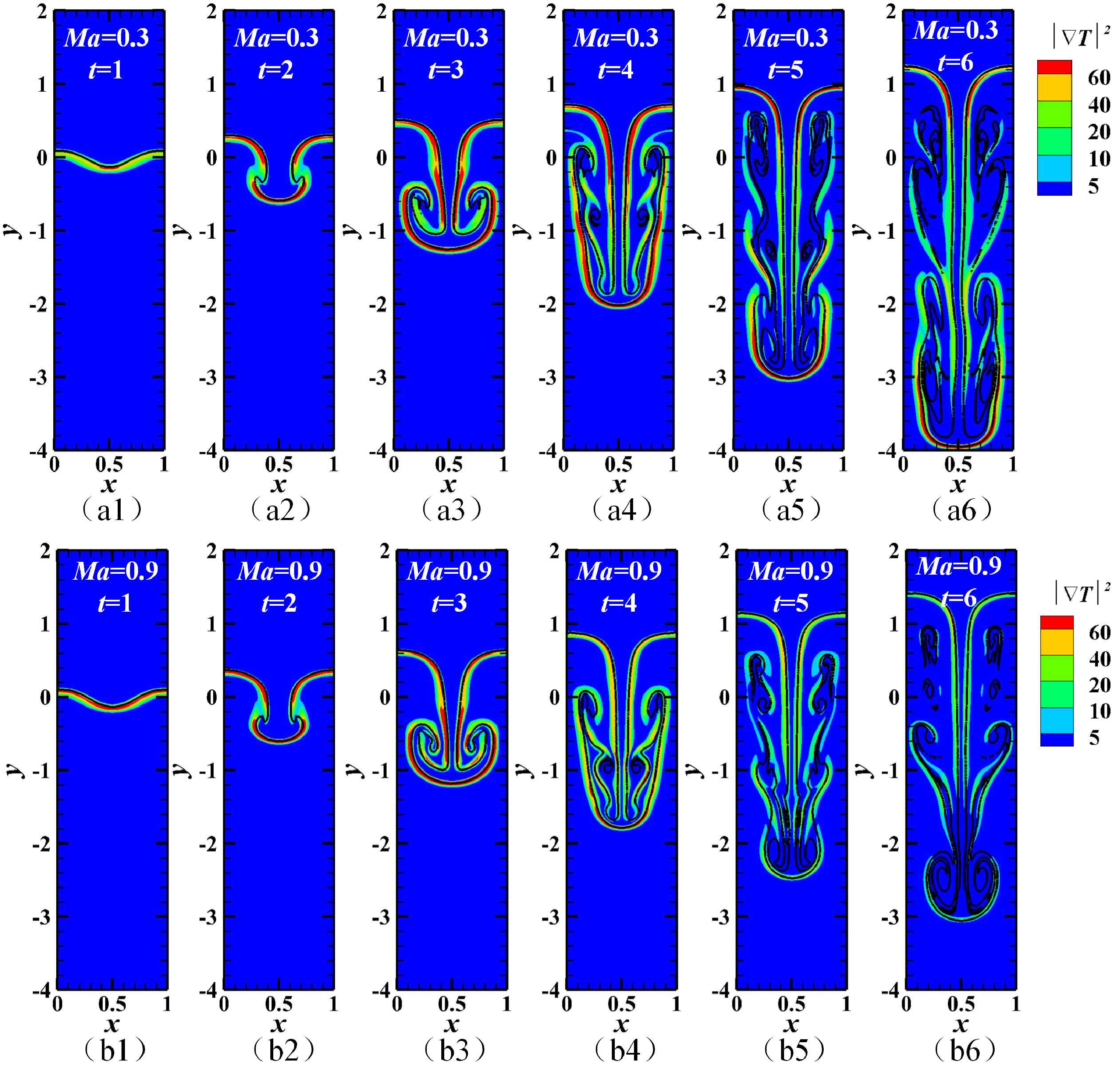,
width=1.0\textwidth,trim=0 0 0 0, clip }}
\caption{The contour plots of ${{\left| \nabla T \right|}^{2}}$ for $At=0.7$ under two $Ma$ conditions: (a) $Ma=0.3$ and (b) $Ma=0.9$. From (1) to (6): $t=1, 2, 3, 4, 5$, and 6, respectively. The dark lines represent the interface of the heavy and light fluids.} \label{Fig13}
\end{figure}
At the initial time, the heavy fluid is at a uniformly lower temperature, while the light fluid is at a uniformly higher temperature. The temperature gradient is only present at the interface of the light and heavy fluids. As RTI grows, the temperature gradient develops with the development of the interface of the light and heavy fluids. Fig. \ref{Fig13} shows the contour plots of ${{\left| \nabla T \right|}^{2}}$ for $At=0.7$ under $Ma=0.3$ and $Ma=0.9$ conditions. The dark lines represent the interface of the heavy and light fluids. It can be seen that ${{\left| \nabla T \right|}^{2}}$ is mainly present at the interface, which indicates the tight association between $\left| \nabla T \right|_{total}^{2}$ and the evolution of the interface.

\begin{figure}[tbp]
\center{\epsfig{file=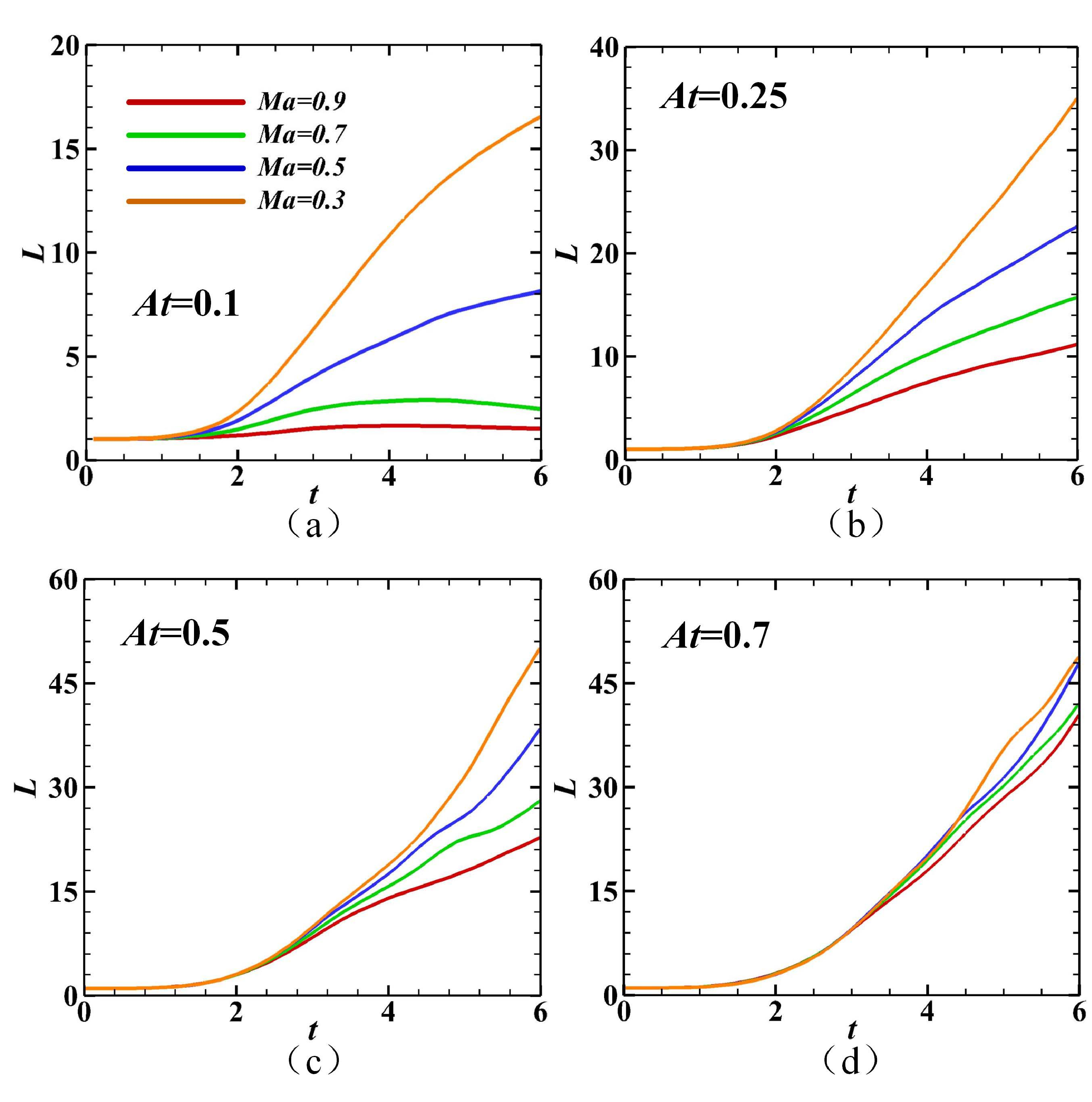,
width=0.8\textwidth,trim=0 0 0 0, clip }}
\caption{The evolution of $L$ for (a) $At=0.1$, (b) $At=0.25$, (c) $At=0.5$, and (d) $At=0.7$.} \label{Fig14}
\end{figure}

\begin{figure}[tbp]
\center{\epsfig{file=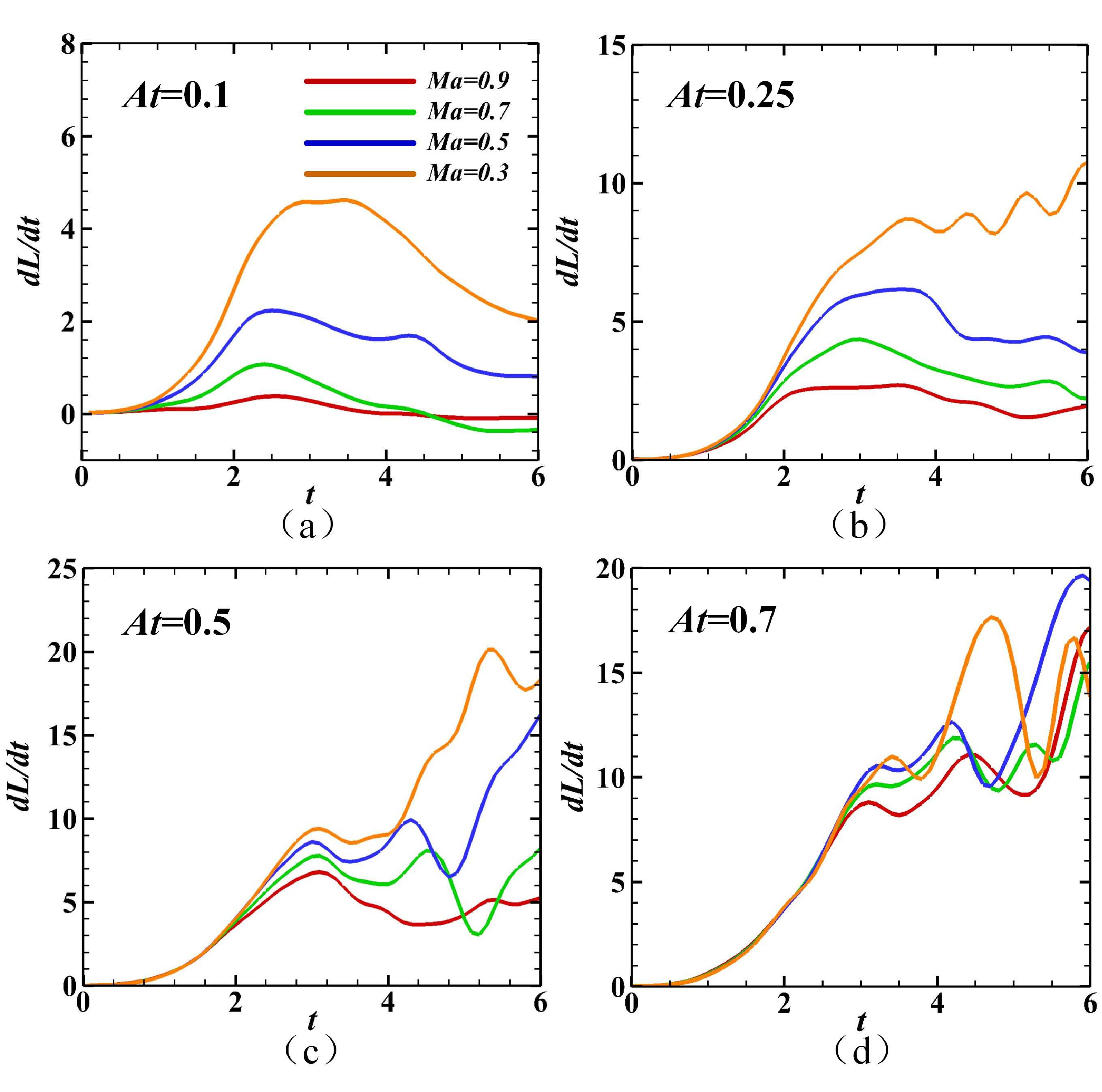,
width=0.8\textwidth,trim=0 0 0 0, clip }}
\caption{The evolution of ${dL}/{dt}$ for (a) $At=0.1$, (b) $At=0.25$, (c) $At=0.5$, and (d) $At=0.7$.} \label{Fig15}
\end{figure}

\begin{figure}[tbp]
\center{\epsfig{file=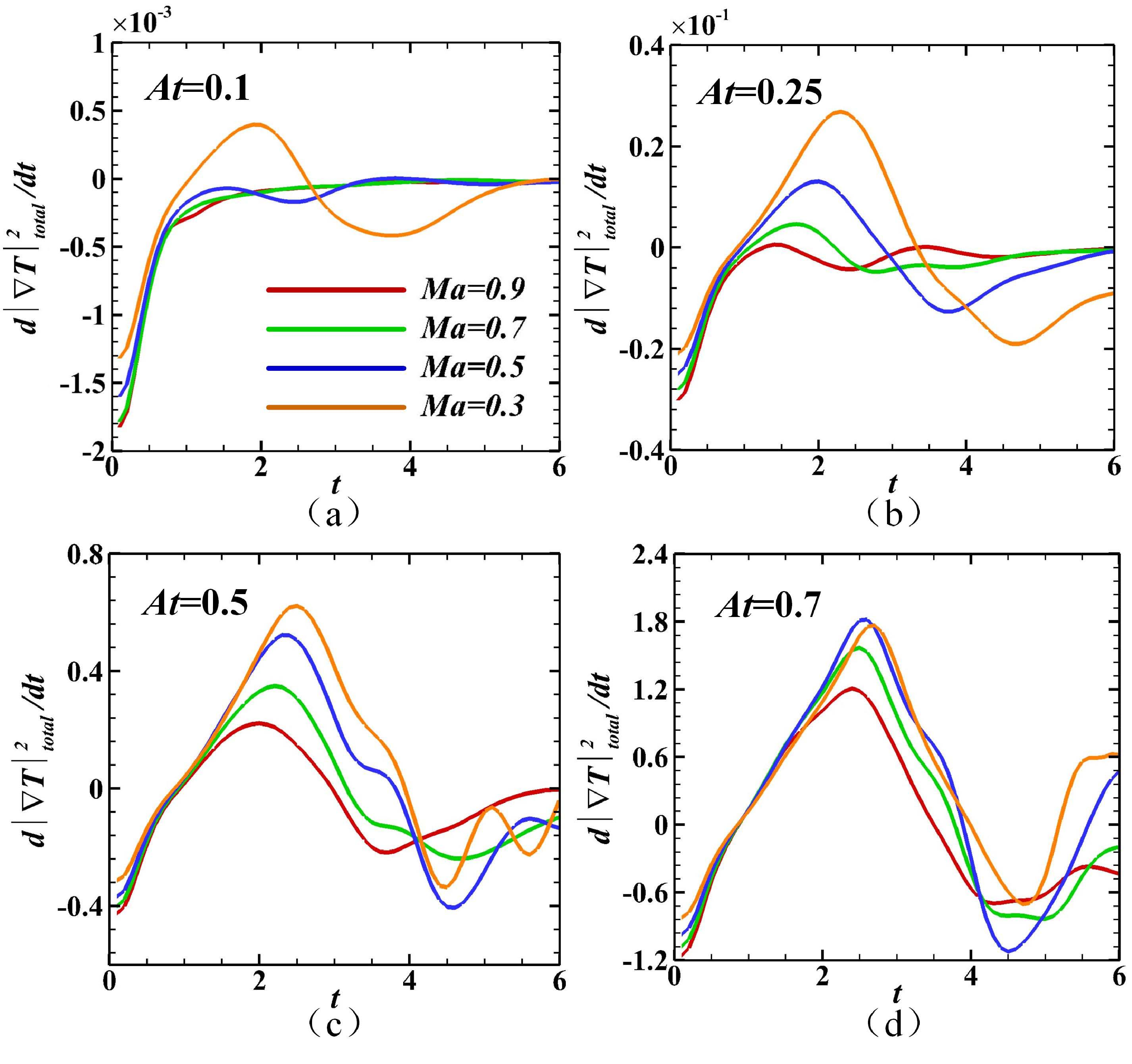,
width=0.8\textwidth,trim=0 0 0 0, clip }}
\caption{The evolution of ${d\left| \nabla T \right|_{total}^{2}}/{dt}$ for (a) $At=0.1$, (b) $At=0.25$, (c) $At=0.5$, and (d) $At=0.7$.} \label{Fig16}
\end{figure}

In addition to the interfacial length $L$, the heat transfer between light and heavy fluids, as well as the energy transfer between the internal energy (${{E}_{b}}$) and the expansion and compression work (${{W}_{CE}}$), will also influence the evolution of ${{\left| \nabla T \right|}^{2}}$. These impacts are coupled. 
Figs. \ref{Fig14}(a)$-$\ref{Fig14}(d) illustrate the evolution of $L$ under different $Ma$ for $At=0.1, 0.25, 0.5,$ and 0.7, respectively. The increase in $L$ has two effects on the evolution of $\left| \nabla T \right|_{total}^{2}$. On the one hand, it increases the area of the system where the temperature gradient exists, which tends to increase $\left| \nabla T \right|_{total}^{2}$; on the other hand, it also increases the area of the system where the heat transfer exists, which increases the total heat transfer rate of the system and thus tends to decrease $\left| \nabla T \right|_{total}^{2}$. The evolution of $\left| \nabla T \right|_{total}^{2}$ results from the competition between increasing temperature gradient area and increasing total heat transfer rate. To better understand how the increase in $L$ affects the evolution of $\left| \nabla T \right|_{total}^{2}$, we calculate the time derivatives for both variables, as depicted in Figs. \ref{Fig15} and \ref{Fig16}, respectively.

Figures \ref{Fig14}(a) and \ref{Fig15}(a) show the evolution of $L$ and ${dL}/{dt}$ under different $Ma$ conditions for $At=0.1$, respectively. It can be seen that before $t=1$, ${dL}/{dt}$ is small, and $L$ grows slowly with time. During this time, $\left| \nabla T \right|_{total}^{2}$ is mainly reduced due to the heat transfer at the interface of the light and heavy fluids. Thus, ${d\left| \nabla T \right|_{total}^{2}}/{dt}$ is negative and gradually increases with time, as shown in Fig. \ref{Fig16}(a). 

Between $t=1$ and 2, in the case of lower compressibility with $Ma=0.3$, ${dL}/{dt}$ increases rapidly. This led to a rapid increase in area where the temperature gradient exists. Therefore, ${d\left| \nabla T \right|_{total}^{2}}/{dt}$ is positive and increases rapidly at this time. This is why $\left| \nabla T \right|_{total}^{2}$ experiences high growth following the initial period of decline. However, for the cases of $Ma=0.5, 0.7$, and 0.9, the growth of ${dL}/{dt}$ is restricted by compressibility. As shown in Fig. \ref{Fig15}(a), the growth of ${dL}/{dt}$ decreases significantly as $Ma$ increases. Because the increase in $L$ is insufficient to induce ${d\left| \nabla T \right|_{total}^{2}}/{dt}$ to increase to a positive value, $\left| \nabla T \right|_{total}^{2}$ is still reduced, mainly due to heat transfer. 
When $Ma = 0.5, 0.7$, and 0.9, ${dL}/{dt}$ gradually decreases between $t=2$ and 6. Thus, $\left| \nabla T \right|_{total}^{2}$ is primarily affected by heat transfer and decreases during this time interval.

For the lower compressibility case with $Ma=0.3$, during the time interval $t=2$ and 4, the growth of ${dL}/{dt}$ gradually slows down and eventually reaches a certain value. During this period, ${d\left| \nabla T \right|_{total}^{2}}/{dt}$ is mainly affected by the increasing rate of heat transfer and decreases gradually.
Between $t=4$  and 6, ${dL}/{dt}$ gradually decreases. During this period, ${d\left| \nabla T \right|_{total}^{2}}/{dt}$ is primarily affected by heat transfer and is still a negative value. However, the gradual decline in ${dL}/{dt}$ also results in a gradual weakening of the trend of increasing the overall heat transfer rate of the system. The growth of the overall heat transfer rate is mainly reduced by the gradual decrease in the temperature gradient. Thus, ${d\left| \nabla T \right|_{total}^{2}}/{dt}$ gradually increases to zero. 

With the increase in $At$, the inhibitory effect of compressibility on RTI's development diminishes. From the perspective of the morphological quantity $L$ (${dL}/{dt}$), this is reflected by the increasing growth rate of $L$ (${dL}/{dt}$) with the increase in $At$ and the diminishing gap between $L$ (${dL}/{dt}$) under different $Ma$ conditions with increasing $At$, as shown in Figs. 14 and 15. This can be used to explain why the evolutionary trend of $\left| \nabla T \right|_{total}^{2}$ varies with the increase in $At$. As shown in Fig. \ref{Fig12}, in the case of $At=0.1$, only under the condition of $Ma=0.3$ does $\left| \nabla T \right|_{total}^{2}$ exhibit an increase after the initial decrease. With the increase in $At$, the growth rate of $L$ (${dL}/{dt}$) increases obviously. Thus, in the cases of $At=0.5$ and 0.7, $\left| \nabla T \right|_{total}^{2}$ exhibits an increase after the initial decrease across all $Ma$ conditions. 

Figures \ref{Fig12}(b) and \ref{Fig12}(c) demonstrate that the impact of compressibility on $\left| \nabla T \right|_{total}^{2}$ can be categorized into two distinct stages in the cases of $At=0.25$ and 0.5. In the early stage, when $\left| \nabla T \right|_{total}^{2}$ decreases with time, $\left| \nabla T \right|_{total}^{2}$ increases as $Ma$ increases, and after that, $\left| \nabla T \right|_{total}^{2}$ decreases as $Ma$ increases. 
Before $t=1$, ${dL}/{dt}$ is very small. Thus, $L$ grows slowly and is almost the same across various $Ma$ conditions. The difference in $\left| \nabla T \right|_{total}^{2}$ under different $Ma$ conditions is mainly due to the difference in ${{\left| \nabla T \right|}^{2}}$ at the interface. 
In the previous analysis, we knew that the pressure of the fluid inside and below the spike was higher than that of the light fluid on both sides at the initial moment, as shown in Fig. \ref{Fig7}. This causes the fluid in the spike and below the spike to press the light fluid on both sides.
The greater the $Ma$, the greater the pressure difference and the stronger the tendency to press. The heavy fluid inside the spike expands and presses the surrounding light fluid, and the degree of expansion and press increases with the increase in $Ma$.
The expansion of the heavy fluid resulted in a decrease in its temperature, whereas the compression of the light fluid led to an increase in its temperature. Thus, ${{\left| \nabla T \right|}^{2}}$ at the interface increases with increasing $Ma$, as shown in Figs. \ref{Fig13}(a1) and \ref{Fig13}(b1). This is why $\left| \nabla T \right|_{total}^{2}$ increases as $Ma$ increases before $t=1$. 
After $t=1$, due to the development of RTI, ${dL}/{dt}$ increases rapidly. Affected by the rapid increase in ${dL}/{dt}$, $\left| \nabla T \right|_{total}^{2}$ begins to increase. At this time, due to the inhibitory effect of compressibility on ${dL}/{dt}$, $\left| \nabla T \right|_{total}^{2}$ decreases with increasing $Ma$.

In fact, as $At$ increases, the bubble's and spike's motion range increases. This led to an increase in ${{W}_{CE}}$ of the heavy and light fluids. This, in turn, makes the effect of the energy transformation between ${{E}_{b}}$ and ${{W}_{CE}}$ on the evolution of $\left| \nabla T \right|_{total}^{2}$ evident. As shown in Figs. \ref{Fig13}(a3)$-$\ref{Fig13}(a5) and Figs. \ref{Fig13}(b3)$-$\ref{Fig13}(b5), it can be seen that the magnitude of ${{\left| \nabla T \right|}^{2}}$ at the interface is obviously different. This resulted from the difference in the transformation between ${{E}_{b}}$ and ${{W}_{CE}}$ in the cases of $Ma=0.3$ and $Ma=0.9$. The initial rise and subsequent decline of $\left| \nabla T \right|_{total}^{2}$ with $Ma$ during $t=2.5$ to 4.5, as depicted in Fig. \ref{Fig12}(d), is also due to the difference in the transformations between ${{E}_{b}}$ and ${{W}_{CE}}$ under different $Ma$ conditions. To figure out how $Ma$ affects the value of $\left| \nabla T \right|_{total}^{2}$, let's use the scenario depicted in Fig. \ref{Fig12}(d) at $t=4$ as an example. This is when $\left| \nabla T \right|_{total}^{2}$ shows a pronounced non-monotonic variation evolution with $Ma$.

\begin{figure}[tbp]
\center{\epsfig{file=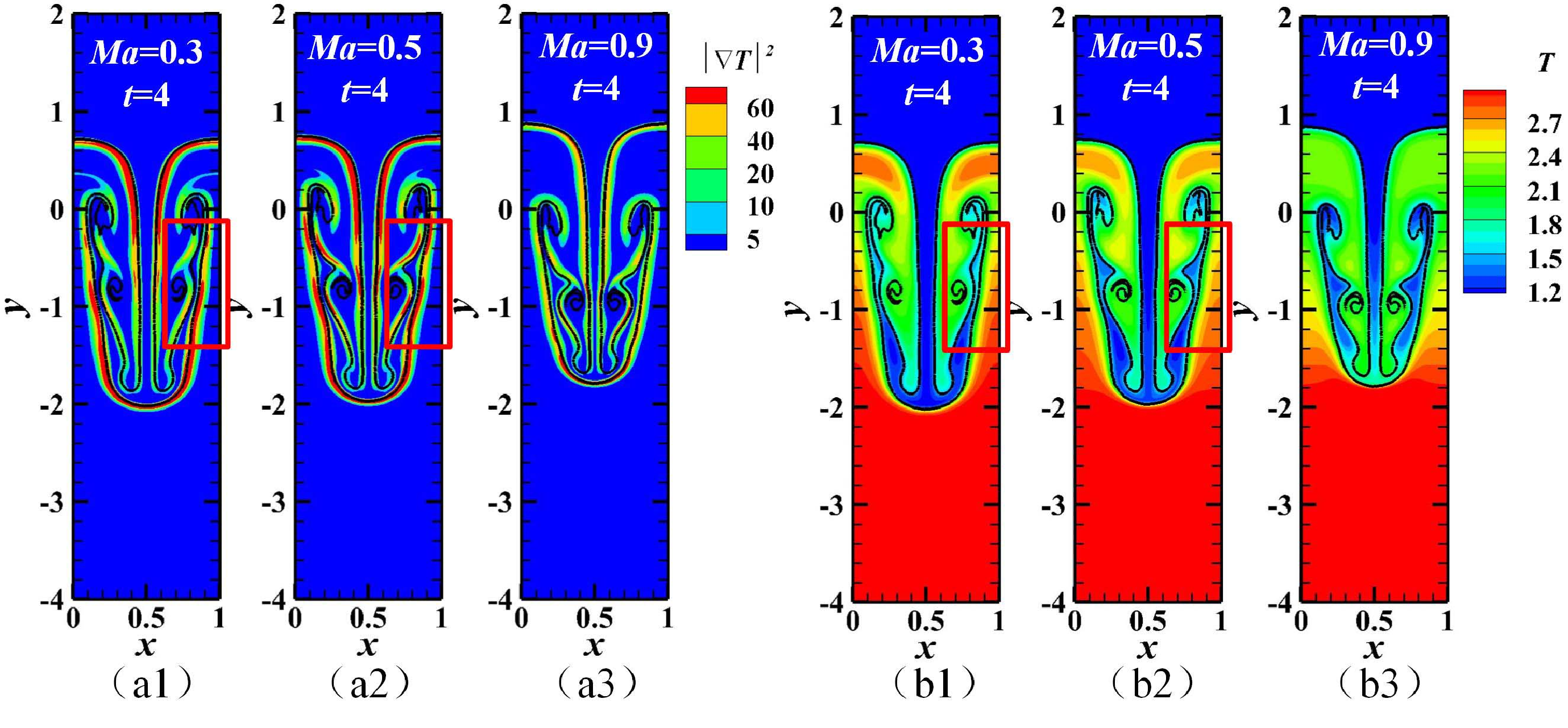,
width=1.0\textwidth,trim=0 0 0 0, clip }}
\caption{The contour plots of (a) ${{\left| \nabla T \right|}^{2}}$ and (b) $T$ for $At=0.7$ at $t=4$. From (1) to (3): $Ma=0.3, 0.5$, and 0.9, respectively.} \label{Fig17}
\end{figure}

Figures \ref{Fig17}(a1)$-$\ref{Fig17}(a3) show the contour plots of ${{\left| \nabla T \right|}^{2}}$ at moment $t=4$ for $At=0.7$ under the conditions of $Ma=0.3, 0.5$, and 0.9, respectively. It can be seen that the length $L$ of the interface is almost the same for $Ma=0.3$ and $Ma=0.9$. The difference in $\left| \nabla T \right|_{total}^{2}$ is mainly caused by the difference in ${{\left| \nabla T \right|}^{2}}$ at the interface, as shown in the red boxes in Figs. \ref{Fig17}(a1) and \ref{Fig17}(a2). Figs. \ref{Fig17}(b1)$-$\ref{Fig17}(b3) show the contour plots of $T$ at $t=4$ for $Ma=0.3, 0.5$, and 0.9, respectively. It is observed that as $Ma$ increases, the compression of the heavy fluid inside the spike head increases during the downward movement of the spike. This leads to a higher internal energy and temperature for the heavy fluid inside the spike head. Meanwhile, the heavy fluid in the spike head curls and expands upward through the KH vortex. The upward expansion of heavy fluid through the KH vortex is primarily governed by two causes. On the one hand, compressibility has an inhibitory effect on the development of the KH vortex, which restricts the upward expansion; on the other hand, as compressibility increases, the degree of upward expansion of the light fluid below the bubble increases, which promotes the expansion of the heavy fluid within the KH vortex into the surrounding light fluid.

When $Ma=0.3$, compressibility has a relatively weak inhibitory effect on the development of the KH vortex; however, at the same time, the degree of upward expansion of the light fluid through the bubble is also limited. Thus, during the upward expansion, the heavy fluid is compressed by the light fluid. As a result, the temperature of the heavy fluid in the middle and tail of the KH vortex increases. As $Ma$ increases to 0.5, the upward expansion of the light fluid through the bubble intensifies. This results in the light fluid exerting less compression on the upwardly expanded heavy fluid. As a result, the temperature increase of the heavy fluid in the middle and tail of the KH vortex decreases relative to that under the condition of $Ma=0.3$. This lower increase in temperature leads to an increase in ${{\left| \nabla T \right|}^{2}}$ at the interface. As $Ma$ increases to 0.9, the expansion degree of the light fluid in the bubble further increases. Thus, the compression effect of the light fluid on the upwardly expanded heavy fluid further decreases, which results in a further decrease in the temperature increase of the heavy fluid in the middle and tail of the KH vortex. However, the expansion of the light fluid leads to a more noticeable decrease in its temperature. This results in lower ${{\left| \nabla T \right|}^{2}}$ at the KH vortex interface compared to the case with $Ma=0.5$.
In addition, at $Ma=0.9$, it is evident that increasing compressibility greatly suppresses the formation of KH vortices, leading to a significant reduction in the size of the KH vortex. This also contributes to the considerable decrease in $\left| \nabla T \right|_{total}^{2}$.
\begin{figure}[tbp]
\center{\epsfig{file=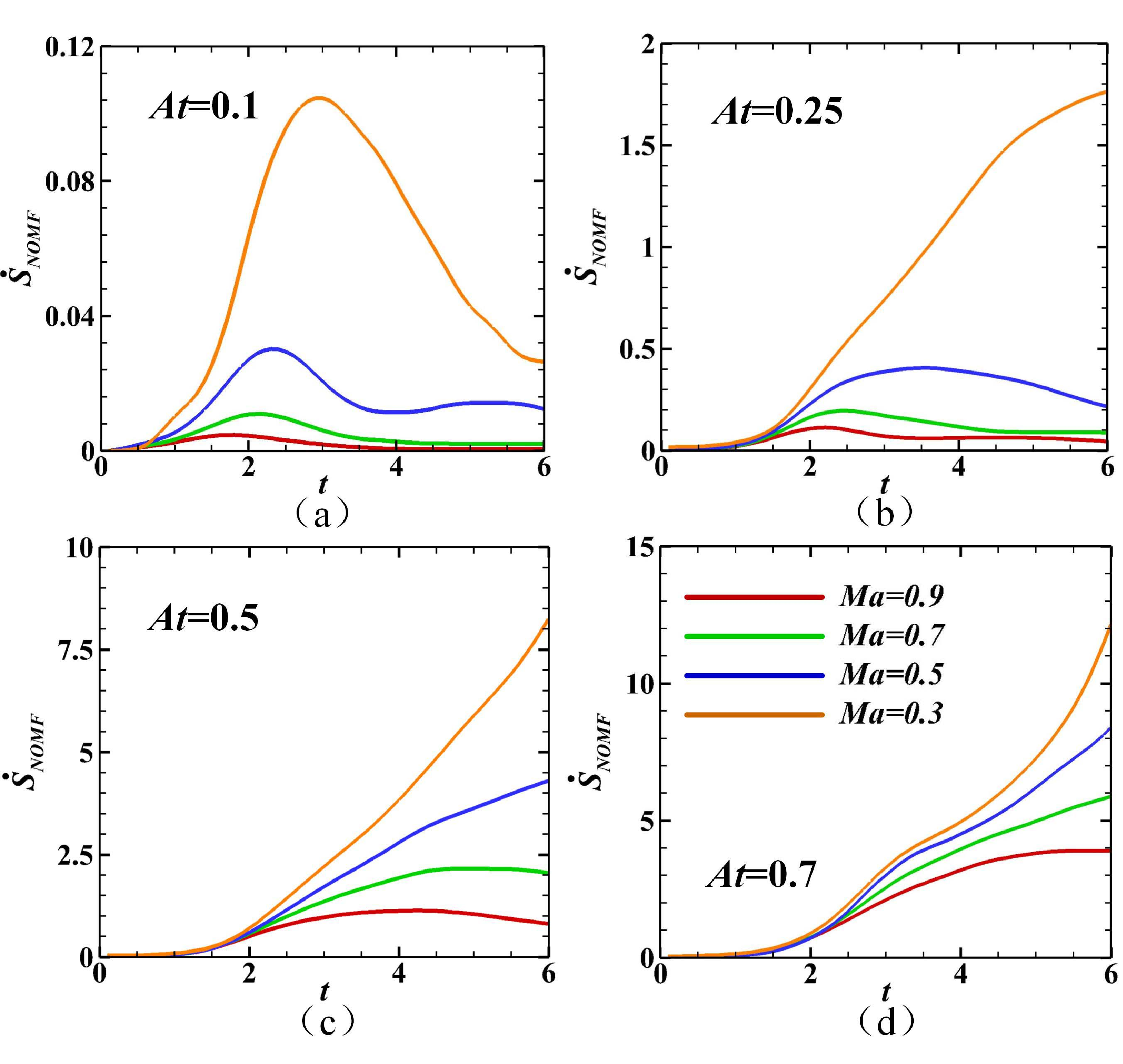,
width=0.8\textwidth,trim=0 0 0 0, clip }}
\caption{The evolution of ${{\dot{S}}_{NOMF}}$ for (a) $At=0.1$, (b) $At=0.25$, (c) $At=0.5$, and (d) $At=0.7$.} \label{Fig18}
\end{figure}

Figures \ref{Fig18}(a)$-$\ref{Fig18}(d) show the evolution of ${{\dot{S}}_{NOMF}}$ under different $Ma$ conditions for $At=0.1, 0.25, 0.5$, and 0.7, respectively. It can be seen that for $At=0.1$, ${{\dot{S}}_{NOMF}}$ across various $Ma$ conditions all exhibit an evolutionary pattern of first increasing and then decreasing with time. As $At$ increases, the evolutionary pattern of ${{\dot{S}}_{NOMF}}$ gradually changes to increase all the time. However, with the increase in $At$, the effect of $Ma$ on ${{\dot{S}}_{NOMF}}$ does not change; ${{\dot{S}}_{NOMF}}$ decreases with the increase in $Ma$ for all $At$.

\begin{figure}[tbp]
\center{\epsfig{file=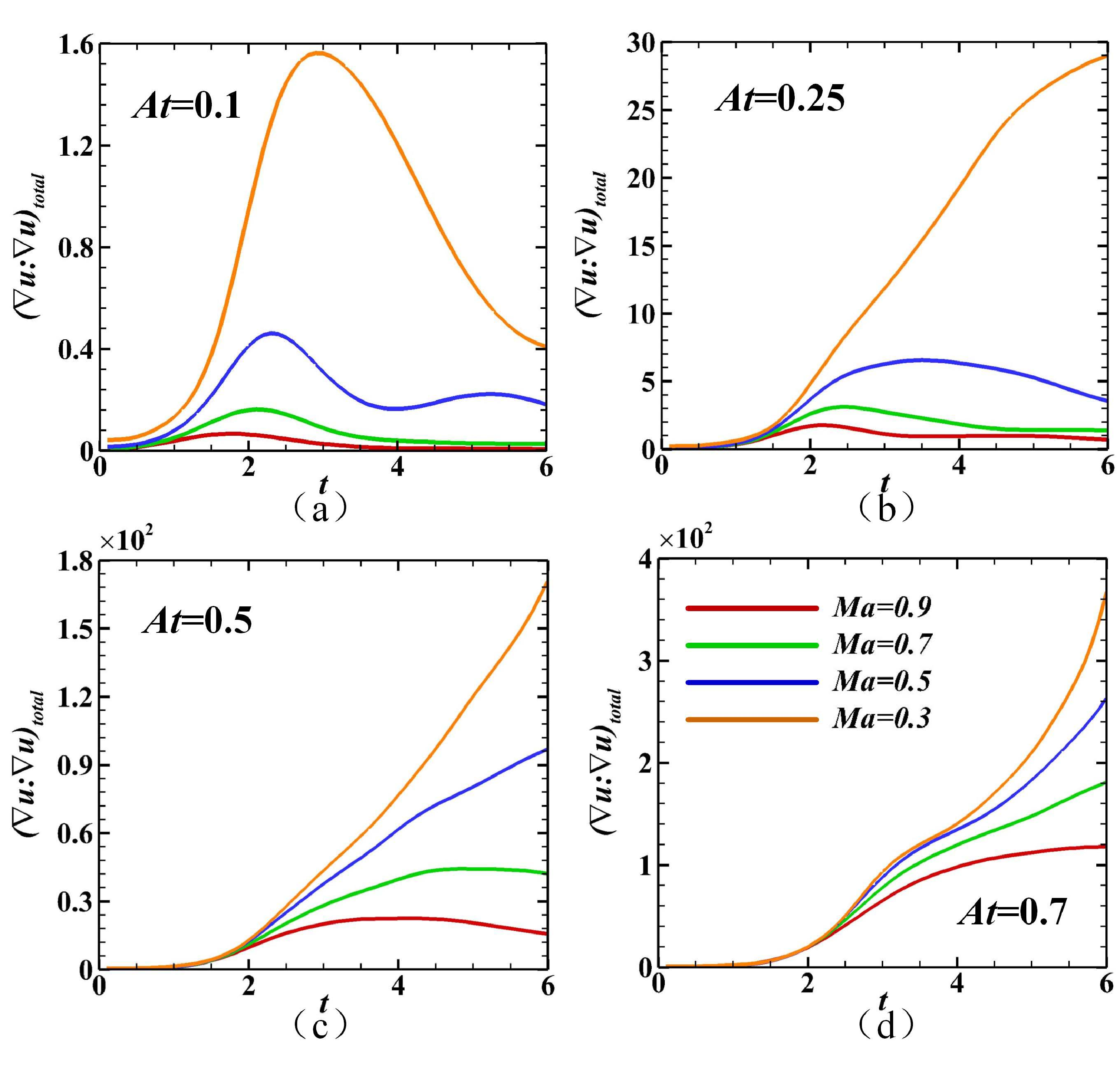,
width=0.8\textwidth,trim=0 0 0 0, clip }}
\caption{The evolution of ${{\left( \nabla \mathbf{u}:\nabla \mathbf{u} \right)}_{total}}$ for (a) $At=0.1$, (b) $At=0.25$, (c) $At=0.5$, and (d) $At=0.7$.} \label{Fig19}
\end{figure}

\begin{figure}[tbp]
\center{\epsfig{file=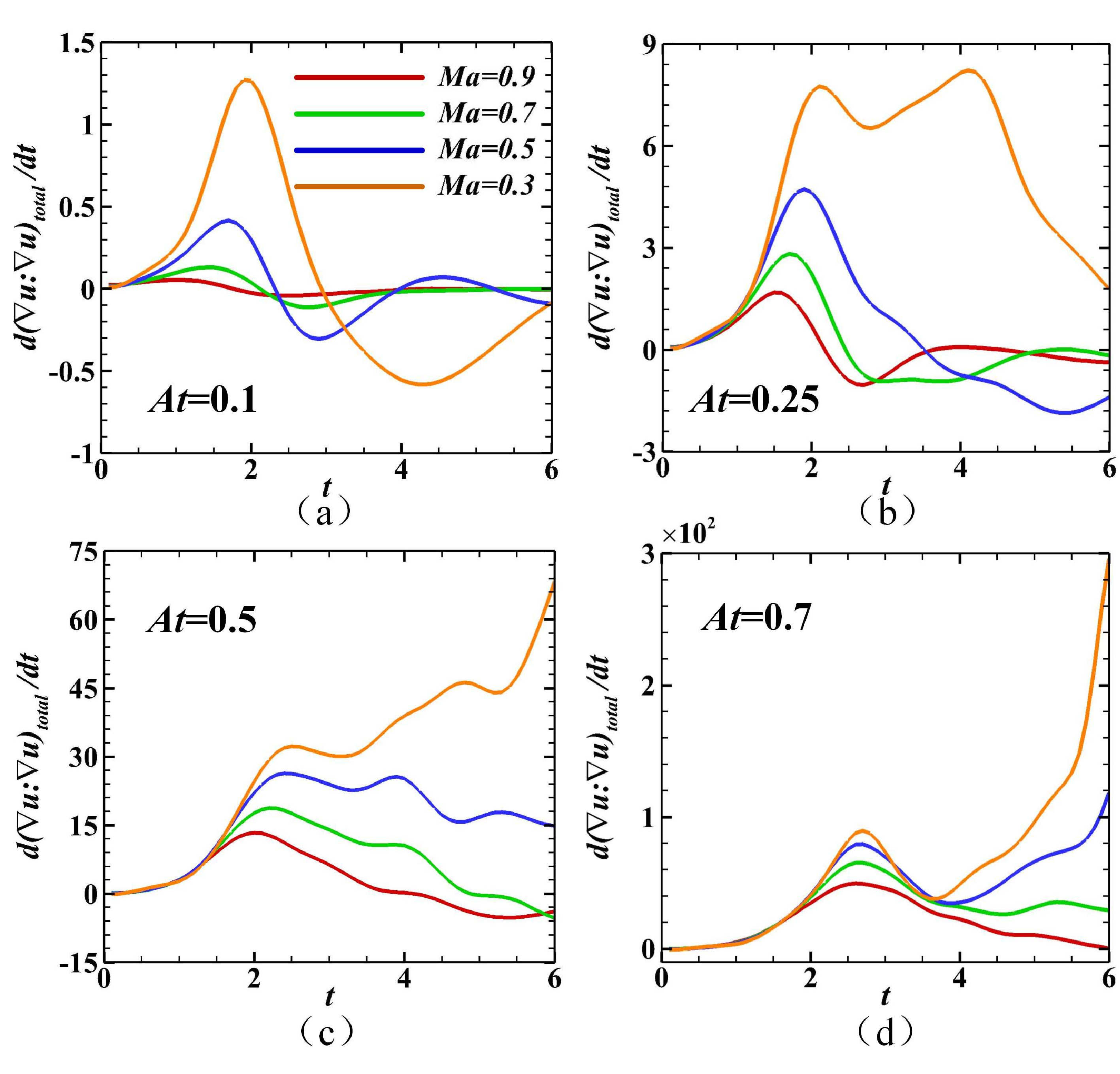,
width=0.8\textwidth,trim=0 0 0 0, clip }}
\caption{The evolution of ${d{{\left( \nabla \mathbf{u}:\nabla \mathbf{u} \right)}_{total}}}/{dt}$ for (a) $At=0.1$, (b) $At=0.25$, (c) $At=0.5$, and (d) $At=0.7$.} \label{Fig20}
\end{figure}

As shown in Eq.(\ref{Eq.19}), ${{\dot{S}}_{NOMF}}$ can be written as a function of the square of the second-order reduction of the velocity gradient, denoted as $\nabla \mathbf{u}:\nabla \mathbf{u}$. 
So, the evolution of the total flow field $\nabla \mathbf{u}:\nabla \mathbf{u}$ can, to a certain extent, represent the evolution of ${{\dot{S}}_{NOMF}}$. Figs \ref{Fig19}(a)$-$\ref{Fig19}(d) show the evolution of ${{\left( \nabla \mathbf{u}:\nabla \mathbf{u} \right)}_{total}}$ under different $Ma$ conditions for $At=0.1, 0.25, 0.5$, and 0.7, respectively. The correlation between the evolution of ${{\dot{S}}_{NOMF}}$ and the evolution of ${{\left( \nabla \mathbf{u}:\nabla \mathbf{u} \right)}_{total}}$. 
To gain a better understanding of the evolution of ${{\left( \nabla \mathbf{u}:\nabla \mathbf{u} \right)}_{total}}$, we calculate the derivative of ${{\left( \nabla \mathbf{u}:\nabla \mathbf{u} \right)}_{total}}$ with respect to time, as depicted in Fig. \ref{Fig20}.

The velocity gradient in the evolution of RTI is mostly a result of the relative motion between the light and heavy fluids, which is predominantly observed at the interface between the two fluids. Therefore, in addition to the magnitude of the relative motion velocity, the evolution of $L$ also has an important effect on the evolution of  ${{\left( \nabla \mathbf{u}:\nabla \mathbf{u} \right)}_{total}}$. Firstly, we take the case with $At=0.1$ and $Ma=0.3$ as an example to provide a preliminary understanding of the physical process involved in the evolution of  ${{\left( \nabla \mathbf{u}:\nabla \mathbf{u} \right)}_{total}}$.

\begin{figure}[tbp]
\center{\epsfig{file=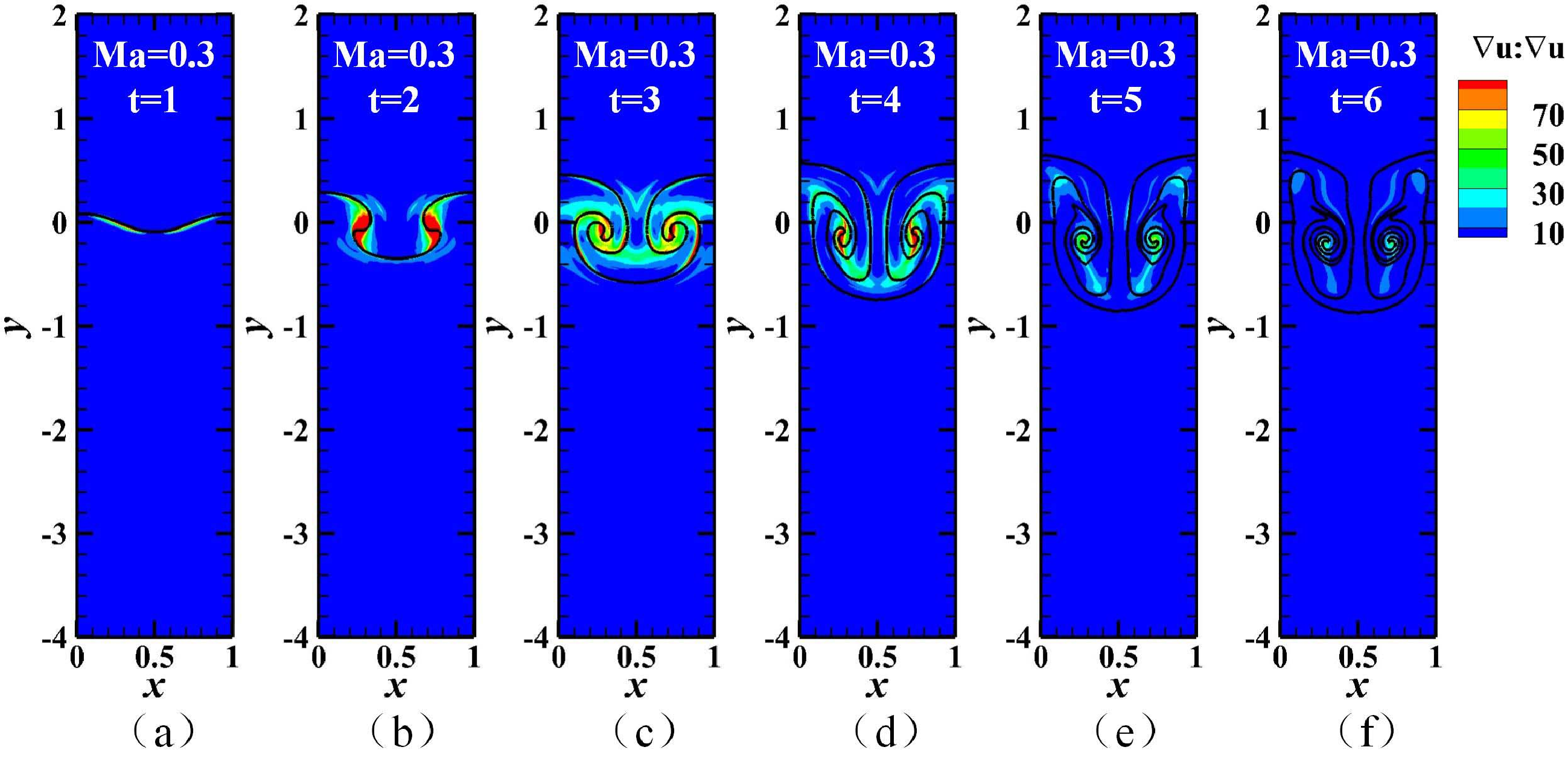,
width=1.0\textwidth,trim=0 0 0 0, clip }}
\caption{The contour plots of $\nabla \mathbf{u}:\nabla \mathbf{u}$ for $At=0.1$ and $Ma=0.3$. From (a) to (f): $t=1, 2, 3, 4, 5$, and 6, respectively.} \label{Fig21}
\end{figure}

Figs. \ref{Fig21}(a)$-$\ref{Fig21}(f) show the contour plots of $\nabla \mathbf{u}:\nabla \mathbf{u}$ under the conditions of $At=0.1$ and $Ma=0.3$ at $t=1, 2, 3, 4, 5$, and 6, respectively. It is clear that $\nabla \mathbf{u}:\nabla \mathbf{u}$ exists primarily where there is a large tangential velocity difference between the light and heavy fluids at the interface. Between t = 0 and 1, ${dL}/{dt}$ grows very slowly. Hence, the gradual increase in $\nabla \mathbf{u}:\nabla \mathbf{u}$ is mainly due to the slow increase in tangential velocity difference between the light and heavy fluids at the interface, as shown in Fig. \ref{Fig21}(a). During the time interval of $t=1$ and 2, the heavy fluid undergoes a downward acceleration to enter the light fluid. In contrast, the light fluid accelerates upward to enter the heavy fluid. Consequently, the tangential velocity difference between light and heavy fluids experiences a rapid increase, and simultaneously, ${dL}/{dt}$ also experiences a rapid increase, resulting in an acceleration increase of ${d{{\left( \nabla \mathbf{u}:\nabla \mathbf{u} \right)}_{total}}}/{dt}$, as depicted in Fig. \ref{Fig20}(a).

After $t=2$, both the bubble and spike decelerate, leading to a weakening of the relative motion between the light and heavy fluids. Consequently, there is a gradual decrease in ${d{{\left( \nabla \mathbf{u}:\nabla \mathbf{u} \right)}_{total}}}/{dt}$ with time. However, as can be seen in Fig. \ref{Fig15}(a), ${dL}/{dt}$ continues to increase during $t=2$ to 3 due to the upward curling of the KH vortex at the spike, which causes ${d{{\left( \nabla \mathbf{u}:\nabla \mathbf{u} \right)}_{total}}}/{dt}$ to remain positive at this period, although it gradually decreases with time. Between $t=3$ and 4, ${dL}/{dt}$ ceases to increase, and the motion of the bubble and spike continues to diminish, which causes ${d{{\left( \nabla \mathbf{u}:\nabla \mathbf{u} \right)}_{total}}}/{dt}$ to continue to decrease to a value less than zero. After $t=4$, ${d{{\left( \nabla \mathbf{u}:\nabla \mathbf{u} \right)}_{total}}}/{dt}$ gradually increases again. This is mainly attributed to the formation of numerous small-scale vortex structures near the interface, as shown in Figs. \ref{Fig21}(e)$-$\ref{Fig21}(f). The decreasing scale of the vortex structures retards the decay of the $\nabla \mathbf{u}:\nabla \mathbf{u}$.

As $At$ increases to 0.25, the evolution of ${d{{\left( \nabla \mathbf{u}:\nabla \mathbf{u} \right)}_{total}}}/{dt}$ under the condition of $Ma=0.3$ shows some differences in the later stages, as shown in Fig. \ref{Fig20}(b). It can be seen that in the case of $At=0.25$ and $Ma=0.3$, ${d{{\left( \nabla \mathbf{u}:\nabla \mathbf{u} \right)}_{total}}}/{dt}$ gradually increases again during the time interval of $t=3$ and 4, after a brief decline during $t=2$ to 3. This can be understood with the assistance of Fig. \ref{Fig3}(b), which shows that after a short period of decrease in the spike velocity during $t=2$ to 3, there is a reacceleration behavior of the spike during $t=3$ to 4. The downward reacceleration of the spike causes the velocity difference between the heavy and light fluids to increase again, which is the main reason why ${d{{\left( \nabla \mathbf{u}:\nabla \mathbf{u} \right)}_{total}}}/{dt}$ increases again during this period.

\begin{figure}[tbp]
\center{\epsfig{file=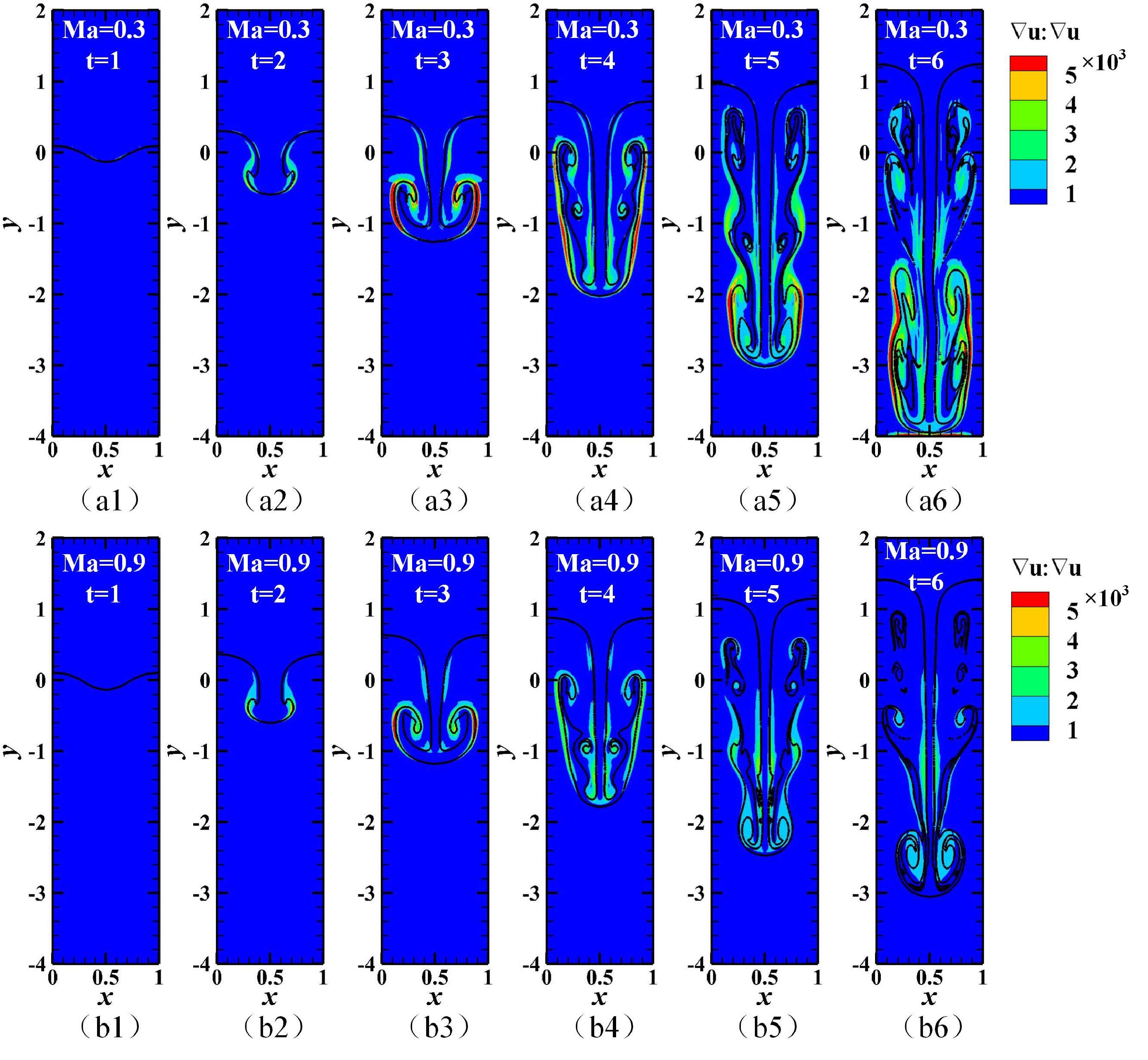,
width=1.0\textwidth,trim=0 0 0 0, clip }}
\caption{The contour plots of $\nabla \mathbf{u}:\nabla \mathbf{u}$ for $At=0.7$ under two $Ma$ conditions: (a) $Ma=0.3$, (b) $Ma=0.9$. From (1)$-$(6): $t=1, 2, 3, 4, 5$, and 6, respectively. The dark lines represent the interface of the heavy and light fluids.} \label{Fig22}
\end{figure}
Figures \ref{Fig20}(c) and \ref{Fig20}(d) show the evolution of ${d{{\left( \nabla \mathbf{u}:\nabla \mathbf{u} \right)}_{total}}}/{dt}$ for $At=0.5$ and 0.7, respectively. It can be seen that as $At$ further increases, ${d{{\left( \nabla \mathbf{u}:\nabla \mathbf{u} \right)}_{total}}}/{dt}$ appears to increase rapidly again at later times at low $Ma$, as shown in Fig. \ref{Fig20}(c) for $Ma=0.3$ and in Fig. \ref{Fig20}(d) for $Ma=0.3$ and $Ma=0.5$. 
From Figs. \ref{Fig3} and \ref{Fig15}, it can be seen that when $At=0.5$ and 0.7, the reacceleration of the spike and a rapid increase in ${dL}/{dt}$ do not only appear in the case under the lower $Ma$ conditions. For example, when $At=0.7$, there is a reacceleration of the spike and a rapid increase in ${dL}/{dt}$ under all $Ma$ conditions. However, the rapid increase in ${d{{\left( \nabla \mathbf{u}:\nabla \mathbf{u} \right)}_{total}}}/{dt}$ at later times appears only under the conditions of $Ma=0.3$ and 0.5. So, in this circumstance, the rapid increase in ${d{{\left( \nabla \mathbf{u}:\nabla \mathbf{u} \right)}_{total}}}/{dt}$ at a later stage is not only determined by the reacceleration of the spike and the rapid increase in ${dL}/{dt}$.
To understand the underlying reason for this behavior, we make the contour plots of $\nabla \mathbf{u}:\nabla \mathbf{u}$ for $At=0.7$ under $Ma=0.3$ and $Ma=0.9$ conditions, as depicted in Fig. \ref{Fig22}. It can be seen that, although the difference in $L$ between the two cases is small, it is evident that there is a significant difference in the value of $\nabla \mathbf{u}:\nabla \mathbf{u}$ at the interface after $t=3$. Due to the inhibition effect of compressibility on the deposition of vorticity at the interface, the value of $\nabla \mathbf{u}:\nabla \mathbf{u}$ under higher $Ma$ is smaller than that under lower $Ma$. This phenomenon explains the absence of reacceleration of ${d{{\left( \nabla \mathbf{u}:\nabla \mathbf{u} \right)}_{total}}}/{dt}$ under high $Ma$ conditions.

\section{Conclusions and discussions}

The relatively little research on strongly compressible fluid systems within the field of RTI has resulted in a deficiency in understanding their physical processes and mechanics. 
In the case of strong compressibility, the density of the fluid from the upper layer (originally heavy fluid) may become smaller than that of the surrounding (originally light) fluid, thus invalidating the early method of distinguishing light and heavy fluids based on density. In this paper, tracer particles are incorporated into a single-fluid discrete Boltzmann method (DBM) model that considers the van der Waals potential. By using tracer particles to label the matter-particle sources (where the matter particle comes from: the originally heavy fluid or the originally light fluid), the study of the compressible RTI's evolution is realized in a single-fluid framework.
The focus of DBM is on physical modeling before simulation and complex physical field analysis after simulation. Our simulation results contain two parts: one that can be given by the NS model and one that cannot be given or is not conveniently given by the NS model. A series of methods for analyzing complex physical fields are provided by DBM, allowing the study to be carried out in depth.

It is found that compressibility has an inhibitory effect on the spike velocity, and the inhibitory effect decreases with increasing $At$. Unlike spike velocity, the influence of compressibility on bubble velocity shows a staged behavior with increasing $At$. There exists a critical Atwood number ${At}_C$. For $At$ values below ${At}_C$, compressibility exhibits an inhibitory effect on bubble velocity, and the strength of the inhibition decreases with increasing $At$. However, when $At={At}_C$, compressibility is observed to promote the bubble velocity before the bubble reacceleration stage. Interestingly, due to the competition between the inhibitory effect of compressibility on the downward motion of the spike and its inhibitory effect on the development of the KH vortex, the bubble does not exhibit reaccelerating behavior for either higher or lower compressibility. For the specific scenario studied in this paper, the ${At}_C$ is about 0.7. The amplitude and velocity of the bubble and spike are a convenient and intuitive way to describe the evolution of the RTI. However, they represent the evolution of RTI in a single direction and dimension; their depiction of the system is constrained. It is worth pointing out that these results can also be given using the NS model.

The non-equilibrium evolution of RTI systems, especially compressible RTI systems, is complex and multidimensional. 
It has been discovered that for compressible RTI systems, the variation in fluid density is closely related to the stages of RTI evolution. In the two-dimensional case, we can express the change in the densities of the light and heavy fluids in relation to the evolution of the area proportion occupied by the heavy fluid, denoted as ${{A}_{h}}$. It is found that the first minimum point of ${d{{A}_{h}}}/{dt}$ can serve as a criterion for identifying the moment when the bubble velocity reaches its initial maximum value.
The evolution of entropy production rate related to heat conduction (${{\dot{S}}_{NOEF}}$) is a complex process influenced by multiple factors, including (\romannumeral 1) length of the interface between light and heavy fluids $L$, (\romannumeral 2) heat transfer rate, and (\romannumeral 3) the interconversion between the work performed by expansion and compression of light and heavy fluids ${{W}_{CE}}$ and internal energy ${{E}_{b}}$. The $At$, $Ma$, and the stage of the evolution of RTI collectively determine the evolution of ${{\dot{S}}_{NOEF}}$. Under conditions of low $At$ and high $Ma$, the bubble and spike movements are confined to a narrow region around the initial interface, so ${{\dot{S}}_{NOEF}}$ is mainly affected by heat transfer and gradually decreases. As $At$ increases, the inhibitory effect of compressibility on the motions of the bubble and spike diminishes. This results in an increase in $L$ and ${{W}_{CE}}$, which in turn causes an increase in the impact of ${{W}_{CE}}$ and the variation of ${{W}_{CE}}$ on ${{\dot{S}}_{NOEF}}$. Under these circumstances, the evolutionary trend of the rate of ${{\dot{S}}_{NOEF}}$ is characterized by a decrease, followed by an increase, and then another decrease over time. Entropy production rate related to viscous stress (${{\dot{S}}_{NOMF}}$) is mainly influenced by the interface length and the tangential velocity difference between light and heavy fluids. Compressibility exerts a suppressive effect on ${{\dot{S}}_{NOMF}}$ for all $At$ values. For low $At$, due to the inhibitory effect on bubble and spike movements, ${{\dot{S}}_{NOMF}}$ demonstrates an evolutionary trend of initially growing and subsequently declining over time. As the value of $At$ grows, the movements of the bubble and spike grow more pronounced; thus, ${{\dot{S}}_{NOMF}}$ tends to exhibit continuous growth over time. The RTI system is a complex, non-equilibrium system, various perspectives on non-equilibrium evolution are interrelated and complementary, and they can form a more complete image of the system by combining each other. The incorporation of these complex physical field analysis techniques is a good complement to the description of RTI evolution. However, these results cannot be or are not conveniently given by the NS model. The DBM's complex physical field analysis method makes this part of the study possible or easier.

In this work, we started with a relatively simple case, a two-dimensional single-mode RTI. The two-dimensional problem can be considered a special case of the three-dimensional problem, in which all physical quantities are uniform and unchanged in the third dimension (Z).
Due to the additional third dimension, there will be some differences between the three- and two-dimensional cases. In three-dimensional problems, the change in densities of light and heavy fluids will be expressed as the change in volume occupied by the heavy fluid. The interface between light and heavy fluids is described by the interface area, rather than the interface length.
For a three-dimensional single-mode RTI problem, there should also be a turning point in the change rate of the proportion of the volume occupied by the heavy fluid, which can serve as a criterion for the bubble velocity reaching its initial maximum value. Non-equilibrium quantities, such as the system's temperature gradient and velocity gradient and their change rates, will also show different variation patterns at various stages of RTI development. The factors that determine their evolutionary patterns are the same as in the two-dimensional case. These are consistent with the two-dimensional, single-mode RTI.
However, even in simple three-dimensional, single-mode scenarios, the coupling between perturbations in various directions makes the evolution of the heavy fluid volume and the interface more complex, necessitating further study.

The evolution of RTI systems in late time may exhibit different behavior from that in early time, but late-time behavior studies require either using sufficiently large systems or considering the effects of various boundaries. The former significantly increases the demands on computation resources and workload, while the latter requires the design of reasonable kinetic boundary conditions for different scenarios. The simulation results presented in this paper are not sufficient to answer the characteristics of the RTI systems' late-time behavior, such as under what conditions the RTI will enter the turbulent stage. Furthermore, they cannot address how material properties, environmental, and other factors influence the mixing of matter and energy in the turbulence stage if it occurs. 
The simulation result in this paper shows that with the increase in $At$, the inhibitory effect of compressibility on the spike velocity diminishes. Under the condition of $At = 0.7$ and $Ma = 0.3$, at $t = 6$, the spike nearly reaches the lower boundary. When $t > 6$, the spike will hit the lower boundary, which will significantly impact the flow field. To avoid considering the influence of the boundary on the flow field, the results after $t = 6$ were not analyzed in our work. For convenient comparison in all cases, we have only taken the results before $t = 6$ for all the calculations in our work. Nevertheless, we can make some remarks for extended late-time calculations for different Atwood number conditions.
In cases with $At = 0.1$, compressibility strongly inhibits bubble and spike motion due to the small density difference between the light and heavy fluids. After $t = 5$, the velocities of the bubbles and spikes remain near zero. The bubbles and spikes are confined near the initial interface due to the effect of compressibility. As time evolves, the systems will slowly reach equilibrium as heat transfers between the heavy and light fluids and vorticity dissipates. This is consistent with the evolutionary pattern observed at $t = 6$. Under this circumstance, no major changes are expected in the evolution during extended later-time calculations.
However, for cases with $At = 0.25$ and 0.5, it is evident that the relative motion of the bubble and the spike remains strong at $t = 6$. Therefore, further studies are needed to determine whether the vortex's deposition occurring after $t = 6$ and before the spike reaches the bottom boundary will cause the bubble and spike to accelerate again, and whether these reaccelerations will alter the evolution of the system's non-equilibrium qualities.

In addition to the DBM modeling and analysis method for complex fluids with strong compressibility, the series of physical cognitions in this paper provides a more accurate understanding of the RTI kinetics and a helpful reference for the development of corresponding regulation techniques.

\section*{Acknowledgements}

The authors thank Yanbiao Gan, Feng Chen, Chuandong Lin, Huilin Lai, Zhipeng Liu, Ge Zhang, Yiming Shan, Dejia Zhang, Jiahui Song, Hanwei Li, Yingqi Jia, and Xuan Zhang for helpful discussions on DBM. This work was supported by the National Natural Science Foundation of China (under Grant Nos. 12172061,12102397 and 12101064), the Opening Project of State Key Laboratory of Explosion Science and Safety Protection (Beijing Institute of Technology) (Grant No. KFJJ23-02M), the Foundation of National Key Laboratory of Shock Wave and Detonation Physics (Grant No. JCKYS2023212003),  and the 2023 Computational Physics Key Laboratory Youth Fund Sponsored Project (under Grant No. 6241A05QN23001).

\section*{Appendix}

\begin{figure}[tbp]
\center{\epsfig{file=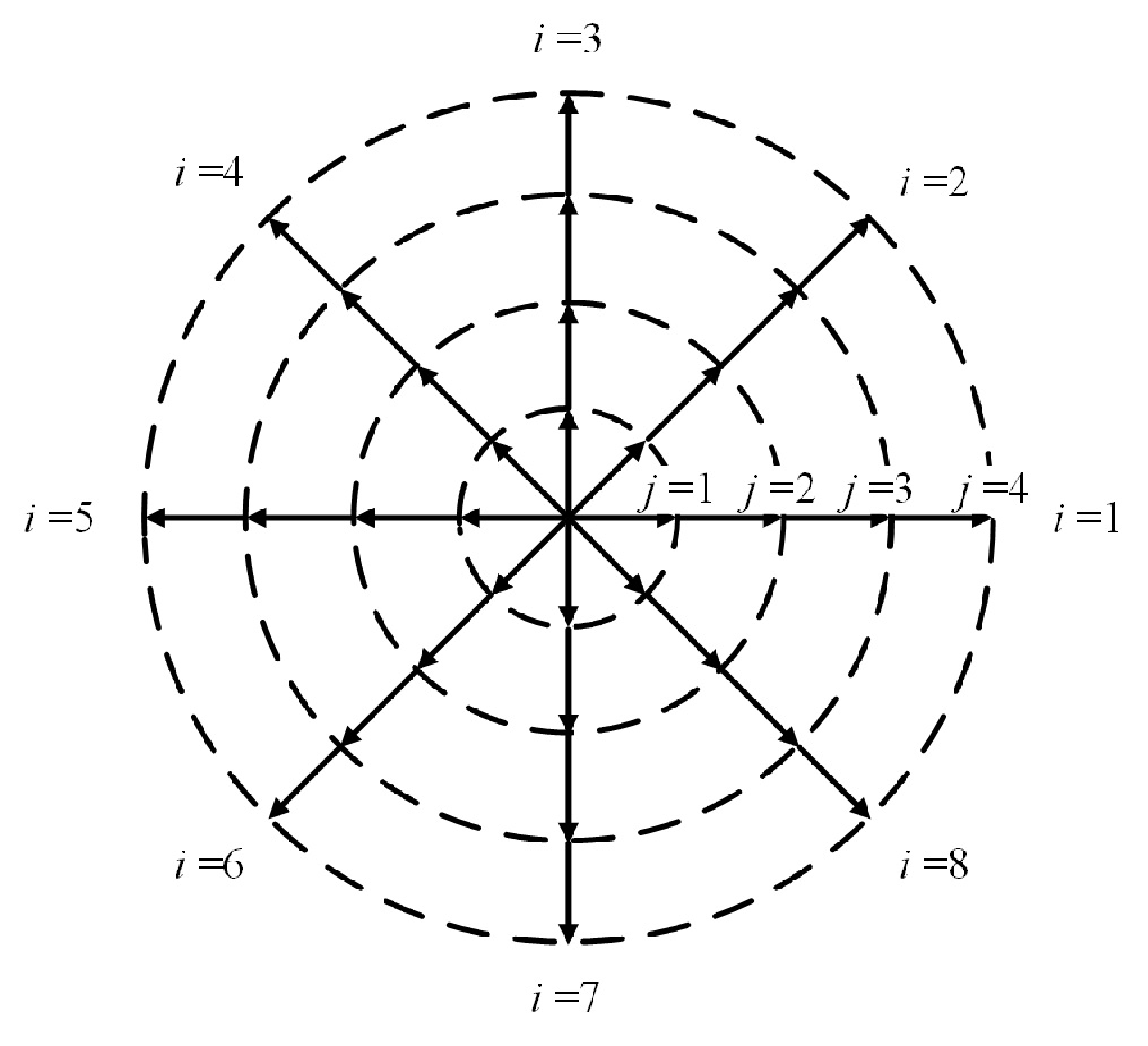,
width=0.8\textwidth,trim=10 10 10 10, clip }}
\caption{Schematic of discrete velocities.} \label{FigA1}
\end{figure}

Figure \ref{FigA1} shows the schematic of discrete velocities used in this work, which contains a zero velocity and four sets of 8 velocities in each direction. Each set of discrete velocities has the same velocity value ${{v}_{j}}$, and discrete velocity ${{\mathbf{v}}_{ji}}$ can be written as:
\begin{equation}
{{\mathbf{v}}_{ji}}={{v}_{j}}\left[ \cos (\frac{i-1}{4}\pi ),\sin(\frac{i-1}{4}\pi ) \right]\text{,}  \label{Eq.32}
\end{equation}%
where $j=0,1,2,3,4;i=1,2,\cdots ,8$, $j=0$ is the zero velocity. The equilibrium distribution function corresponding to the discrete velocity can be written as:
\begin{equation}
f_{ji}^{eq}=\rho {{F}_{j}}\left[ \left( 1-\frac{{{u}^{2}}}{2T}+\frac{{{u}^{4}}}{8{{T}^{2}}} \right)+\frac{{{\mathbf{v}}_{ji}}\cdot \mathbf{u}}{T}\left( 1-\frac{{{u}^{2}}}{2T} \right)+\frac{{{({{\mathbf{v}}_{ji}}\cdot \mathbf{u})}^{2}}}{2{{T}^{2}}}\left( 1-\frac{{{u}^{2}}}{2T} \right) \right.\left. +\frac{{{({{\mathbf{v}}_{ji}}\cdot \mathbf{u})}^{3}}}{6{{T}^{3}}}+\frac{{{({{\mathbf{v}}_{ji}}\cdot \mathbf{u})}^{4}}}{24{{T}^{4}}} \right]\text{,}  \label{Eq.33}
\end{equation}%
where ${{F}_{j}}$ is the weighting factor:
\begin{equation}
{{F}_{1}}=\frac{48{{T}^{4}}-6(v_{2}^{2}+v_{3}^{2}+v_{4}^{2}){{T}^{3}}+(v_{2}^{2}v_{3}^{2}+v_{2}^{2}v_{4}^{2}+v_{3}^{2}v_{4}^{2}){{T}^{2}}-\frac{1}{4}v_{2}^{2}v_{3}^{2}v_{4}^{2}T}{v_{1}^{2}(v_{1}^{2}-v_{2}^{2})(v_{1}^{2}-v_{3}^{2})(v_{1}^{2}-v_{4}^{2})}\text{,}  \label{Eq.34}
\end{equation}%
\begin{equation}
{{F}_{2}}=\frac{48{{T}^{4}}-6(v_{1}^{2}+v_{3}^{2}+v_{4}^{2}){{T}^{3}}+(v_{1}^{2}v_{3}^{2}+v_{1}^{2}v_{4}^{2}+v_{3}^{2}v_{4}^{2}){{T}^{2}}-\frac{1}{4}v_{1}^{2}v_{3}^{2}v_{4}^{2}T}{v_{2}^{2}(v_{2}^{2}-v_{1}^{2})(v_{2}^{2}-v_{3}^{2})(v_{2}^{2}-v_{4}^{2})}\text{,}  \label{Eq.35}
\end{equation}%
\begin{equation}
{{F}_{3}}=\frac{48{{T}^{4}}-6(v_{1}^{2}+v_{2}^{2}+v_{4}^{2}){{T}^{3}}+(v_{1}^{2}v_{2}^{2}+v_{1}^{2}v_{4}^{2}+v_{2}^{2}v_{4}^{2}){{T}^{2}}-\frac{1}{4}v_{1}^{2}v_{2}^{2}v_{4}^{2}T}{v_{3}^{2}(v_{3}^{2}-v_{1}^{2})(v_{3}^{2}-v_{2}^{2})(v_{1}^{2}-v_{4}^{2})}\text{,}  \label{Eq.36}
\end{equation}%
\begin{equation}
{{F}_{4}}=\frac{48{{T}^{4}}-6(v_{1}^{2}+v_{2}^{2}+v_{3}^{2}){{T}^{3}}+(v_{1}^{2}v_{2}^{2}+v_{1}^{2}v_{3}^{2}+v_{2}^{2}v_{3}^{2}){{T}^{2}}-\frac{1}{4}v_{1}^{2}v_{2}^{2}v_{3}^{2}T}{v_{4}^{2}(v_{4}^{2}-v_{1}^{2})(v_{4}^{2}-v_{2}^{2})(v_{4}^{2}-v_{3}^{2})}\text{,}  \label{Eq.37}
\end{equation}%
\begin{equation}
{{F}_{0}}=1-8({{F}_{1}}+{{F}_{2}}+{{F}_{3}}+{{F}_{4}})\text{.}  \label{Eq.38}
\end{equation}%

\section*{References}

\begin{thebibliography}{62}
\expandafter\ifx\csname natexlab\endcsname\relax\def\natexlab#1{#1}\fi
\expandafter\ifx\csname bibnamefont\endcsname\relax
  \def\bibnamefont#1{#1}\fi
\expandafter\ifx\csname bibfnamefont\endcsname\relax
  \def\bibfnamefont#1{#1}\fi
\expandafter\ifx\csname citenamefont\endcsname\relax
  \def\citenamefont#1{#1}\fi
\expandafter\ifx\csname url\endcsname\relax
  \def\url#1{\texttt{#1}}\fi
\expandafter\ifx\csname urlprefix\endcsname\relax\def\urlprefix{URL }\fi
\providecommand{\bibinfo}[2]{#2}
\providecommand{\eprint}[2][]{\url{#2}}

\bibitem[{\citenamefont{Rayleigh}(1882)}]{rayleigh1882investigation}
\bibinfo{author}{\bibfnamefont{L.}~\bibnamefont{Rayleigh}},
  \bibinfo{journal}{Proceedings of the London mathematical society}
  \textbf{\bibinfo{volume}{1}}, \bibinfo{pages}{170} (\bibinfo{year}{1882}).

\bibitem[{\citenamefont{Taylor}(1950)}]{taylor1950instability}
\bibinfo{author}{\bibfnamefont{G.}~\bibnamefont{Taylor}},
  \bibinfo{journal}{Proceedings of the Royal Society of London.Series
  A.Mathematical and Physical Sciences} \textbf{\bibinfo{volume}{201}},
  \bibinfo{pages}{192} (\bibinfo{year}{1950}).

\bibitem[{\citenamefont{Zhou}(2024)}]{zhou2024hydrodynamic}
\bibinfo{author}{\bibfnamefont{Y.}~\bibnamefont{Zhou}}, in
  \emph{\bibinfo{booktitle}{Hydrodynamic Instabilities and Turbulence:
  Rayleigh-Taylor, Richtmyer-Meshkov, and Kelvin-Helmholtz Mixing}}
  (\bibinfo{publisher}{Cambridge University Press},
  \bibinfo{address}{Cambridge}, \bibinfo{year}{2024}).

\bibitem[{\citenamefont{Zhou}(2017{\natexlab{a}})}]{ZHOU20171}
\bibinfo{author}{\bibfnamefont{Y.}~\bibnamefont{Zhou}},
  \bibinfo{journal}{Physics Reports} \textbf{\bibinfo{volume}{720-722}},
  \bibinfo{pages}{1} (\bibinfo{year}{2017}{\natexlab{a}}).

\bibitem[{\citenamefont{Zhou}(2017{\natexlab{b}})}]{zhou2017rayleigh}
\bibinfo{author}{\bibfnamefont{Y.}~\bibnamefont{Zhou}},
  \bibinfo{journal}{Physics Reports} \textbf{\bibinfo{volume}{723}},
  \bibinfo{pages}{1} (\bibinfo{year}{2017}{\natexlab{b}}).

\bibitem[{\citenamefont{Zhou et~al.}(2021)\citenamefont{Zhou, Williams,
  Ramaprabhu, Groom, Thornber, Hillier, Mostert, Rollin, Balachandar, Powell
  et~al.}}]{zhou2021rayleigh}
\bibinfo{author}{\bibfnamefont{Y.}~\bibnamefont{Zhou}},
  \bibinfo{author}{\bibfnamefont{R.~J.~R.} \bibnamefont{Williams}},
  \bibinfo{author}{\bibfnamefont{P.}~\bibnamefont{Ramaprabhu}},
  \bibinfo{author}{\bibfnamefont{M.}~\bibnamefont{Groom}},
  \bibinfo{author}{\bibfnamefont{B.}~\bibnamefont{Thornber}},
  \bibinfo{author}{\bibfnamefont{A.}~\bibnamefont{Hillier}},
  \bibinfo{author}{\bibfnamefont{W.}~\bibnamefont{Mostert}},
  \bibinfo{author}{\bibfnamefont{B.}~\bibnamefont{Rollin}},
  \bibinfo{author}{\bibfnamefont{S.}~\bibnamefont{Balachandar}},
  \bibinfo{author}{\bibfnamefont{P.~D.} \bibnamefont{Powell}},
  \bibnamefont{et~al.}, \bibinfo{journal}{Physica D}
  \textbf{\bibinfo{volume}{423}}, \bibinfo{pages}{132838}
  (\bibinfo{year}{2021}).

\bibitem[{\citenamefont{Bernstein and Book}(1983)}]{Bernstein1983Effect}
\bibinfo{author}{\bibfnamefont{I.~B.} \bibnamefont{Bernstein}}
  \bibnamefont{and} \bibinfo{author}{\bibfnamefont{D.~L.} \bibnamefont{Book}},
  \bibinfo{journal}{Physics of Fluids} \textbf{\bibinfo{volume}{26}},
  \bibinfo{pages}{453} (\bibinfo{year}{1983}).

\bibitem[{\citenamefont{Yang and Zhang}(1993)}]{yang1993general}
\bibinfo{author}{\bibfnamefont{Y.~M.} \bibnamefont{Yang}} \bibnamefont{and}
  \bibinfo{author}{\bibfnamefont{Q.}~\bibnamefont{Zhang}},
  \bibinfo{journal}{Physics of Fluids A: Fluid Dynamics}
  \textbf{\bibinfo{volume}{5}}, \bibinfo{pages}{1167} (\bibinfo{year}{1993}).

\bibitem[{\citenamefont{Blake}(1972)}]{blake1972fluid}
\bibinfo{author}{\bibfnamefont{G.~M.} \bibnamefont{Blake}},
  \bibinfo{journal}{Monthly Notices of the Royal Astronomical Society}
  \textbf{\bibinfo{volume}{156}}, \bibinfo{pages}{67} (\bibinfo{year}{1972}).

\bibitem[{\citenamefont{Baker}(1983)}]{baker1983compressible}
\bibinfo{author}{\bibfnamefont{L.}~\bibnamefont{Baker}}, \bibinfo{journal}{The
  Physics of Fluids} \textbf{\bibinfo{volume}{26}}, \bibinfo{pages}{950}
  (\bibinfo{year}{1983}).

\bibitem[{\citenamefont{Livescu}(2004)}]{livescu2004compressibility}
\bibinfo{author}{\bibfnamefont{D.}~\bibnamefont{Livescu}},
  \bibinfo{journal}{Physics of fluids} \textbf{\bibinfo{volume}{16}},
  \bibinfo{pages}{118} (\bibinfo{year}{2004}).

\bibitem[{\citenamefont{Xue and Ye}(2010)}]{xue2010destabilizing}
\bibinfo{author}{\bibfnamefont{C.}~\bibnamefont{Xue}} \bibnamefont{and}
  \bibinfo{author}{\bibfnamefont{W.~H.} \bibnamefont{Ye}},
  \bibinfo{journal}{Physics of Plasmas} \textbf{\bibinfo{volume}{17}}
  (\bibinfo{year}{2010}).

\bibitem[{\citenamefont{Lafay et~al.}(2007)\citenamefont{Lafay, Creurer, and
  Gauthier}}]{lafay2007compressibility}
\bibinfo{author}{\bibfnamefont{M.~A.} \bibnamefont{Lafay}},
  \bibinfo{author}{\bibfnamefont{B.~L.} \bibnamefont{Creurer}},
  \bibnamefont{and} \bibinfo{author}{\bibfnamefont{S.}~\bibnamefont{Gauthier}},
  \bibinfo{journal}{Europhysics Letters} \textbf{\bibinfo{volume}{79}},
  \bibinfo{pages}{64002} (\bibinfo{year}{2007}).

\bibitem[{\citenamefont{Reckinger et~al.}(2016)\citenamefont{Reckinger,
  Livescu, and Vasilyev}}]{reckinger2016comprehensive}
\bibinfo{author}{\bibfnamefont{S.~J.} \bibnamefont{Reckinger}},
  \bibinfo{author}{\bibfnamefont{D.}~\bibnamefont{Livescu}}, \bibnamefont{and}
  \bibinfo{author}{\bibfnamefont{O.~V.} \bibnamefont{Vasilyev}},
  \bibinfo{journal}{Journal of Computational Physics}
  \textbf{\bibinfo{volume}{313}}, \bibinfo{pages}{181} (\bibinfo{year}{2016}).

\bibitem[{\citenamefont{Gauthier}(2017)}]{gauthier2017compressible}
\bibinfo{author}{\bibfnamefont{S.}~\bibnamefont{Gauthier}},
  \bibinfo{journal}{Journal of Fluid Mechanics} \textbf{\bibinfo{volume}{830}},
  \bibinfo{pages}{211} (\bibinfo{year}{2017}).

\bibitem[{\citenamefont{Luo et~al.}(2020)\citenamefont{Luo, Wang, Xie, Wan, and
  Chen}}]{luo2020effects}
\bibinfo{author}{\bibfnamefont{T.~F.} \bibnamefont{Luo}},
  \bibinfo{author}{\bibfnamefont{J.~C.} \bibnamefont{Wang}},
  \bibinfo{author}{\bibfnamefont{C.~Y.} \bibnamefont{Xie}},
  \bibinfo{author}{\bibfnamefont{M.~P.} \bibnamefont{Wan}}, \bibnamefont{and}
  \bibinfo{author}{\bibfnamefont{S.~Y.} \bibnamefont{Chen}},
  \bibinfo{journal}{Physics of fluids} \textbf{\bibinfo{volume}{32}}
  (\bibinfo{year}{2020}).

\bibitem[{\citenamefont{Fu et~al.}(2022)\citenamefont{Fu, Zhao, Xu, Wang, Liu,
  Wang, and Lu}}]{fu2022nonlinear}
\bibinfo{author}{\bibfnamefont{C.~Q.} \bibnamefont{Fu}},
  \bibinfo{author}{\bibfnamefont{Z.~Y.} \bibnamefont{Zhao}},
  \bibinfo{author}{\bibfnamefont{X.}~\bibnamefont{Xu}},
  \bibinfo{author}{\bibfnamefont{P.}~\bibnamefont{Wang}},
  \bibinfo{author}{\bibfnamefont{N.~S.} \bibnamefont{Liu}},
  \bibinfo{author}{\bibfnamefont{Z.~H.} \bibnamefont{Wang}}, \bibnamefont{and}
  \bibinfo{author}{\bibfnamefont{X.~Y.} \bibnamefont{Lu}},
  \bibinfo{journal}{Physical Review Fluids} \textbf{\bibinfo{volume}{7}},
  \bibinfo{pages}{023902} (\bibinfo{year}{2022}).

\bibitem[{\citenamefont{Qin et~al.}(2001)\citenamefont{Qin, Zhang, and
  Li}}]{QinChengsen2001}
\bibinfo{author}{\bibfnamefont{C.~S.} \bibnamefont{Qin}},
  \bibinfo{author}{\bibfnamefont{F.~G.} \bibnamefont{Zhang}}, \bibnamefont{and}
  \bibinfo{author}{\bibfnamefont{Y.}~\bibnamefont{Li}},
  \bibinfo{journal}{EXPLOSION AND SHOCK WAVES} \textbf{\bibinfo{volume}{21}},
  \bibinfo{pages}{5} (\bibinfo{year}{2001}).

\bibitem[{\citenamefont{Qin and Wang}(2004)}]{QinChengsen2004}
\bibinfo{author}{\bibfnamefont{C.~S.} \bibnamefont{Qin}} \bibnamefont{and}
  \bibinfo{author}{\bibfnamefont{P.}~\bibnamefont{Wang}},
  \bibinfo{journal}{EXPLOSION AND SHOCK WAVES} \textbf{\bibinfo{volume}{24}},
  \bibinfo{pages}{1} (\bibinfo{year}{2004}).

\bibitem[{\citenamefont{Olson and Cook}(2007)}]{Olson2007}
\bibinfo{author}{\bibfnamefont{B.~J.} \bibnamefont{Olson}} \bibnamefont{and}
  \bibinfo{author}{\bibfnamefont{A.~W.} \bibnamefont{Cook}},
  \bibinfo{journal}{Physics of Fluids} \textbf{\bibinfo{volume}{19}},
  \bibinfo{pages}{128108} (\bibinfo{year}{2007}).

\bibitem[{\citenamefont{Reckinger et~al.}(2012)\citenamefont{Reckinger,
  Livescu, and Vasilyev}}]{reckinger2012simulations}
\bibinfo{author}{\bibfnamefont{S.~J.} \bibnamefont{Reckinger}},
  \bibinfo{author}{\bibfnamefont{D.}~\bibnamefont{Livescu}}, \bibnamefont{and}
  \bibinfo{author}{\bibfnamefont{O.~V.} \bibnamefont{Vasilyev}}, in
  \emph{\bibinfo{booktitle}{Seventh International Conference on Computational
  Fluid Dynamics (ICCFD7), Big Island, Hawaii}} (\bibinfo{year}{2012}).

\bibitem[{\citenamefont{Wieland et~al.}(2019)\citenamefont{Wieland, Hamlington,
  Reckinger, and Livescu}}]{wieland2019effects}
\bibinfo{author}{\bibfnamefont{S.~A.} \bibnamefont{Wieland}},
  \bibinfo{author}{\bibfnamefont{P.~E.} \bibnamefont{Hamlington}},
  \bibinfo{author}{\bibfnamefont{S.~J.} \bibnamefont{Reckinger}},
  \bibnamefont{and} \bibinfo{author}{\bibfnamefont{D.}~\bibnamefont{Livescu}},
  \bibinfo{journal}{Physical review fluids} \textbf{\bibinfo{volume}{4}},
  \bibinfo{pages}{093905} (\bibinfo{year}{2019}).

\bibitem[{\citenamefont{Fu et~al.}(2023)\citenamefont{Fu, Zhao, Wang, Liu, Wan,
  and Lu}}]{fu2023bubble}
\bibinfo{author}{\bibfnamefont{C.~G.} \bibnamefont{Fu}},
  \bibinfo{author}{\bibfnamefont{Z.~Y.} \bibnamefont{Zhao}},
  \bibinfo{author}{\bibfnamefont{P.}~\bibnamefont{Wang}},
  \bibinfo{author}{\bibfnamefont{N.~S.} \bibnamefont{Liu}},
  \bibinfo{author}{\bibfnamefont{Z.~H.} \bibnamefont{Wan}}, \bibnamefont{and}
  \bibinfo{author}{\bibfnamefont{X.~Y.} \bibnamefont{Lu}},
  \bibinfo{journal}{Journal of Fluid Mechanics} \textbf{\bibinfo{volume}{954}},
  \bibinfo{pages}{A16} (\bibinfo{year}{2023}).

\bibitem[{\citenamefont{Chen et~al.}(2020)\citenamefont{Chen, Xu, Zhang, and
  Zeng}}]{chen2020morphological}
\bibinfo{author}{\bibfnamefont{F.}~\bibnamefont{Chen}},
  \bibinfo{author}{\bibfnamefont{A.~G.} \bibnamefont{Xu}},
  \bibinfo{author}{\bibfnamefont{Y.~D.} \bibnamefont{Zhang}}, \bibnamefont{and}
  \bibinfo{author}{\bibfnamefont{Q.~K.} \bibnamefont{Zeng}},
  \bibinfo{journal}{Physics of Fluids} \textbf{\bibinfo{volume}{32}},
  \bibinfo{pages}{104111} (\bibinfo{year}{2020}).

\bibitem[{\citenamefont{Xu and Zhang}(2022)}]{Xu2022-Complex}
\bibinfo{author}{\bibfnamefont{A.~G.} \bibnamefont{Xu}} \bibnamefont{and}
  \bibinfo{author}{\bibfnamefont{Y.~D.} \bibnamefont{Zhang}}, in
  \emph{\bibinfo{booktitle}{Complex Media Kinetics}}
  (\bibinfo{publisher}{Science Press}, \bibinfo{address}{BeiJing},
  \bibinfo{year}{2022}).

\bibitem[{\citenamefont{Xu et~al.}(2024)\citenamefont{Xu, Zhang, and
  Gan}}]{xu2024advances}
\bibinfo{author}{\bibfnamefont{A.~G.} \bibnamefont{Xu}},
  \bibinfo{author}{\bibfnamefont{J.~D.} \bibnamefont{Zhang}}, \bibnamefont{and}
  \bibinfo{author}{\bibfnamefont{Y.~B.} \bibnamefont{Gan}},
  \bibinfo{journal}{Frontiers of Physics} \textbf{\bibinfo{volume}{19}},
  \bibinfo{pages}{42500} (\bibinfo{year}{2024}).

\bibitem[{\citenamefont{Succi}(2001)}]{succi2001lattice}
\bibinfo{author}{\bibfnamefont{S.}~\bibnamefont{Succi}}, in
  \emph{\bibinfo{booktitle}{The lattice Boltzmann equation: for fluid dynamics
  and beyond}} (\bibinfo{publisher}{Oxford university press},
  \bibinfo{address}{Oxford}, \bibinfo{year}{2001}).

\bibitem[{\citenamefont{Osborn et~al.}(1995)\citenamefont{Osborn, Orlandini,
  Swift, Yeomans, and Banavar}}]{osborn1995lattice}
\bibinfo{author}{\bibfnamefont{W.}~\bibnamefont{Osborn}},
  \bibinfo{author}{\bibfnamefont{E.}~\bibnamefont{Orlandini}},
  \bibinfo{author}{\bibfnamefont{M.~R.} \bibnamefont{Swift}},
  \bibinfo{author}{\bibfnamefont{J.}~\bibnamefont{Yeomans}}, \bibnamefont{and}
  \bibinfo{author}{\bibfnamefont{J.~R.} \bibnamefont{Banavar}},
  \bibinfo{journal}{Physical review letters} \textbf{\bibinfo{volume}{75}},
  \bibinfo{pages}{4031} (\bibinfo{year}{1995}).

\bibitem[{\citenamefont{Swift et~al.}(1995)\citenamefont{Swift, Osborn, and
  Yeomans}}]{swift1995lattice}
\bibinfo{author}{\bibfnamefont{M.~R.} \bibnamefont{Swift}},
  \bibinfo{author}{\bibfnamefont{W.}~\bibnamefont{Osborn}}, \bibnamefont{and}
  \bibinfo{author}{\bibfnamefont{J.}~\bibnamefont{Yeomans}},
  \bibinfo{journal}{Physical review letters} \textbf{\bibinfo{volume}{75}},
  \bibinfo{pages}{830} (\bibinfo{year}{1995}).

\bibitem[{\citenamefont{Liang et~al.}(2014)\citenamefont{Liang, Shi, Guo, and
  Chai}}]{liang2014phase}
\bibinfo{author}{\bibfnamefont{H.}~\bibnamefont{Liang}},
  \bibinfo{author}{\bibfnamefont{B.~C.} \bibnamefont{Shi}},
  \bibinfo{author}{\bibfnamefont{Z.~L.} \bibnamefont{Guo}}, \bibnamefont{and}
  \bibinfo{author}{\bibfnamefont{Z.~H.} \bibnamefont{Chai}},
  \bibinfo{journal}{Physical review E} \textbf{\bibinfo{volume}{89}},
  \bibinfo{pages}{053320} (\bibinfo{year}{2014}).

\bibitem[{\citenamefont{Liang et~al.}(2016)\citenamefont{Liang, Li, Shi, and
  Chai}}]{liang2016lattice}
\bibinfo{author}{\bibfnamefont{H.}~\bibnamefont{Liang}},
  \bibinfo{author}{\bibfnamefont{Q.~X.} \bibnamefont{Li}},
  \bibinfo{author}{\bibfnamefont{B.~C.} \bibnamefont{Shi}}, \bibnamefont{and}
  \bibinfo{author}{\bibfnamefont{Z.~H.} \bibnamefont{Chai}},
  \bibinfo{journal}{Physical review E} \textbf{\bibinfo{volume}{93}},
  \bibinfo{pages}{033113} (\bibinfo{year}{2016}).

\bibitem[{\citenamefont{Liang et~al.}(2021)\citenamefont{Liang, Xia, and
  Huang}}]{liang2021late}
\bibinfo{author}{\bibfnamefont{H.}~\bibnamefont{Liang}},
  \bibinfo{author}{\bibfnamefont{Z.~H.} \bibnamefont{Xia}}, \bibnamefont{and}
  \bibinfo{author}{\bibfnamefont{H.~W.} \bibnamefont{Huang}},
  \bibinfo{journal}{Physics of Fluids} \textbf{\bibinfo{volume}{33}},
  \bibinfo{pages}{082103} (\bibinfo{year}{2021}).

\bibitem[{\citenamefont{Chen et~al.}(2018)\citenamefont{Chen, Xu, and
  Zhang}}]{chen2018collaboration}
\bibinfo{author}{\bibfnamefont{F.}~\bibnamefont{Chen}},
  \bibinfo{author}{\bibfnamefont{A.~G.} \bibnamefont{Xu}}, \bibnamefont{and}
  \bibinfo{author}{\bibfnamefont{G.~C.} \bibnamefont{Zhang}},
  \bibinfo{journal}{Physics of Fluids} \textbf{\bibinfo{volume}{30}},
  \bibinfo{pages}{102105} (\bibinfo{year}{2018}).

\bibitem[{\citenamefont{Gan et~al.}(2019)\citenamefont{Gan, Xu, Zhang, Lin,
  Lai, and Liu}}]{gan2019nonequilibrium}
\bibinfo{author}{\bibfnamefont{Y.~B.} \bibnamefont{Gan}},
  \bibinfo{author}{\bibfnamefont{A.~G.} \bibnamefont{Xu}},
  \bibinfo{author}{\bibfnamefont{G.~C.} \bibnamefont{Zhang}},
  \bibinfo{author}{\bibfnamefont{C.~D.} \bibnamefont{Lin}},
  \bibinfo{author}{\bibfnamefont{H.~L.} \bibnamefont{Lai}}, \bibnamefont{and}
  \bibinfo{author}{\bibfnamefont{Z.~P.} \bibnamefont{Liu}},
  \bibinfo{journal}{Frontiers of Physics} \textbf{\bibinfo{volume}{14}},
  \bibinfo{pages}{1} (\bibinfo{year}{2019}).

\bibitem[{\citenamefont{Lai et~al.}(2016)\citenamefont{Lai, Xu, Zhang, Gan,
  Ying, and Succi}}]{lai2016nonequilibrium}
\bibinfo{author}{\bibfnamefont{H.~L.} \bibnamefont{Lai}},
  \bibinfo{author}{\bibfnamefont{A.~G.} \bibnamefont{Xu}},
  \bibinfo{author}{\bibfnamefont{G.~C.} \bibnamefont{Zhang}},
  \bibinfo{author}{\bibfnamefont{Y.~B.} \bibnamefont{Gan}},
  \bibinfo{author}{\bibfnamefont{Y.~J.} \bibnamefont{Ying}}, \bibnamefont{and}
  \bibinfo{author}{\bibfnamefont{S.}~\bibnamefont{Succi}},
  \bibinfo{journal}{Physical Review E} \textbf{\bibinfo{volume}{94}},
  \bibinfo{pages}{023106} (\bibinfo{year}{2016}).

\bibitem[{\citenamefont{Lin et~al.}(2017)\citenamefont{Lin, Xu, Zhang, Luo, and
  Li}}]{lin2017discrete}
\bibinfo{author}{\bibfnamefont{C.~D.} \bibnamefont{Lin}},
  \bibinfo{author}{\bibfnamefont{A.~G.} \bibnamefont{Xu}},
  \bibinfo{author}{\bibfnamefont{G.~C.} \bibnamefont{Zhang}},
  \bibinfo{author}{\bibfnamefont{K.~H.} \bibnamefont{Luo}}, \bibnamefont{and}
  \bibinfo{author}{\bibfnamefont{Y.~J.} \bibnamefont{Li}},
  \bibinfo{journal}{Physical Review E} \textbf{\bibinfo{volume}{96}},
  \bibinfo{pages}{053305} (\bibinfo{year}{2017}).

\bibitem[{\citenamefont{Zhang et~al.}(2021)\citenamefont{Zhang, Xu, Zhang, Li,
  Lai, and Hu}}]{zhang2021delineation}
\bibinfo{author}{\bibfnamefont{G.}~\bibnamefont{Zhang}},
  \bibinfo{author}{\bibfnamefont{A.~G.} \bibnamefont{Xu}},
  \bibinfo{author}{\bibfnamefont{D.~J.} \bibnamefont{Zhang}},
  \bibinfo{author}{\bibfnamefont{Y.~J.} \bibnamefont{Li}},
  \bibinfo{author}{\bibfnamefont{H.~L.} \bibnamefont{Lai}}, \bibnamefont{and}
  \bibinfo{author}{\bibfnamefont{X.~M.} \bibnamefont{Hu}},
  \bibinfo{journal}{Physics of Fluids} \textbf{\bibinfo{volume}{33}},
  \bibinfo{pages}{076105} (\bibinfo{year}{2021}).

\bibitem[{\citenamefont{Li et~al.}(2022)\citenamefont{Li, Xu, Zhang, and
  Shan}}]{li2022rayleigh}
\bibinfo{author}{\bibfnamefont{H.~W.} \bibnamefont{Li}},
  \bibinfo{author}{\bibfnamefont{A.~G.} \bibnamefont{Xu}},
  \bibinfo{author}{\bibfnamefont{G.}~\bibnamefont{Zhang}}, \bibnamefont{and}
  \bibinfo{author}{\bibfnamefont{Y.~M.} \bibnamefont{Shan}},
  \bibinfo{journal}{Communications in Theoretical Physics}
  \textbf{\bibinfo{volume}{74}}, \bibinfo{pages}{115601}
  (\bibinfo{year}{2022}).

\bibitem[{\citenamefont{Miles}(1966)}]{miles1966taylor}
\bibinfo{author}{\bibfnamefont{J.~W.} \bibnamefont{Miles}},
  \bibinfo{type}{Tech. Rep.}, \bibinfo{institution}{General Dynamics San Diego
  Ca General Atomic Div} (\bibinfo{year}{1966}).

\bibitem[{\citenamefont{Piriz et~al.}(2009)\citenamefont{Piriz, Cela, and
  Tahir}}]{piriz2009linear}
\bibinfo{author}{\bibfnamefont{A.~R.} \bibnamefont{Piriz}},
  \bibinfo{author}{\bibfnamefont{J.~J.~L.} \bibnamefont{Cela}},
  \bibnamefont{and} \bibinfo{author}{\bibfnamefont{N.~A.} \bibnamefont{Tahir}},
  \bibinfo{journal}{Physical Review E} \textbf{\bibinfo{volume}{80}},
  \bibinfo{pages}{046305} (\bibinfo{year}{2009}).

\bibitem[{\citenamefont{Li et~al.}(2021)\citenamefont{Li, Peng, Gu, and
  He}}]{bi2020experimental}
\bibinfo{author}{\bibfnamefont{B.~Y.} \bibnamefont{Li}},
  \bibinfo{author}{\bibfnamefont{J.~X.} \bibnamefont{Peng}},
  \bibinfo{author}{\bibfnamefont{Y.}~\bibnamefont{Gu}}, \bibnamefont{and}
  \bibinfo{author}{\bibfnamefont{L.~H.} \bibnamefont{He}},
  \bibinfo{journal}{Acta Physica Sinica} \textbf{\bibinfo{volume}{70}},
  \bibinfo{pages}{114701} (\bibinfo{year}{2021}).

\bibitem[{\citenamefont{Xu et~al.}(2015)\citenamefont{Xu, Zhang, and
  Ying}}]{2015Progess}
\bibinfo{author}{\bibfnamefont{A.~G.} \bibnamefont{Xu}},
  \bibinfo{author}{\bibfnamefont{G.~C.} \bibnamefont{Zhang}}, \bibnamefont{and}
  \bibinfo{author}{\bibfnamefont{Y.~J.} \bibnamefont{Ying}},
  \bibinfo{journal}{Acta Physica Sinica} \textbf{\bibinfo{volume}{64}}
  (\bibinfo{year}{2015}).

\bibitem[{\citenamefont{Xu et~al.}(2018)\citenamefont{Xu, Zhang, and
  Zhang}}]{Xu2018-Chap2}
\bibinfo{author}{\bibfnamefont{A.~G.} \bibnamefont{Xu}},
  \bibinfo{author}{\bibfnamefont{G.~C.} \bibnamefont{Zhang}}, \bibnamefont{and}
  \bibinfo{author}{\bibfnamefont{Y.~D.} \bibnamefont{Zhang}}, in
  \emph{\bibinfo{booktitle}{Kinetic Theory}}, edited by
  \bibinfo{editor}{\bibfnamefont{G.}~\bibnamefont{Kyzas}} \bibnamefont{and}
  \bibinfo{editor}{\bibfnamefont{A.}~\bibnamefont{Mitropoulos}}
  (\bibinfo{publisher}{InTech}, \bibinfo{address}{Rijeka},
  \bibinfo{year}{2018}), chap.~\bibinfo{chapter}{02}.

\bibitem[{\citenamefont{Xu et~al.}(2021{\natexlab{a}})\citenamefont{Xu, Chen,
  Song, Chen, and Chen}}]{xu2021progress}
\bibinfo{author}{\bibfnamefont{A.~G.} \bibnamefont{Xu}},
  \bibinfo{author}{\bibfnamefont{J.}~\bibnamefont{Chen}},
  \bibinfo{author}{\bibfnamefont{J.~H.} \bibnamefont{Song}},
  \bibinfo{author}{\bibfnamefont{D.~W.} \bibnamefont{Chen}}, \bibnamefont{and}
  \bibinfo{author}{\bibfnamefont{Z.~H.} \bibnamefont{Chen}},
  \bibinfo{journal}{Acta Aerodyn.Sin} \textbf{\bibinfo{volume}{39}},
  \bibinfo{pages}{138} (\bibinfo{year}{2021}{\natexlab{a}}).

\bibitem[{\citenamefont{Xu et~al.}(2021{\natexlab{b}})\citenamefont{Xu, Shan,
  Chen, Gan, and Lin}}]{xu2021Progressofmesoscale}
\bibinfo{author}{\bibfnamefont{A.~G.} \bibnamefont{Xu}},
  \bibinfo{author}{\bibfnamefont{Y.~M.} \bibnamefont{Shan}},
  \bibinfo{author}{\bibfnamefont{F.}~\bibnamefont{Chen}},
  \bibinfo{author}{\bibfnamefont{Y.~B.} \bibnamefont{Gan}}, \bibnamefont{and}
  \bibinfo{author}{\bibfnamefont{C.~D.} \bibnamefont{Lin}},
  \bibinfo{journal}{Acta Aeronauticaet Astronautica Sinica}
  \textbf{\bibinfo{volume}{42}}, \bibinfo{pages}{46}
  (\bibinfo{year}{2021}{\natexlab{b}}).

\bibitem[{\citenamefont{Xu et~al.}(2021{\natexlab{c}})\citenamefont{Xu, Song,
  Chen, Xie, and Ying}}]{xu2021modeling}
\bibinfo{author}{\bibfnamefont{A.~G.} \bibnamefont{Xu}},
  \bibinfo{author}{\bibfnamefont{J.~H.} \bibnamefont{Song}},
  \bibinfo{author}{\bibfnamefont{F.}~\bibnamefont{Chen}},
  \bibinfo{author}{\bibfnamefont{K.}~\bibnamefont{Xie}}, \bibnamefont{and}
  \bibinfo{author}{\bibfnamefont{Y.~J.} \bibnamefont{Ying}},
  \bibinfo{journal}{Chinese Journal of Computational Physics}
  \textbf{\bibinfo{volume}{38}}, \bibinfo{pages}{631}
  (\bibinfo{year}{2021}{\natexlab{c}}).

\bibitem[{\citenamefont{Gan et~al.}(2022{\natexlab{a}})\citenamefont{Gan, Xu,
  Lai, Li, Sun, and Succi}}]{GanXuLai2022}
\bibinfo{author}{\bibfnamefont{Y.~B.} \bibnamefont{Gan}},
  \bibinfo{author}{\bibfnamefont{A.~G.} \bibnamefont{Xu}},
  \bibinfo{author}{\bibfnamefont{H.~L.} \bibnamefont{Lai}},
  \bibinfo{author}{\bibfnamefont{W.}~\bibnamefont{Li}},
  \bibinfo{author}{\bibfnamefont{G.~L.} \bibnamefont{Sun}}, \bibnamefont{and}
  \bibinfo{author}{\bibfnamefont{S.}~\bibnamefont{Succi}},
  \bibinfo{journal}{Journal of Fluid Mechanics} \textbf{\bibinfo{volume}{951}},
  \bibinfo{pages}{A8} (\bibinfo{year}{2022}{\natexlab{a}}).

\bibitem[{\citenamefont{Zhang et~al.}(2023)\citenamefont{Zhang, Xu, Gan, Zhang,
  Song, and Li}}]{Zhang2023SBI}
\bibinfo{author}{\bibfnamefont{D.~J.} \bibnamefont{Zhang}},
  \bibinfo{author}{\bibfnamefont{A.~G.} \bibnamefont{Xu}},
  \bibinfo{author}{\bibfnamefont{Y.~B.} \bibnamefont{Gan}},
  \bibinfo{author}{\bibfnamefont{Y.~D.} \bibnamefont{Zhang}},
  \bibinfo{author}{\bibfnamefont{J.~H.} \bibnamefont{Song}}, \bibnamefont{and}
  \bibinfo{author}{\bibfnamefont{Y.~J.} \bibnamefont{Li}},
  \bibinfo{journal}{Physics of Fluids} \textbf{\bibinfo{volume}{35}},
  \bibinfo{pages}{106113} (\bibinfo{year}{2023}).

\bibitem[{\citenamefont{Wu and Liu}(2011)}]{LiuZongh2011}
\bibinfo{author}{\bibfnamefont{X.}~\bibnamefont{Wu}} \bibnamefont{and}
  \bibinfo{author}{\bibfnamefont{Z.~H.} \bibnamefont{Liu}},
  \bibinfo{journal}{Complex systems and complexity scienc}
  \textbf{\bibinfo{volume}{8}}, \bibinfo{pages}{0039} (\bibinfo{year}{2011}).

\bibitem[{\citenamefont{Wang et~al.}(2010)\citenamefont{Wang, He, and
  Hu}}]{Wang2010PRL}
\bibinfo{author}{\bibfnamefont{L.}~\bibnamefont{Wang}},
  \bibinfo{author}{\bibfnamefont{D.~H.} \bibnamefont{He}}, \bibnamefont{and}
  \bibinfo{author}{\bibfnamefont{B.~B.} \bibnamefont{Hu}},
  \bibinfo{journal}{Phys. Rev. Lett} \textbf{\bibinfo{volume}{105}},
  \bibinfo{pages}{160601} (\bibinfo{year}{2010}).

\bibitem[{\citenamefont{Gonnella et~al.}(2007)\citenamefont{Gonnella, Lamura,
  and Sofonea}}]{gonnella2007lattice}
\bibinfo{author}{\bibfnamefont{G.}~\bibnamefont{Gonnella}},
  \bibinfo{author}{\bibfnamefont{A.}~\bibnamefont{Lamura}}, \bibnamefont{and}
  \bibinfo{author}{\bibfnamefont{V.}~\bibnamefont{Sofonea}},
  \bibinfo{journal}{Physical Review E} \textbf{\bibinfo{volume}{76}},
  \bibinfo{pages}{036703} (\bibinfo{year}{2007}).

\bibitem[{\citenamefont{Gan et~al.}(2022{\natexlab{b}})\citenamefont{Gan, Xu,
  Lai, Li, Sun, and Succi}}]{Gan2022JFM}
\bibinfo{author}{\bibfnamefont{Y.~B.} \bibnamefont{Gan}},
  \bibinfo{author}{\bibfnamefont{A.~G.} \bibnamefont{Xu}},
  \bibinfo{author}{\bibfnamefont{H.~L.} \bibnamefont{Lai}},
  \bibinfo{author}{\bibfnamefont{W.}~\bibnamefont{Li}},
  \bibinfo{author}{\bibfnamefont{G.~L.} \bibnamefont{Sun}}, \bibnamefont{and}
  \bibinfo{author}{\bibfnamefont{S.}~\bibnamefont{Succi}}, \bibinfo{journal}{J.
  Fluid Mech.} \textbf{\bibinfo{volume}{951}}, \bibinfo{pages}{A8}
  (\bibinfo{year}{2022}{\natexlab{b}}).

\bibitem[{\citenamefont{Chen et~al.}(2022)\citenamefont{Chen, Xu, Chen, Zhang,
  and Chena}}]{chen2022discrete}
\bibinfo{author}{\bibfnamefont{J.}~\bibnamefont{Chen}},
  \bibinfo{author}{\bibfnamefont{A.~G.} \bibnamefont{Xu}},
  \bibinfo{author}{\bibfnamefont{D.~W.} \bibnamefont{Chen}},
  \bibinfo{author}{\bibfnamefont{Y.~D.} \bibnamefont{Zhang}}, \bibnamefont{and}
  \bibinfo{author}{\bibfnamefont{Z.~H.} \bibnamefont{Chena}},
  \bibinfo{journal}{Physical Review E} \textbf{\bibinfo{volume}{106}},
  \bibinfo{pages}{015102} (\bibinfo{year}{2022}).

\bibitem[{\citenamefont{Song et~al.}(2024)\citenamefont{Song, Xu, Miao, Chen,
  Liu, Wang, and Hou}}]{Song2024}
\bibinfo{author}{\bibfnamefont{J.~H.} \bibnamefont{Song}},
  \bibinfo{author}{\bibfnamefont{A.~G.} \bibnamefont{Xu}},
  \bibinfo{author}{\bibfnamefont{L.}~\bibnamefont{Miao}},
  \bibinfo{author}{\bibfnamefont{F.}~\bibnamefont{Chen}},
  \bibinfo{author}{\bibfnamefont{Z.~P.} \bibnamefont{Liu}},
  \bibinfo{author}{\bibfnamefont{L.~F.} \bibnamefont{Wang}}, \bibnamefont{and}
  \bibinfo{author}{\bibfnamefont{X.}~\bibnamefont{Hou}},
  \bibinfo{journal}{Physics of Fluids} \textbf{\bibinfo{volume}{36}},
  \bibinfo{pages}{016107} (\bibinfo{year}{2024}).

\bibitem[{\citenamefont{Cai et~al.}(2021)\citenamefont{Cai, Yan, Yao, and
  Zhu}}]{2021Hybrid}
\bibinfo{author}{\bibfnamefont{H.~B.} \bibnamefont{Cai}},
  \bibinfo{author}{\bibfnamefont{X.~X.} \bibnamefont{Yan}},
  \bibinfo{author}{\bibfnamefont{P.~L.} \bibnamefont{Yao}}, \bibnamefont{and}
  \bibinfo{author}{\bibfnamefont{S.~P.} \bibnamefont{Zhu}},
  \bibinfo{journal}{Matter and Radiation at Extremes}
  \textbf{\bibinfo{volume}{6}}, \bibinfo{pages}{9} (\bibinfo{year}{2021}).

\bibitem[{\citenamefont{Shan et~al.}(2018)\citenamefont{Shan, Cai, Zhang, Tang,
  Zhang, Song, Bi, Ge, Chen, Liu et~al.}}]{2018Experimental}
\bibinfo{author}{\bibfnamefont{L.~Q.} \bibnamefont{Shan}},
  \bibinfo{author}{\bibfnamefont{H.~B.} \bibnamefont{Cai}},
  \bibinfo{author}{\bibfnamefont{W.~S.} \bibnamefont{Zhang}},
  \bibinfo{author}{\bibfnamefont{Q.}~\bibnamefont{Tang}},
  \bibinfo{author}{\bibfnamefont{F.}~\bibnamefont{Zhang}},
  \bibinfo{author}{\bibfnamefont{Z.~F.} \bibnamefont{Song}},
  \bibinfo{author}{\bibfnamefont{B.}~\bibnamefont{Bi}},
  \bibinfo{author}{\bibfnamefont{F.~J.} \bibnamefont{Ge}},
  \bibinfo{author}{\bibfnamefont{J.~B.} \bibnamefont{Chen}},
  \bibinfo{author}{\bibfnamefont{D.~X.} \bibnamefont{Liu}},
  \bibnamefont{et~al.}, \bibinfo{journal}{Physical review letters}
  \textbf{\bibinfo{volume}{120}}, \bibinfo{pages}{195001}
  (\bibinfo{year}{2018}).

\bibitem[{\citenamefont{Qiu et~al.}(2024)\citenamefont{Qiu, Yan, Bao, You, and
  J}}]{e26030200}
\bibinfo{author}{\bibfnamefont{R.~F.} \bibnamefont{Qiu}},
  \bibinfo{author}{\bibfnamefont{X.~Y.} \bibnamefont{Yan}},
  \bibinfo{author}{\bibfnamefont{Y.}~\bibnamefont{Bao}},
  \bibinfo{author}{\bibfnamefont{Y.~C.} \bibnamefont{You}}, \bibnamefont{and}
  \bibinfo{author}{\bibfnamefont{H.}~\bibnamefont{J}},
  \bibinfo{journal}{Entropy} \textbf{\bibinfo{volume}{26}},
  \bibinfo{pages}{200} (\bibinfo{year}{2024}).

\bibitem[{\citenamefont{Onuki}(2005)}]{onuki2005dynamic}
\bibinfo{author}{\bibfnamefont{A.}~\bibnamefont{Onuki}},
  \bibinfo{journal}{Physical review letters} \textbf{\bibinfo{volume}{94}},
  \bibinfo{pages}{054501} (\bibinfo{year}{2005}).

\bibitem[{\citenamefont{Bian et~al.}(2020)\citenamefont{Bian, Aluie, Zhao,
  Zhang, and Livescu}}]{bian2020revisiting}
\bibinfo{author}{\bibfnamefont{X.}~\bibnamefont{Bian}},
  \bibinfo{author}{\bibfnamefont{H.}~\bibnamefont{Aluie}},
  \bibinfo{author}{\bibfnamefont{D.~X.} \bibnamefont{Zhao}},
  \bibinfo{author}{\bibfnamefont{H.~S.} \bibnamefont{Zhang}}, \bibnamefont{and}
  \bibinfo{author}{\bibfnamefont{D.}~\bibnamefont{Livescu}},
  \bibinfo{journal}{Physica D: Nonlinear Phenomena}
  \textbf{\bibinfo{volume}{403}}, \bibinfo{pages}{132250}
  (\bibinfo{year}{2020}).

\bibitem[{\citenamefont{Zhang et~al.}(2019)\citenamefont{Zhang, Xu, Zhang, Gan,
  Chen, and Succi}}]{zhang2019entropy}
\bibinfo{author}{\bibfnamefont{Y.~D.} \bibnamefont{Zhang}},
  \bibinfo{author}{\bibfnamefont{A.~G.} \bibnamefont{Xu}},
  \bibinfo{author}{\bibfnamefont{G.~C.} \bibnamefont{Zhang}},
  \bibinfo{author}{\bibfnamefont{Y.~B.} \bibnamefont{Gan}},
  \bibinfo{author}{\bibfnamefont{Z.~H.} \bibnamefont{Chen}}, \bibnamefont{and}
  \bibinfo{author}{\bibfnamefont{S.}~\bibnamefont{Succi}},
  \bibinfo{journal}{Soft matter} \textbf{\bibinfo{volume}{15}},
  \bibinfo{pages}{2245} (\bibinfo{year}{2019}).

\bibitem[{\citenamefont{Tiribocchi et~al.}(2009)\citenamefont{Tiribocchi,
  Stella, Gonnella, and Lamura}}]{tiribocchi2009hybrid}
\bibinfo{author}{\bibfnamefont{A.}~\bibnamefont{Tiribocchi}},
  \bibinfo{author}{\bibfnamefont{N.}~\bibnamefont{Stella}},
  \bibinfo{author}{\bibfnamefont{G.}~\bibnamefont{Gonnella}}, \bibnamefont{and}
  \bibinfo{author}{\bibfnamefont{A.}~\bibnamefont{Lamura}},
  \bibinfo{journal}{Physical Review E} \textbf{\bibinfo{volume}{80}},
  \bibinfo{pages}{026701} (\bibinfo{year}{2009}).

\bibitem[{\citenamefont{Wang et~al.}(2013)\citenamefont{Wang, Yang, Shi, Xiao,
  He, and Li}}]{wang2013cascade}
\bibinfo{author}{\bibfnamefont{J.~C.} \bibnamefont{Wang}},
  \bibinfo{author}{\bibfnamefont{Y.~T.} \bibnamefont{Yang}},
  \bibinfo{author}{\bibfnamefont{Y.~P.} \bibnamefont{Shi}},
  \bibinfo{author}{\bibfnamefont{Z.~L.} \bibnamefont{Xiao}},
  \bibinfo{author}{\bibfnamefont{X.~T.} \bibnamefont{He}}, \bibnamefont{and}
  \bibinfo{author}{\bibfnamefont{Y.~J.} \bibnamefont{Li}},
  \bibinfo{journal}{Physical review letters} \textbf{\bibinfo{volume}{110}},
  \bibinfo{pages}{214505} (\bibinfo{year}{2013}).
\end{thebibliography}

\end{document}